\documentclass[longauth]{aa} 
\usepackage{amsmath,amsfonts,amssymb}
\usepackage[dvipsnames]{xcolor}
\usepackage{color}
\usepackage{url}\usepackage[varg]{txfonts}
\usepackage{graphicx}
\usepackage{enumitem} 
\usepackage{array}
\usepackage{dblfloatfix}
\usepackage{multicol}
\usepackage[normalem]{ulem} 
\usepackage{float}  % load float package first!
\usepackage{caption} 
\usepackage{mhchem}
\usepackage{siunitx} % \usepackage{siunitx-v2} to be changed when local

\definecolor{mulberry}{rgb}{0.77, 0.29, 0.55}

\usepackage{natbib,twoopt}
\usepackage[breaklinks=true]{hyperref} %% to avoid \citeads line fills
\bibpunct{(}{)}{;}{a}{}{,}             %% natbib format for A&A and ApJ
\makeatletter
  \newcommandtwoopt{\citeads}[3][][]{\href{http://adsabs.harvard.edu/abs/#3}%
    {\def\hyper@linkstart##1##2{}%
     \let\hyper@linkend\@empty\citealp[#1][#2]{#3}}}
  \newcommandtwoopt{\citepads}[3][][]{\href{http://adsabs.harvard.edu/abs/#3}%
    {\def\hyper@linkstart##1##2{}%
     \let\hyper@linkend\@empty\citep[#1][#2]{#3}}}
  \newcommandtwoopt{\citetads}[3][][]{\href{http://adsabs.harvard.edu/abs/#3}%
    {\def\hyper@linkstart##1##2{}%
     \let\hyper@linkend\@empty\citet[#1][#2]{#3}}}
  \newcommandtwoopt{\citeyearads}[3][][]%
    {\href{http://adsabs.harvard.edu/abs/#3}
    {\def\hyper@linkstart##1##2{}%
     \let\hyper@linkend\@empty\citeyear[#1][#2]{#3}}}
\makeatother
\usepackage{color}
\hypersetup{colorlinks=true,linkcolor=blue,citecolor=blue,urlcolor=blue}

\usepackage[breaklinks=true]{hyperref}

\makeatletter
\renewcommand*\aa@pageof{, page \thepage{} of \pageref*{LastPage}}
\makeatother

\begin{document} 

%\title{Precise measurement of planetary system parameters in the presence of stellar activity: the case of HD 73344}
\title{{A low-mass sub-Neptune planet transiting\\ the bright active star HD 73344}}

 \author{ S. Sulis \inst{1} 
 \and     I. J. M. Crossfield \inst{2} 
 \and     A. Santerne \inst{1}
 \and     M. Saillenfest \inst{3}
 \and     S. Sousa \inst{4}
 \and     D. Mary \inst{5}
 \and     A. Aguichine \inst{6}
 \and     M. Deleuil \inst{1}
 \and     E. Delgado Mena \inst{4}
 \and     S. Mathur \inst{7,8}
 \and     A. Polanski \inst{2}
 \and     V. Adibekyan \inst{4,9}
 \and     I. Boisse \inst{1}
 \and     J. C. Costes \inst{1}
 \and      M. Cretignier \inst{10}
 \and     N. Heidari \inst{11}
 \and     C. Lebarbé \inst{3}
 \and      T. Forveille \inst{12}
 \and     N. Hara \inst{1} 
 \and     N. Meunier  \inst{12}
 \and     N. Santos \inst{4,9}
 \and     S. Balcarcel-Salazar \inst{13}
 \and      P. Cortés-Zuleta \inst{14}
 \and      S. Dalal \inst{15}
 \and     V. Gorjian \inst{16} 
 \and     S. Halverson \inst{16} 
 \and     A. W. Howard \inst{17} 
 \and     M.R. Kosiarek \inst{6}  
 \and      T. A. Lopez \inst{1}
 \and      D. V. Martin \inst{18}
 \and     O. Mousis \inst{1,19} 
 \and     B. Rajkumar \inst{5} 
 \and      P. A. Str{\o}m \inst{20}
 \and      S. Udry \inst{21}
 \and      O. Venot \inst{22}
 \and      E. Willett \inst{23}
         }

\authorrunning{S. Sulis et al.}

 \institute{ 
%\label{inst:1} 
Universit\'e Aix Marseille, CNRS, CNES, LAM, Marseille, France, \email{sophia.sulis@lam.fr} \and
%\label{inst:2} 
Department of Physics and Astronomy, University of   Kansas, Lawrence, KS, USA \and
%\label{inst:3} 
IMCCE, Observatoire de Paris, PSL Research University, CNRS, Sorbonne Universit\'e, Universit\'e de Lille, 75014 Paris, France \and
%\label{inst:4}
Instituto de Astrof\'isica e Ci\^encias do Espa\c{c}o, Universidade do Porto, CAUP, Rua das Estrelas, 4150-762 Porto, Portugal \and
%\label{inst:5}
Université Côte d’Azur, Observatoire de la Côte d’Azur, CNRS, Laboratoire Lagrange, Bd de l’Observatoire, CS 34229, 06304 Nice Cedex 4, France \and
%\label{inst:6}
Department of Astronomy and Astrophysics, University of California, Santa Cruz, CA, USA \and
%\label{inst:7}
Instituto de Astrof\'isica de Canarias (IAC), E-38205 La Laguna, Tenerife, Spain \and
%\label{inst:8} 
Universidad de La Laguna (ULL), Departamento de Astrof\'isica, E-38206 La Laguna, Tenerife, Spain \and
 %\label{inst:9} 
Departamento de F\'isica e Astronomia, Faculdade de Ci\^encias, Universidade do Porto, Rua do Campo Alegre, 4169-007 Porto, Portugal \and
Sub-department of Astrophysics, Department of Physics, University of Oxford, Oxford, OX1 3RH, UK \and
Institut d'astrophysique de Paris, UMR 7095 CNRS Université Pierre et Marie curie, 98 bis, Bd Arago, 75014, Paris \and
%\label{inst:10}
Universit\'e Grenoble Alpes, CNRS, IPAG, F-38000 Grenoble, France \and
%\label{inst:11} 
Department of Physics, Massachusetts Institute of Technology, Cambridge, MA 02139, USA \and
SUPA, School of Physics \& Astronomy, University of St Andrews, North Haugh, St Andrews, KY169SS, UK \and
Astrophysics Group, University of Exeter, Exeter EX4 2QL, UK \and
Jet Propulsion Laboratory,  California Institute of Technology, Pasadena, CA, 91109, USA \and
%\label{inst:13} 
Department of Astronomy, California Institute of Technology, Pasadena, CA 91125, USA \and
%\label{inst:14} 
Department of Physics \& Astronomy, Tufts University, 574 Boston Avenue, Medford, MA 02155, USA \and
Institut Universitaire de France (IUF), France \and
Department of Physics, University of Warwick, Gibbet Hill Road, Coventry, CV4 7AL, UK \and
Geneva Observatory, University of Geneva, chemin des Maillettes 51, CH-1290 Versoix \and
Université Paris Cité and Univ Paris Est Creteil, CNRS, LISA, F-75013 Paris, France \and
School of Physics and Astronomy, University of Birmingham, Edgbaston, Birmingham B15 2TT, UK 
 }
 
 \date{To be submitted to A\&A}

%%%%%%%%%%%%%%%%%%%%%%%%%%%%%%%%%%%%%%%%
% ABSTRACT AND KEYWORDS
%%%%%%%%%%%%%%%%%%%%%%%%%%%%%%%%%%%%%%%%

% 5 {} token are mandatory

  \abstract
   {Planets with radii of between $2$ and $4~R_\oplus$  closely orbiting solar-type stars are of significant importance for studying the transition from rocky to giant planets, and are prime targets for atmospheric characterization by missions such as JWST and ARIEL. Unfortunately, only a handful of examples with precise mass measurements are known to orbit bright stars.}
  {Our goal is to determine the mass of a transiting planet around the very bright F6 star HD 73344 (Vmag=6.9). This star exhibits high activity and has a rotation period that is close to the orbital period of the planet ($P_\mathrm{b}=15.6$ days).}
   {The transiting planet, initially a K2 candidate, is confirmed through TESS observations (TOI 5140.01). We refined its parameters using TESS data and rule out a false positive with \textit{Spitzer} observations. 
    We analyzed high-precision radial velocity (RV) data from the SOPHIE and HIRES spectrographs.
    We conducted separate and joint analyses of K2, TESS, SOPHIE, and HIRES data using the \texttt{PASTIS} software. Given the star's early type and high activity, we used a novel observing strategy, targeting the star at high cadence for two consecutive nights with SOPHIE to understand the short-term stellar variability. We modeled stellar noise with two Gaussian processes: one for rotationally modulated stellar processes, and one for short-term stellar variability.}
   {High-cadence RV observations provide better constraints on stellar variability and precise orbital parameters for the transiting planet: a radius of $R_\mathrm{b} = 2.88^{+0.08}_{-0.07}~R_\oplus$ and a mass of $M_\mathrm{b}=2.98^{+2.50}_{-1.90}~M_\oplus$ {(upper-limit at $3\sigma$ is $<10.48~M_\oplus$)}. The derived mean density suggests a sub-Neptune-type composition, but uncertainties in the planet's mass prevent a detailed characterization.
   {In addition, we find a periodic signal in the RV data that we attribute to the signature of a nontransiting exoplanet, without totally excluding the possibility of a nonplanetary origin. This planetary candidate would have a minimum mass of about  $M_\mathrm{c}\sin i_\mathrm{c} = 116.3\pm ^{+12.8}_{-13.0}~M_\oplus$ and a period of $P_\mathrm{c}= 66.45^{+0.10}_{-0.25}$ days.}
   %This planetary candidate with a minimum mass of about  $M_\mathrm{c}\sin i_\mathrm{c} = 116.3\pm ^{+12.8}_{-13.0}~M_\oplus$ and a period of $P_\mathrm{c}= 66.45^{+0.10}_{-0.25}$ days in the RV data.  
   Dynamical analyses confirm the stability of the two-planet system and provide constraints on the inclination of the candidate planet; these findings favor a near-coplanar system.}
    {While the transiting planet orbits the bright star at a short period, stellar activity prevented us from precise mass measurements despite intensive RV follow-up. Long-term RV tracking of this planet could improve this measurement, as well as our understanding of the activity of the host star. The latter will be essential if we are to characterize the atmosphere of planets around F-type stars using transmission spectroscopy.
    }


   \maketitle

%-------------------------------------------------------------------

\section{Introduction}

To date, $790$ exoplanets have been characterized by combining photometric ---transits--- and spectroscopic ---radial velocity (RV)--- observations\footnote{Statistics from the NASA Exoplanet Archive (September 2023, \url{https://exoplanetarchive.ipac.caltech.edu})} \citep{2022NatAs...6..516C}. 
Among them, only $20$ orbit stars of magnitudes of $<8$ and most are short-period planets ($<30$ days).  In the context of new and future space missions, such as the JWST \citepads{2006SSRv..123..485G} and ARIEL (2029; \citeads{2018ExA....46..135T}), exoplanets orbiting bright stars are priority targets for atmospheric characterization. 

Most of the known exoplanets are sub-Neptunes\footnote{For reference, Neptune's radius is $\sim 3.8~R_\oplus$.} and Super-Earths; that is, planets with a radius of around $2.0-4.0~R_\oplus$. These planet populations are not present in our Solar System. However, because they lie in the transition regime between rocky planets and gas giants, they can provide strong constraints on planet-formation models \citepads{2010Sci...330..653H}. To conduct statistical studies of these planets at a population level, we require precise knowledge of the physical properties of individual targets. In this context, our goal is to characterize the candidate sub-Neptune planet HD 73344b.

\noindent In this paper, we present analyses of new photometric and spectroscopic data for this candidate planet, which orbits the bright F star HD 73344 ($V= 6.9$ mag) with a period of  $\approx 15$ days. This planet was first discovered by  \citetads{2018AJ....156...22Y} based on six transits observed in the K2 data. While the detection was challenging due to the high activity level of this early-type star, we confirm the detection of this planet by combining K2 data with new \textit{Spitzer} and TESS photometric data, as well as a set of SOPHIE and HIRES RV observations. In addition, our RV analyses reveal a new sub-Jupiter-mass planet candidate, which is nontransiting and has a period of $\approx 66$ days. 
We propose a new observing strategy to identify and overcome the different sources of stellar activity that impact the characterization of planetary systems (see e.g., \citeads{2011A&A...525A.140D}; \citeads{2004A&A...414.1139A}; \citeads{2020A&A...636A..70S}; \citeads{2023A&A...676A..82M}). In particular, we demonstrate the benefits of tracking the star at high cadence for whole nights in order to characterize its short-timescale stellar variability (p-mode oscillations, granulation, supergranulation), which is of very large RV amplitude.

The paper is structured as follows. In Sect.~\ref{sec2} we present the observations. In Sect.~\ref{sec3} we derive the fundamental parameters of the star and characterize various sources of stellar activity. 
We study the planetary system around HD 73344 in Sect.~\ref{sec4}. In Sect.~\ref{dis} we discuss the stability of the system and the internal composition of the transiting planet. We conclude in Sect.~\ref{ccl}.

%--------------------------------------------------------------------
\section{Observations}
\label{sec2}

In this section, we present the various sets of photometric and spectroscopic observations of HD 73344 we used in this study\footnote{The observations used in this work are  available in electronic form at the CDS via anonymous ftp to cdsarc.u-strasbg.fr (130.79.128.5) or via \url{http://cdsweb.u-strasbg.fr/cgi-bin/qcat?J/A+A/}.}. The main information is summarized in Table~\ref{tab_data}.

\begin{table*}[h!]
    \centering
    \def\arraystretch{1.1}%
    \caption{Summary of photometric (top) and spectroscopic (bottom) observations of the HD 73344 system. }% 

    \begin{tabular}{|c|c|c|c|c|c|}
        \hline
        Instrument& Starting date & T & $dt$ & N & Comments \\
                   & [BJD] & [days] & [s] &  &\\
        \hline
         K2 (C16) &  2458095.47& 79.55 & 1765& 3684 & 6 transits \\ % 2017-12-07 to 2018-02-25
         \textit{Spitzer}  & 2458704.65 & $0.35$ & 0.1 & 230400 & 1 transit    \\ % 2019 -08-09
         TESS (S45) & 2459525.73& 24.90 & 120& 16867 & 2 transits \\ % 2021-11 -  2022-01
         TESS (S46) & 2459552.01& 26.7 & 120 & 18190 & 1 transit \\ % 2021-11 -  2022-01
         \hline
    \end{tabular}
    
\vspace{0.5cm}

    \begin{tabular}{|c|c|c|c|c|c|c|c|c|}
        \hline
        Instrument & Starting date & T      & $\tau_{exp}$ & N &  Mean & [Min, Max] & Comments \\
                   & [BJD]         & [days] &  [min]       &   & $\sigma_{RV}$ [m/s] & $\sigma_{RV}$ [m/s] &\\
        \hline
         %Hamilton/Lick & 2450831.84 & 4015.09 & --- & 23  & $6.1$ & $[3.6, 9.6]$ & Data not used in this paper \\ %  1998-01-18 and 2009-01-15
         HIRES/Keck & 2458194.89 & 314.18 & $\sim1.0$ & 238 & $1.2$ & $[1.07, 2.2]$ & Observing strategy: $3\times5$ pts/night\\ % 2018-03-17 to 2021-06-03.
         %APF/Lick & 2458207.64 & 301.17 & $\sim 7$& 24 & $4.5$ & $[3.4, 8.2]$& Data not used in this paper \\ % 2018-03-10 to 2019-01-25 
         SOPHIE/OHP & 2458425.66 & 484.68 & $\sim 15$ & 312 &  $2.7$ & $[2.4, 5.0]$ & Observing strategy: 3 pts/night\\ % 2018-11-02 and 2020-03-01 
         SOPHIE/OHP & 2459591.34 & 0.38 & $[10.4, 16.4]$ & 51 &  $4.64$ & $[4.5, 4.8]$ & Full night 1: moderate-cadence\\ % DDT (2022-01-11)
         SOPHIE/OHP & 2459592.34 & 0.39 & $[3.4, 8.3]$ &  152 &  $2.6$ & $[2.0, 3.7]$ & Full night 2: high-cadence\\ % DDT (2022-01-12)
         \hline
    \end{tabular}
\vspace{0.2cm}\\
\parbox{7in}{
\footnotesize 
Notes. Columns are: instrument, starting date of observation (BJD); total observation duration since the starting date (T); temporal sampling ($dt$, top only); exposure time ($\tau_{exp}$, bottom only); total number of observations after detrending (N); Mean, Min and Max RV errorbars ($\sigma_{RV}$, bottom only); and comment. 
}
\vspace{-0.1cm}
    \label{tab_data}
\end{table*}

\subsection{Photometry}
\label{sec_photom}

\subsubsection{K2}

K2 \citepads{2014PASP..126..398H} observed along the ecliptic a series of $100$-square degree zones, each lasting approximately up to $\sim80$ days. The broad bandpass of K2 was ranging from $420$ to $900$ nm. K2 observed HD 73344 (EPIC 212178066) during campaign 16 (C16), which ran from December 07, 2017 to February 25, 2018, and during campaign 18 (C18), which ran from May 12, 2018 to July 02, 2018. Observations were taken at a long cadence, with an integration time of $30$ minutes. In this work, we used only the data acquired during C16 since the observations are affected by strong systematics during C18. This dataset contains $6$ transits of HD 73344 b, originally identified by \citetads{2018AJ....156...22Y}. 

We detrended the C16 light curve using the software \texttt{EVEREST}\footnote{\url{https://github.com/rodluger/everest}} (\citeads{2016AJ....152..100L}; \citeads{ 2018AJ....156...99L}). We started by masking the transit events with a window taken as twice the transit duration ($T_\mathrm{dur} \sim 3.3$ hours). We then corrected the light curve with a single cotrending basis vector (CBV). We obtained the CBV-corrected  detrended flux from which we first removed the $3\sigma$ outliers. We then used a second sigma clipping step to remove the remaining outliers that look like "flares" in the dataset. For this second step, we used a median filter of $5$-hours to smooth the light curve and identified the data points at $1\sigma$ above this smoothed light curve. 
The resulting light curve is shown in Fig.~\ref{fig_lcs_tot} (top), and the individual transits in Appendix~\ref{app_transits}.

\begin{figure}[t!]
\centering
\resizebox{\hsize}{!}{\includegraphics{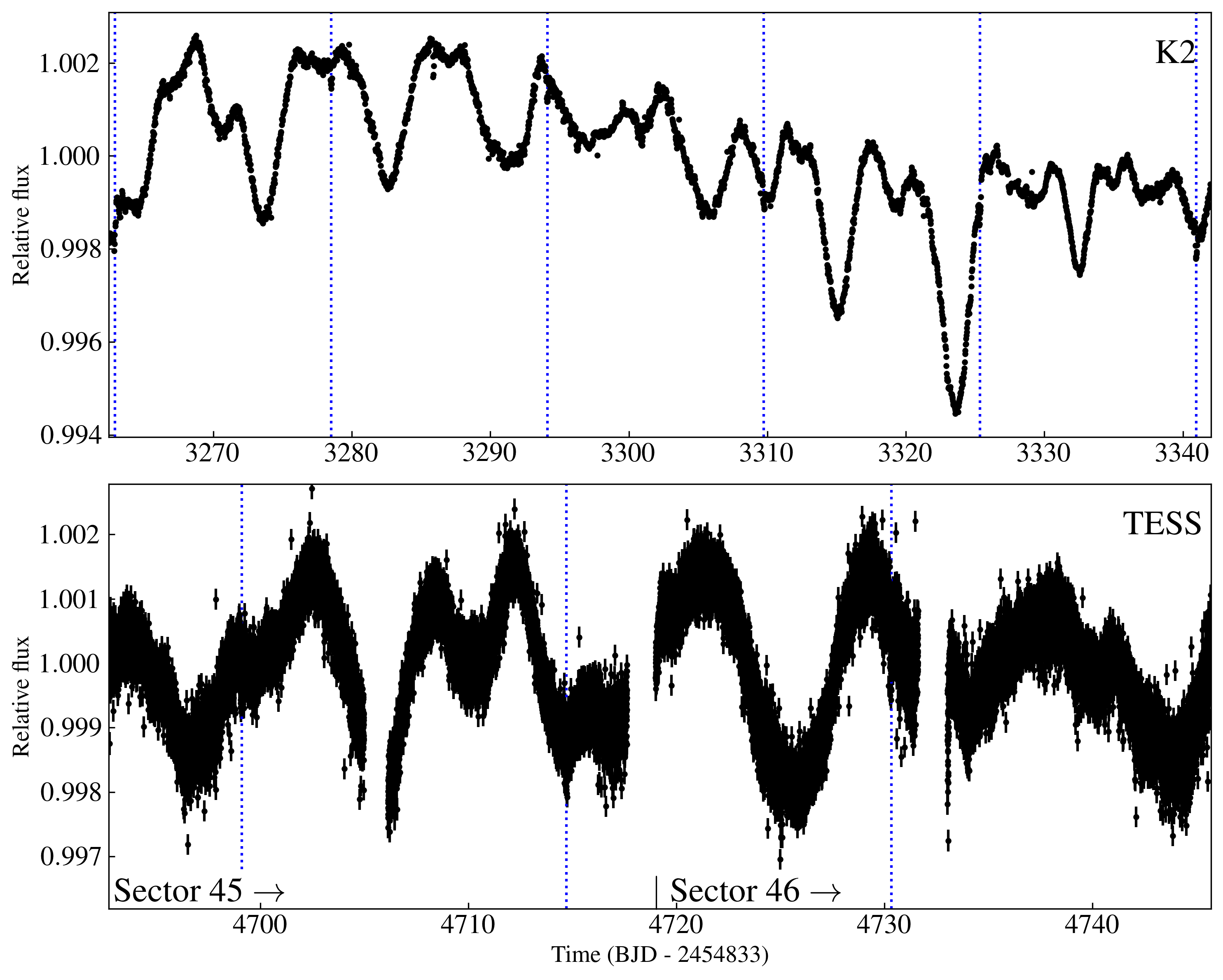}}
\caption{Light curves of HD 73344. The transit mid-times of planet b in K2 (top) and TESS (bottom) observations are shown with the dotted vertical lines. 
}
\label{fig_lcs_tot}  
\end{figure}

% =====================================
\subsubsection{TESS}

The Transiting Exoplanets Survey Satellite (TESS; \citeads{2015JATIS...1a4003R}) observed HD 73344 (TIC 175193677) in the red-optical bandpass ($600-1100$ nm) during sector 45 (November 2021 - December 2021) and sector 46 (December 2021 - January 2022).
The two sectors contain three transits of the planet, which has been identified as TOI 5140.01\footnote{\url{https://exofop.ipac.caltech.edu/tess}}. % --> https://exofop.ipac.caltech.edu/tess/target.php?id=175193677#
 The short cadence of these observations ($120$ seconds) allows a detailed characterization of the transits. 

In this work, we used the Simple Aperture Photometry (SAP) flux released by the TESS team on MAST\footnote{\url{https://mast.stsci.edu/}}. The resulting light curve (normalized by the median flux) is shown in Fig.~\ref{fig_lcs_tot} (bottom), and the individual transits in Appendix~\ref{app_transits}.

% =====================================
\subsubsection{Spitzer}
% -- 1 transit 

We also observed one transit of HD~73344b with the {\em Spitzer} space telescope as part of program 14292 (PI: I. Crossfield; \citeads{crossfield:2019spitz}).  On August 9, 2019, we obtained 3600$\times$64 0.1\,s subarray frames of HD~73344 with the IRAC2 4.5\,$\mu$m channel \citep{fazio:2005}, spanning 8.5\,hours and covering one transit of planet b.  The raw and calibrated {\em Spitzer} data products are available at the {\em Spitzer} Heritage Archive; the analysis is presented in Sect.~\ref{sec_spitzer}.

% =====================================
\subsection{High-resolution spectroscopy }

We carried out the RV follow-up observations of HD 73344 with SOPHIE and HIRES spectrographs over a total time span covering $\sim 715$ days.  The SOPHIE and HIRES RV are analyzed to get the mass of the transiting planet in Sect.~\ref{sec4}.

% =====================================
\subsubsection{SOPHIE}
\label{sec_sophie}

We observed HD 73344 with the high-resolution echelle spectrograph SOPHIE \citepads{2008SPIE.7014E..0JP} at the Haute-Provence Observatory (OHP, France) as part of the program dedicated to the RV follow-up of K2 planet candidates\footnote{Programme IDs: 18B.PNP.LOPE,  19A.DISC.LOPE, 19B.PNP.LOPE, 21B.DISC.SULIS}. The target was observed between 2018-11-02 and 2020-03-01, gathering $345$ high-resolution spectra.

The observations were carried out using SOPHIE high resolution (HR) mode (resolving power of $\lambda / \Delta \lambda \approx 75,000$ at 550 nm), with simultaneous Fabry-Perot (FP) calibration lamp measurements. The latter enabled us to monitor instrumental drift, ensuring precise and accurate RV measurements. The exposure time was set at 900 seconds with the classic observational strategy of $3$ points per night to average the stellar variability, resulting in a median signal-to-noise ratio (S/N; measured on each points) of $149$ per pixel at $550$ nm.

Radial velocity calculations were performed using the SOPHIE data reduction system \citepads[DRS,][]{bouchy2009sophie}, employing a G2 mask to extract RVs. To enhance the accuracy of SOPHIE measurements, we implemented the optimized procedure outlined in \citetads{heidari:tel-04043297} and \citepads{2024A&A...681A..55H}. This procedure in particular encompasses: (1) CCD charge transfer inefficiency correction 
 \citep{bouchy2013sophie+}; (2) atmospheric dispersion correction \citep{modigliani2019espresso}; and (3) RV master constant correction to correct long-term instrumental drifts \citep{courcol2015sophie}.
In addition to the RV observations, using the DRS we also calculated some useful spectroscopic activity indicators such as the Full Width at Half Maximum (FWHM) and the bisector inverse slope (BIS, \citeads{2001A&A...379..279Q}). We then calculated the $\rm log R'_{\rm HK}$ following \citetads{1984ApJ...279..763N} and \citetads{2010A&A...523A..88B}, and the H$\alpha$ index following \citetads{2011A&A...528A...4B}. 
From the raw RV, we removed the $3\sigma$ outliers, and the data points with RV uncertainties $>5$ m/s ($7$ points removed in total). The final RV time series contain $312$ data points (hereafter: the ``unbinned'' dataset), spread over $137$ individual nights (used to generate the ``binned'' dataset). The mean RV uncertainty on all measurements is $2.7$ m/s. \\

In complement to this long RV campaign, we observed HD 73344 continuously for two consecutive nights to monitor the short timescale stellar variability (dominated by p-mode oscillations, granulation, and supergranulation). The first night (2022-01-11) contains $N=51$ data points, taken with an exposure time between $\tau_{exp} =[10.4, 16.4]$ min, during a total of $T\sim 9.12$ hours. The RV shows a significant dispersion, with an RMS of $4.65$ m/s. RV uncertainties on each measurements range from $4.5$ to $4.8$ m/s over the night, and are therefore similar to the observed dispersion. 
{To investigate in more detail this short-term variability}, we observed HD 73344 during a second night (2022-01-12) at a shorter temporal cadence ($\tau_{exp} = [3.4, 8.3]$ min, $T\sim 9.36$ hours, $N=152$). The RMS of this second RV dataset is $8.72$ m/s, confirming the strong amplitude of the {short-term variability. }
RV uncertainties on each measurements are significantly lower (from $2$ to $3.7$ m/s over the night) compared to the first night due, in particular, to very good atmospheric condition (seeing).

%For both nights, we note an airmass $<1.7$ {for X\% of the data points (except at the very begginning and end of the nights)}, which is the requirement for high precision RV programs.

%The two sets of observations are shown in Fig.~\ref{Fig_gr}. We note a visual similarity between them.  Although this may seem intriguing, we do not favor an instrumental origin (pressure, temperature), since the indicators show a good stability over the nights and no systematic effect (e.g., color-effect) is known on the timescales of several hours that would be so strong to question the stellar origin of this variability (the peak-to-peak RV variations are $>10$ m/s). We therefore suggest that it is a matter of some coherence in the stellar variability observed over consecutive nights. However, to firmly confirm this interpretation, we would need to obtain additional observations separated in time.
%We suggest, then, that it's a matter of some coherence in the stellar variability observed over consecutive nights. However, to firmly confirm this interpretation, we would need to obtain additional observations separated in time.

\begin{figure}[t!]
\centering
\resizebox{\hsize}{!}{\includegraphics{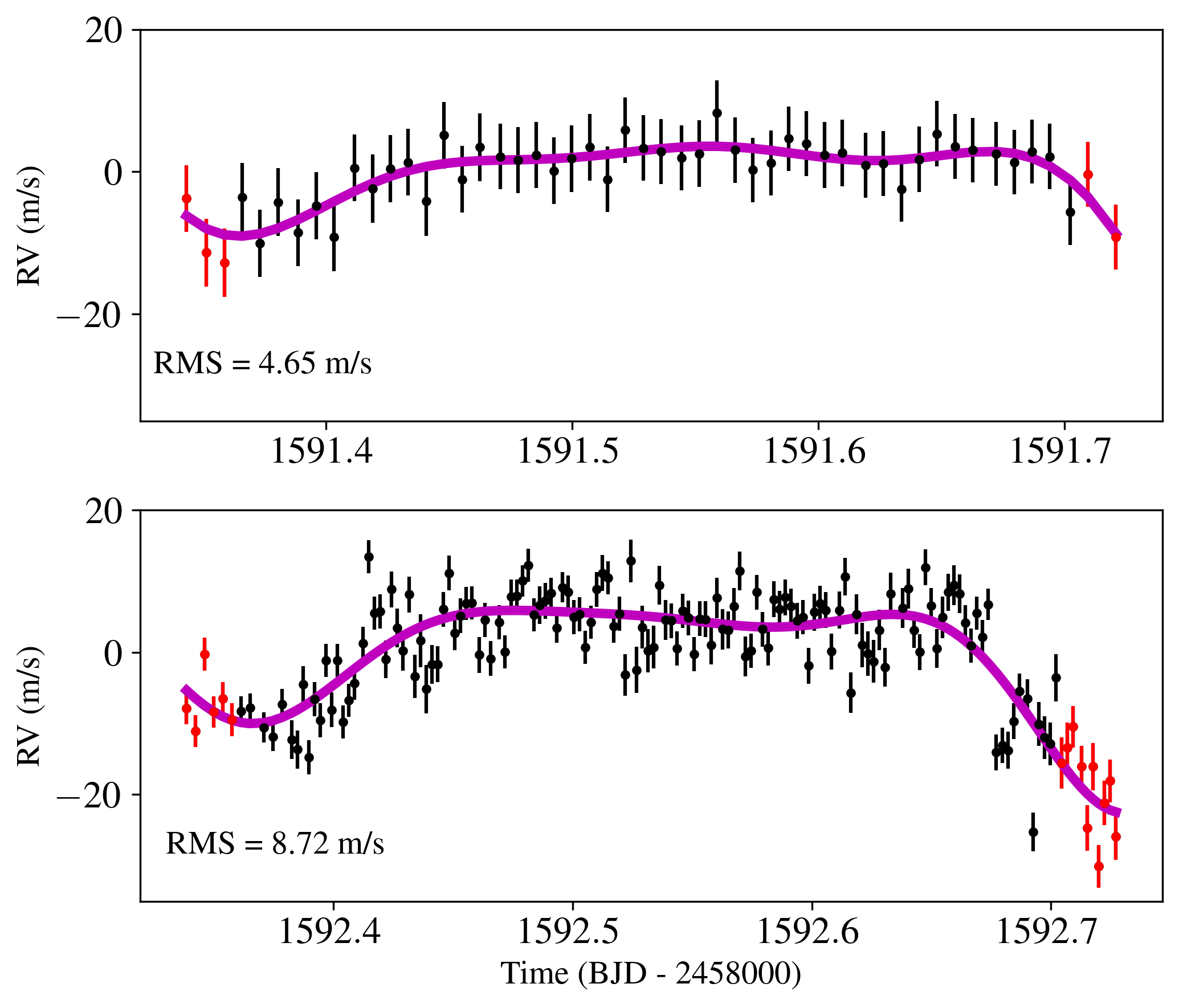}}
\caption{Radial velocity of HD 73344 obtained over two consecutive nights with the SOPHIE spectrograph. Best fitting GP model from their joint analysis is shown in purple. {Observations taken at airmass $>1.7$ are shown in red.}}
\label{Fig_gr}  
\end{figure}

{
The two sets of observations are shown in Fig.~\ref{Fig_gr}. We note a similar pattern across both nights, marked by flux drops at the beginning and end of each night. While these flux drops may indeed stem from instrumental systematics (in particular, we identified a potential issue with the ADC used for observations taken at airmass$>1.7$), the exact source remains uncertain. When considering only the data points obtained during the middle of the nights (airmass$<1.7$), we still observe a considerable RV dispersion (exceeding $7$ m/s for the second night). Similar RV amplitudes are also independently observed in the nightly observations taken with the HIRES/Keck spectrograph. We are therefore confident that the dominant RV variability observed in the high cadence SOPHIE dataset is of stellar origin. The characteristics of this variability are given in Sect. \ref{sec_gr}.
} 
Based on these two nights of observations, we expect that the classic observing strategy (which consists of observing the target $3$ times per night and binning these $3$ points) will not be sufficient to significantly reduce the short-term {(}stellar{)} variability. This will be confirmed in Sect.~\ref{sec42}{.}% of the present paper. 

% =================================

\subsubsection{HIRES}
\label{sec_hires}

We obtained $238$ additional RV data points with the HIRES spectrometer \citep{1994SPIE.2198..362V} installed at the Keck~I telescope from 2018-03-17 to 2021-06-03. These observations used the B5 decker, which has a slit width of \SI{0.861}{\arcsecond} and gives an effective resolution of $48~000$, and HIRES iodine cell. They had typical integration times of 40~s (depending on observing conditions). We followed standard procedures of the California Planet Search for the HIRES observations and reductions \citepads{2010ApJ...721.1467H}. The observing strategy was to take $3$ sets of $5$ consecutive observations over the course of the nights to reduce sensitivity to stellar variations. We grouped these observations into $3$ points per night to mimic the sampling of SOPHIE observations. The final, binned time series contains $39$ data points, spread over $19$ nights. The mean RV uncertainty on all measurements is $1.2$ m/s.

%--------------------------------------------------------------------
\section{Stellar properties}
\label{sec3}

In this section, we first describe how we inferred the fundamental stellar parameters of HD 73344 from the SOPHIE spectra, and the stellar abundances from both the SOPHIE and HIRES spectra. Then, we study the stellar activity signatures in both photometric and spectroscopic data.  In particular, we look at the variability modulated with the stellar rotation (spots/faculae), and the variability evolving on short timescales (oscillations, convection). 

\subsection{Fundamental parameters}
\label{sec31}

The stellar spectroscopic parameters ($T_{\mathrm{eff}}$, $\log g$, microturbulence, [Fe/H]) were estimated using the ARES+MOOG methodology. The methodology is described in detail in \citetads{2013A&A...556A.150S}; \citetads{2014dapb.book..297S}; \citetads{2021A&A...656A..53S}. To consistently measure the equivalent widths (EW) we used the latest version of \texttt{ARES}\footnote{The latest version, ARES v2, can be downloaded at \url{https://github.com/sousasag/ARES}.} (\cite{Sousa-07}; \cite{Sousa-15}). The list of iron lines is the same as the one presented in \citetads{Sousa-08}. For this we used the combined SOPHIE spectra: we coadded the spectra until reaching S/N $\sim2000$, where each individual spectra was corrected in RV to the rest frame prior to be coadded. To find the ionization and excitation equilibrium in this analysis we used a minimization process to converge for the best set of spectroscopic parameters. This process makes use of a grid of ATLAS model atmospheres \citep{Kurucz-93} and the radiative transfer code \texttt{MOOG} \citep{Sneden-73}. 
%{To get a more accurate value, we corrected the derived spectroscopic ${\rm log}g$ using a linear function of stellar temperature}, using Eq.(4) of \citetads{2014A&A...572A..95M}. 
We also derived a more accurate trigonometric surface gravity using recent \textit{Gaia} data following the same procedure as described in \citetads{2021A&A...656A..53S}, which provided a consistent value when compared with the spectroscopic surface gravity. In this last process, we also estimated the stellar mass and radius using the calibrations presented in \citetads{Torres-2010}. 
Furthermore, we determine the Li abundance of this star by performing spectral synthesis also using the code \texttt{MOOG} and \texttt{ATLAS} atmospheres, as well as the above derived stellar parameters. We obtained a value of A(Li)\,=\,2.81\,$\pm$\,0.05\,dex, which is typical of young stars of this $T_\mathrm{eff}$. From this analysis we can also get an estimate of the inclined rotational velocity, after considering the instrumental broadening given by the SOPHIE spectral resolution (R $\sim$ 75000) and applying the macrotuburbulence velocity empirical calibration from \cite{Doyle-14} dependent on $T_\mathrm{eff}$ and ${\rm log}g$ (V$_{mac}$\,=\,4.7\,km/s). The measured projected rotational velocity $v\sin{i_\star}$ is 5.3\,km/s.
We report the stellar parameters in Table~\ref{tab_params}.

In addition, we measured the stellar abundances for multiple chemical elements using both SOPHIE and HIRES spectra. For SOPHIE spectra, using the aforementioned stellar atmospheric parameters (we considered the trigonometric surface gravity), we determined the abundances of refractory elements following the classical curve-of-growth analysis method described in \citet[e.g.,][]{Adibekyan-12, Adibekyan-15, Delgado-17}. Similar to the stellar parameter determination, we used \texttt{ARES} to measure the EWs of the spectral lines of these elements, and used a grid of Kurucz model atmospheres along with the radiative transfer code \texttt{MOOG} to convert the EWs into abundances, assuming local thermodynamic equilibrium. Although the EWs of the spectral lines were automatically measured with \texttt{ARES}, for Mg which has only three lines available we performed careful visual inspection of the EWs measurements. Abundances of the volatile elements, C and O, were derived following the method of \citet{Delgado-21}; \citet{Bertrandelis-15} and using the same code and model atmospheres. All the abundance ratios [X/H] are obtained by doing a differential analysis with respect to a high S/N solar (Vesta) spectrum. The final abundances, shown in Appendix~\ref{App_abun}, are typical of a galactic thin-disk star. Moreover, we used the chemical abundances of some elements to derive ages through the so-called chemical clocks (i.e., certain chemical abundance ratios which have a strong correlation for age). We applied the 3D formulas described in Table 10 of \citeads{2019A&A...624A..78D}, which also consider the variation in age produced by the effective temperature and iron abundance. The chemical clocks [Y/Mg], [Y/Zn], [Y/Ti], [Y/Si], [Sr/Zn], [Sr/Ti], [Sr/Mg] and [Sr/Si] were used from which we obtain a weighted average age of 2.0\,$\pm$\,0.2 Gyr. We note that this small uncertainty reflects the high precision of the different chemical clocks for this specific star and is smaller than the true age uncertainty. For HIRES spectra, we measured the stellar abundances following the approach of \citetads[][and in prep.]{2022RNAAS...6..155P}, using the \texttt{Cannon} \citepads{Ness2018}, which was designed to be applied to iodine-free spectra from HIRES on Keck I \citepads{Rice2020}. \texttt{KeckSpec} was trained using a sample of high-quality (S/N $> 100$) HIRES spectra for which abundances of $15$ chemical elements were determined in \citetads{Brewer2016}. We used an iodine-free spectrum that reached an S/N per pixel of $214$. We calculated the $\alpha$ element enhancement and found [$\alpha$/Fe] values of $\sim$-0.03 dex making HD 73344 chemically consistent with the thin disk. We also report the stellar abundances in Appendix~\ref{App_abun}.

\begin{table*}[t] \centering
\caption{Properties of the star HD 73344.  }
\vspace{-0.1cm}
\label{tab_params} 
\begin{tabular}{*4c}
\noalign{\smallskip}\hline\noalign{\smallskip}
Parameters &  Values & Unit &Source \\
\noalign{\smallskip}\hline\hline\noalign{\smallskip}
Target names  & HD 73344, HIP 42403 & & Simbad$^a$ \\
  &  EPIC 212178066,  TIC 175193677  &  & \\
  &  {\small \textit{Gaia} EDR3 666427539629086976}  & & \\
\noalign{\smallskip}\hline\noalign{\smallskip}
Spectral type &  F6V &  &  Simbad$^a$ \\ 
Right Ascension  (ep=J2000)   &  08:38:45.52 & & Simbad$^a$ \\ 
Declination (ep=J2000)   & +23:41:09.25 & & Simbad$^a$ \\ 
V-band magnitude (Vmag) &  6.9 & & Simbad$^a$\\ 
J-band magnitude (Jmag) & 5.8 & & Simbad$^a$\\ 
Distance (d)&     $35.2093_{-0.0361}^{+0.0718}$ & pc &\textit{Gaia} DR3$^b$ \\ 
\noalign{\smallskip}\hline\noalign{\smallskip}
Effective temperature ($T_\mathrm{eff}$) &   $6220 \pm 64$ &  K &This work$^c$  \\
Metallicity ([Fe/H]) &  $0.18 \pm 0.043$ & dex & This work$^c$  \\ 
Surface gravity (${\rm log}g$) &  $4.496 \pm 0.105$  & cgs &  This work$^c$ (spectroscopy)  \\ % &  $4.176 \pm 0.227$  & cgs &  This work$^c$ (spectroscopy)  \\
  &  $4.39 \pm 0.02$  & cgs &  This work$^c$ (trigonometry)  \\
Radius ($R_\star$)  &  $1.22 \pm 0.04$ & $R_\odot$& This work$^c$ \\ 
Mass ($M_\star$) &  $1.20 \pm 0.02$ & $M_\odot$& This work$^c$ \\ 
% Age [Gyr]&  $x \pm x$ &  x \\ --> not found anywhere
Rotational velocity ($v\sin{i_\star}$)  & $ \sim 5.3$  &  km/s & This work$^c$ \\ 
%%%%%
Rotation period ($P_\mathrm{rot}$)  & $ 9.09 \pm 0.04 $ & days  & This work$^d$  \\ % from GP combined analysis
%%%%%
Stellar inclination ($i_\star$) &  $\sim 53$  & degrees  & This work$^c$   \\ % mettre une footnote pour reference section
%%%%%
\noalign{\smallskip}\hline\noalign{\smallskip}
\textit{Activity indicators} & & & \\
Photospheric activity proxy $S_\mathrm{ph}$ &  $(1132; 894)$& ppm & This work$^c$ (K2; TESS) \\ 
${\rm log} R'_{\rm HK}$ &    $\sim -4.6$ && This work$^c$ (SOPHIE) \\ 
\noalign{\smallskip}\hline\noalign{\smallskip}
\end{tabular}
\parbox{7in}{
\footnotesize Notes. $^a$SIMBAD astronomical database from the \textit{Centre de Données astronomiques de Strasbourg} (\href{http://simbad.u-strasbg.fr/simbad/}{http://simbad.u-strasbg.fr/simbad/}). $^b$Archive of the \textit{Gaia} mission of the European Space Agency (\href{https://gea.esac.esa.int/archive/}{https://gea.esac.esa.int/archive/}). $^c$See Sect.~\ref{sec3}. $^d$Values obtained based on the joint analysis of the photometric and spectroscopic observations in Sect.~\ref{sec_combined}.
}
\vspace{-0.1cm}
\end{table*}

%--------------------------------------------------------------------
\subsection{Stellar activity signatures}
\label{sec_star2}

\subsubsection{Magnetic activity modulated with rotation period}
\label{sec_star2_mag}

\begin{figure}[t!]
\centering
\resizebox{\hsize}{!}{\includegraphics{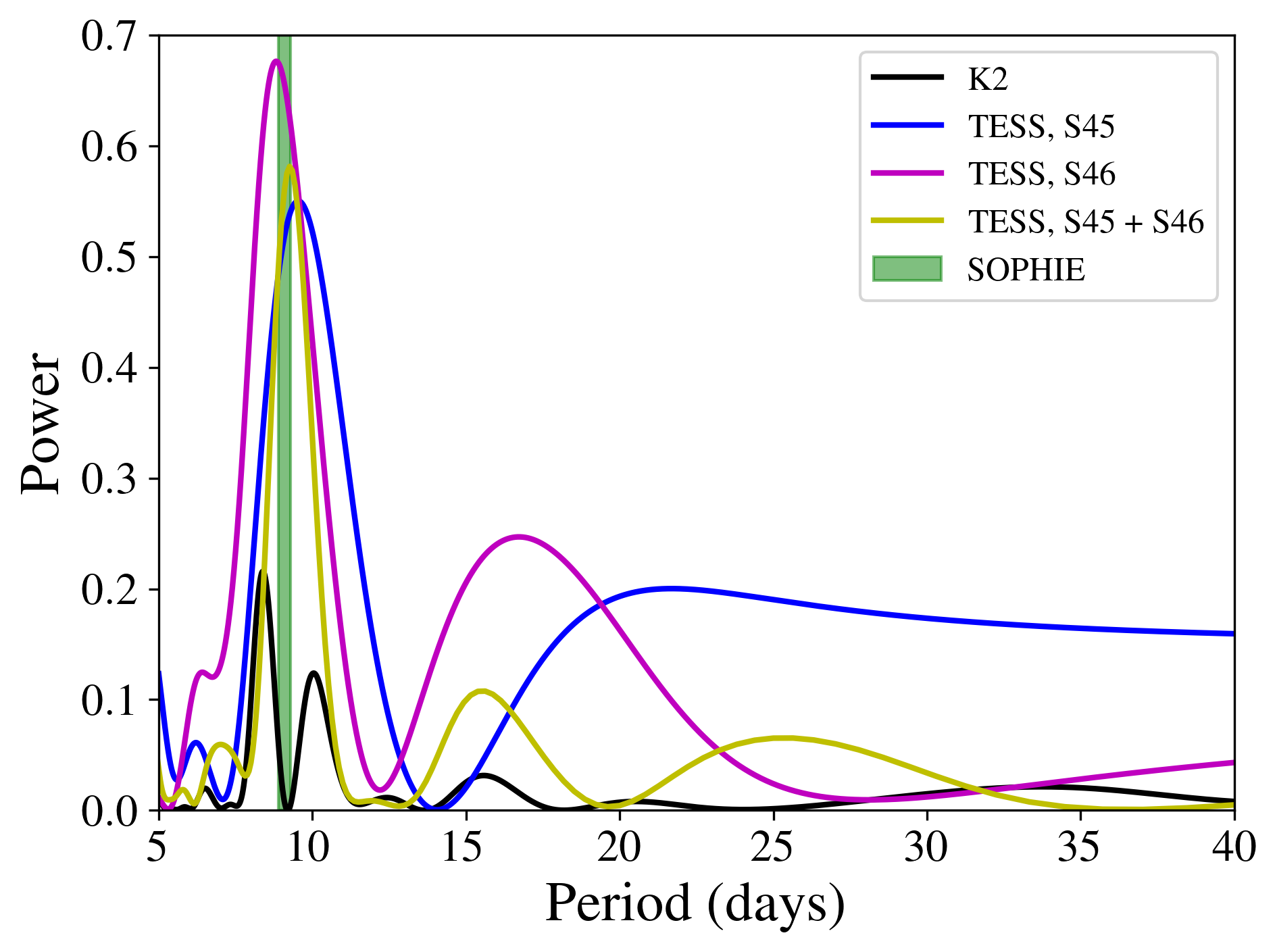}}
\caption{Generalized Lomb-Scargle periodograms of the K2 and TESS photometric data between periods of $5$ and $40$ days. Transit of planet b have been masked. The green vertical lines indicate the period of $9.1 \pm 0.2$ days found with RV data and chromospheric indicators. }
\label{fig_periodo_lcs}  
\end{figure}

Both the light curves (Fig.~\ref{fig_lcs_tot}) and the RV observations  (see Sect.~\ref{sec_combined}) show strong activity signatures. To analyze the frequency content of this variability, we first performed the generalized Lomb-Scargle periodogram (GLSP; \citeads{2008MNRAS.385.1279B}; \citeads{2009A&A...496..577Z}) of the K2 photometric data, after masking out planet transits (see Fig.~\ref{fig_periodo_lcs}). The GLS approach is to fit at each frequency a floating mean (a constant) coupled with a periodic term of unknown phase and amplitude. 
The definition of the GLSP used here is 
\begin{equation}
    P_{GLS}(\nu) := \frac{\chi_0^2-\chi^2(\nu)}{\chi_0^2},
\label{eq_GLS1}
\end{equation} 
with $\chi_0^2$ the residual sum of squares (RSS)  resulting by fitting only a constant,
 and $\chi^2(\nu)$ the RSS by jointly fitting a constant and a sinusoid at frequency $\nu$ (see Eq.(4) of \citeads{2009A&A...496..577Z}). The fit is obtained through a weighted least squares problem, with weights provided by the RV uncertainties.

In Fig.~\ref{fig_periodo_lcs}, we observe two peaks, the highest of which corresponds to the period of $\sim 8.39$~d, close to the value reported by \citetads{2018AJ....156...22Y} for the stellar rotation period ($P_\mathrm{rot}$).
The same analyses on the individual and combined TESS sectors return slightly longer periods with the highest periodogram peak appearing at $\sim 9.54$~d (sector 45), $\sim 8.82$~d (sector 46), and $\sim 9.26$~d (both sectors, see Fig.~\ref{fig_periodo_lcs}). The GLSPs of RV data and chromospheric indicators also show a clear peak at $9.1 \pm 0.2$ days. 
Finally, the joint analysis of photometric and RV observations with quasi-periodic Gaussian process models finds $P_\mathrm{rot} = 9.09 \pm 0.04$ days (see details in Sect.~\ref{sec_combined} and GLSP of the RV data and indicators in Fig.~\ref{fig_indicators}). The latter is the value that we reported in Table~\ref{tab_params}. 
% vsin.i
Combining $P_\mathrm{rot}$ with the $v \sin i_\star \sim 5.3$ km/s measured from SOPHIE observations\footnote{$P_\mathrm{rot} / \sin i_\star = 2\pi R_\star / (v \sin i_\star)$},  we find an inclination of the rotation axis of $i_\star \sim 53 ^\circ$. This suggest that the stellar rotation and the orbit of the transiting planet could be misaligned, which would deserve further investigation for the implication on the system history \citepads{2013Sci...342..331H}. 

% RV indicators
Over the two SOPHIE campaigns, we observe an increase in magnetic activity signatures, with median $\rm log R'_{\rm HK}$ values decreasing from $-4.65 \pm 0.03$ to $-4.56\pm 0.03$ (a variation of $2\%$ over an average duration of $224$ days). This star is definitely more active than the Sun, as indicated by its $\rm log R'_{\rm HK}$ value of approximately $-4.9$ \citepads{2017ApJ...845...79B}. 
Looking in more detail at the temporal variability of the various spectroscopic indicators {in Appendix~\ref{appC}}, we note that these magnetic features remain significant over $3$ to $4~P_\mathrm{rot}$ (i.e., $27-36$ days). {Although it is difficult to identify a precise stellar origin (spot/faculae) of the periodicities observed in the various activity indicators, it is worth noting that the strongest signals at $3-4~P_\mathrm{rot}$ occur in the CCF area indicator, rather than in the $\rm log R'_{\rm HK}$ indicator (see Appendix~\ref{appC}). As suggested by \citetads{10.1093/mnras/stab1183}, this observation may imply that HD 73344 is dominated by faculae.}

% photometric indicators
 We then tracked the photometric signatures of the stellar magnetic structures using the photospheric activity proxy $S_\mathrm{ph}$ (\citeads{2014A&A...562A.124M}) {to place our target into the F-type star population}. This global proxy is defined as the standard deviation calculated over subseries of $5\times P_\mathrm{rot}$ in length (\citeads{2014JSWSC...4A..15M}). 
  The contribution of photon noise ($\sigma_{\phi}$) is subtracted from this value. For K2 data, a direct relationship between stellar magnitude and photon noise has been derived in \citetads{Jenkins_2010}: it gives $\sigma_{\phi}\sim5$ ppm. For the TESS data (both sectors combined), our calculations\footnote{To estimate $\sigma_{\phi}$ in the TESS data, we calculated the power spectrum, evaluated the average power spectrum over high frequencies ($\nu > 3000~\mu$Hz), and converted it to amplitude.} also give a value of $\sigma_{\phi}\sim5$ ppm. 
At the end, we found a mean $<S_\mathrm{ph}> = 1132$ ppm in the K2 data, and $<S_\mathrm{ph}> = 894$ ppm in the TESS data. 
When compared to the sample of 22 F dwarfs studied by \citetads{2014A&A...562A.124M}, our target appears to be much more active. However, we need to keep in mind that the sample is biased because the stars in their sample have detected solar-like oscillations. It is known that strong magnetic activity can lead to smaller mode amplitudes (e.g., \citeads{2010Sci...329.1032G}; \citeads{2011ApJ...732L...5C}; \citeads{2019FrASS...6...46M}) so the F dwarfs sample is mostly constituted of low magnetic activity stars. The recent catalog of rotation periods measured for more than 55,000 {\it Kepler} stars (\citeads{2019ApJS..244...21S}; \citeads{2021ApJS..255...17S}) gives a better representation of the magnetic activity level of main-sequence solar-like stars. From the Fig.~7 (third row) of \citetads{2021ApJS..255...17S}, we can see that F-type stars with rotation periods around 8-9 days can have $S_{\rm ph}$ values similar to the ones obtained for HD~73344. This makes our target less atypical but it is still among the most active stars of the F-dwarf sample. This can be explained by the star's young age (see Sects.~\ref{sec31} and \ref{sec_combined}).

%transit: no spot crossing. Variation light curve: non-crossing spots/plages. 
We note that we do not detect ``obvious'' bright or dark spot crossing events during the planetary transits to study the evolution of such features on the stellar surface. We are therefore most sensitive to the uncrossed magnetic regions that generate the long-term photometric variability observed in Fig.~\ref{fig_lcs_tot}.

% =================================

\subsubsection{Stellar short-term variability }
\label{sec_gr}

\begin{figure}[t!]
\centering
\resizebox{\hsize}{!}{\includegraphics{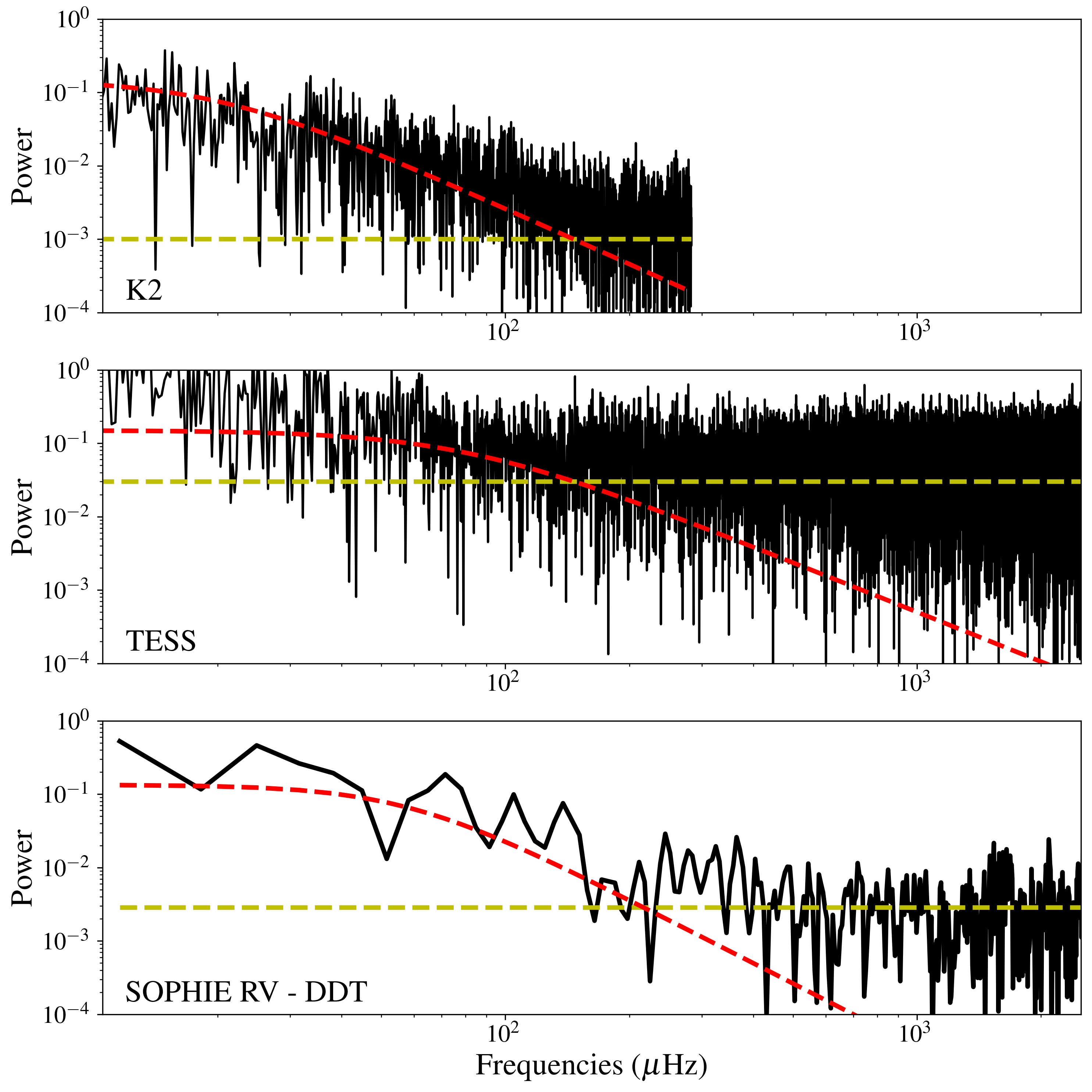}}
\caption{Generalized Lomb-Scargle periodograms of K2, TESS, and SOPHIE RV data for periods of less than one day. The axes are presented in a log-scale format. The yellow and red dashed curves represent best Harvey-function fits to the periodograms, helping for a visual representation of both white Gaussian noise (WGN) and short-term stellar variability.}
\label{fig_ShortTerm}  
\end{figure}

At short timescales, both photometric and RV observations exhibit correlated noise, likely originating from convective phenomena (granulation, supergranulation) and stellar oscillations (p-modes). Below, we analyzed periodograms of both K2 and TESS photometric data (transits masked), and short-cadence SOPHIE RVs within the period range $<1$ day (Fig.~\ref{fig_ShortTerm}).

The K2 periodogram (first panel) shows a typical power increase towards low frequency, indicative of granulation signal \citepads{2014A&A...570A..41K}, but temporal sampling (30 min) hinders precise characterization. The TESS periodogram (middle panel) exhibits a slight power increase, with less significance at low frequency compared to K2, possibly due to the redder wavelength range of TESS observations, reducing granulation amplitude (because the contrast between the rising and falling cells is reduced; e.g., see similar conclusions in \citeads{2023A&A...670A..24S}). A 20s cadence, known to reduce noise and allow the detection of some stellar p-modes signals in TESS data \citepads{2022AJ....163...79H}, could be explored for stellar granulation detection.

The periodogram of SOPHIE RV data (bottom panel) also reveals increased power at low frequency (periods between $50$ min and $6.8$ hours). {Note that this power increase does not change when we remove the RV data that are possibly affected by instrumental systematics (see Sect.~\ref{sec_sophie}).} The short temporal cadence of this RV dataset  (see Table.~\ref{tab_data}) allows to characterize this short-term correlated noise. We then first model the RV periodogram using classical Harvey functions \citepads{1988IAUS..123..497H} with two components \citepads{2014A&A...570A..41K}: a WGN (high-frequency region), and a Lorentzian-like function for the granulation noise. Stellar oscillations, not resolved in our observations, were not modeled\footnote{
While the modes are not resolved, based on predictions from the asteroseismic scaling relations, we expect an oscillation frequency at maximum power of $\nu_\mathrm{max} \sim 2447~\mu$Hz (see Eq.(10) of \citetads{1995A&A...293...87K} with $\nu_\mathrm{max,\odot} = 3150~\mu$Hz). This corresponds to approximately $6.8$ minutes, which is close to the exposure time $\tau_{\text{exp}}$ (see Table \ref{tab_data}). The expected RV amplitude is predicted to be less than $2.8$ m/s (see Eq.(7) of \citeads{2022AJ....164..254G}), which is close to the typical RV errorbars $\sigma_{\text{RV}}$ (see Table \ref{tab_data}).
}.
 Best fitting Harvey functions are shown in Fig.~\ref{fig_ShortTerm} to help the visual inspection. 

For a more accurate modeling of this short timescale stellar variability, we then fitted various GPs with different covariance matrices. The model based on a square exponential (SE) covariance function best reproduced (in terms of likelihood) the RV data (taken individually or jointly). The GP kernel writes as a decreasing function of the time interval $\tau = |t_i-t_j|$:
\begin{equation}
k_\mathrm{SE}(\tau; \Phi_\mathrm{SE}) = \alpha_\mathrm{SE}^2 ~ \exp~ \Big( -\frac{\tau^2}{2 ~\lambda_\mathrm{SE}^2} \Big),
\label{eq_SE}
\end{equation}
with the hyperparameters $\Phi_\mathrm{SE} = \{\alpha_\mathrm{SE}, \lambda_\mathrm{SE}\}$ representing characteristic amplitude and length scale. 
By jointly fitting this GP to both nights, we found $\alpha_\mathrm{SE} = 12.8 \pm 6.0$ m/s and $\lambda_\mathrm{SE} = {2.4 \pm 0.7}$ hours\footnote{We note that when generating synthetic time series from this GP, we find time series with the same RMS as our observations: between $3.9$ and $7.4$ m/s.}. The best-fit model is shown in Fig.~\ref{Fig_gr}. The RMS value of the residuals is $4.3$ m/s. {As a sanity check, we checked the consistency of the inferred GP parameters when excluding the data points taken at the beginning and at the end of the nights (which are suspected to be of instrumental origin).} %Note that we found consistent noise model parameters by analyzing the two nights individually (but the uncertainties are larger).  

In the following sections of this study, we employ these values as priors to model the short timescale stellar signal. However, when applying model \eqref{eq_SE} to the RV data collected during the long RV observation campaign (in Sect.~\ref{sec4}), we anticipate identifying a signal of diminished amplitude. This expectation stems from the longer exposure time ($t_{exp}>15$ min), which attenuates stellar signals, including oscillations \citepads{2019AJ....157..163C}.

% =================================
\begin{table*}[t] \centering
\caption{
Comparison of HD 73344b transit parameters inferred from the K2 data analysis by \citetads{2018AJ....156...22Y}, and from the complete set of photometric data used in this study.  
}
\vspace{-0.1cm}
\label{tab_transit} 
\begin{tabular}{*5c}
\noalign{\smallskip}\hline\noalign{\smallskip}
Parameters & Yu et al. (2018) & K2 & TESS & K2 $+$ TESS \\
\noalign{\smallskip}\hline\hline\noalign{\smallskip}
%%%%%%%%%%%% Fitted parameters
Mid-transit time $T_{0,b}$ (BJD - 2454833)   & $3262.8931^{+0.0020}_{-0.0023}$  & $3262.8958^{+0.0027}_{-0.0030}$ & $3262.9025^{+0.11}_{-0.11}$ & $3262.9003^{+0.0011}_{-0.0010}$\\
Orbital period  $P_\mathrm{b}$ (days) &  $15.61335^{+0.00085}_{-0.00078}$   & $15.61204^{+0.00098}_{-0.00086}$ &  $15.61097^{+0.0011}_{-0.0012}$ & $15.61100^{+0.00017}_{-0.00017}$ \\
Radius ratio $R_{b}/R_\star$ ($\%$)  &  $2.65^{+0.15}_{-0.10}$ & $2.24^{+0.10}_{-0.09}$ & $2.33^{+0.07}_{-0.07}$ & $2.28^{+0.07}_{-0.05}$\\
%%%%%%%%%%%
\noalign{\smallskip}\hline\noalign{\smallskip}
\end{tabular}
\parbox{7in}{
\footnotesize  Notes. Median values and a credibility interval of $68.3\%$ are reported.
}
\end{table*}

%-------------------------------------- 
\section{Characterization of the planetary system}
\label{sec4}

In this section, we carried out a three-step analysis of our data. 
First, we modeled the transits of planet b in the K2 and TESS photometric data to confirm this planet candidate  (Sect.~\ref{sec41}). We then used the planet's ephemeris inferred from this first analysis as priors for the analysis of the SOPHIE RV to evaluate the best strategy to mitigate the stellar activity noise (Sect.~\ref{sec42}). We finally analyzed the photometric and RV data (SOPHIE+HIRES) jointly (Sect.~\ref{sec_combined}). The final adopted parameters result from this last analysis. They are reported in Table~\ref{tab_transit_RV}.

% =================================

\subsection{K2 and TESS transit analyses}
\label{sec41}

We began by jointly analyzing the K2 and TESS observations, containing a total of nine transits of planet b. First, using the Box Least-Square algorithm \citepads{2016ascl.soft07008K}, we performed a transit search analysis but did not detect any transit signatures other than those attributed to planet b.
In the following, we isolated the transit events to save computational time and avoid complex modeling of the stellar activity signals. We took out all data at $3$ and $5$ times the transit duration from the transit center in K2 and TESS light curves, respectively (K2 data are strongly affected by instrumental systematics). The individual transits curves are shown in Appendix~\ref{app_transits}. 

We used the Planet Analysis and Small Transit Investigation Software (\texttt{PASTIS}; \citeads{2014MNRAS.441..983D}) to characterize the nine transits of planet b. 
To account for the different temporal sampling of the K2 (29.6 min) and TESS (2 min) observations, the software oversamples the transit model at the $2$-min rate and then calculates the likelihood over the original rate of the input observations. %This likelihood makes the assumption that the errorbars of each measurement are independent.

The spectral energy distribution (SED) was computed using the \texttt{BT-SETTL} stellar atmosphere models \citepads{2012RSPTA.370.2765A}. The host star was modeled using the Dartmouth stellar-evolution tracks \citepads{2008ApJS..178...89D}. The priors on the stellar parameters ($T_\mathrm{eff}$ , [Fe/H], density $\rho_\star$) were set to follow Gaussian distributions parameterized by the values given in Table~\ref{tab_params}.
{
We used a quadratic law to model the stellar limb darkening for each passband, with parameters ($u_\mathrm{a}$; $u_\mathrm{b}$) interpolated from the \citetads{2011A&A...529A..75C}'s table. 
These interpolations are done for each iteration of the stellar parameters.} %then kept them fixed for both passband. 
%  age determination with stellar isochrones.

Regarding the planet parameters, we used Gaussian priors on the ephemeris from \citetads{2018AJ....156...22Y}, with uncertainties on $P_\mathrm{b}$ and $T_{0,\mathrm{b}}$ enlarged by a factor $100$. For the eccentricity $e_\mathrm{b}$, we used a truncated zero-mean Gaussian distribution with a dispersion of $0.083$ following the recommendation from \citetads{2019AJ....157...61V}. For the other parameters (inclination $i_\mathrm{b}$, radius ratio $R_\mathrm{b}/R_\star$,  argument of periapsis $\omega_\mathrm{b}$), we used  uniform priors.

As the light curves analyzed here are restricted to observations taken in the vicinity of the transits of planet b, we modeled the variability around each of the nine transits by GPs with a SE-type convariance function (see Eq.~\eqref{eq_SE}), with uniform priors on their hyperparameters.

A total of $40$ Markov chains of $500 000$ samples were run. Convergence of each chain was ensured by a Kolmogorov-Smirnov test, and the converged chains were then merged after removing half of the samples as a burn-in phase.

The inferred parameters of planet b are reported in Table~\ref{tab_transit}. The joint analysis indicates a transit depth of $\sim2.2\%$, which corresponds to a mini-Neptune size planet with $R_\mathrm{b} \sim2.8$ R$_\oplus$. The best-fitting model is shown in Appendix~\ref{app_transits}. We reconfirm that the residuals show no signature of spot-crossing events during TESS transits.

By performing a linear propagation of the transit ephemeris from the analysis of the K2 observations alone, we found those estimated from the TESS observations analyzed individually (see Table~\ref{tab_transit}, values compatible within 1$\sigma$). We therefore measure no transit timing variations induced by a nontransiting nearby exoplanet. 

% =================================

\subsection{SOPHIE RV analysis}
\label{sec42}

\begin{figure}[t!]
\centering
\resizebox{\hsize}{!}{\includegraphics{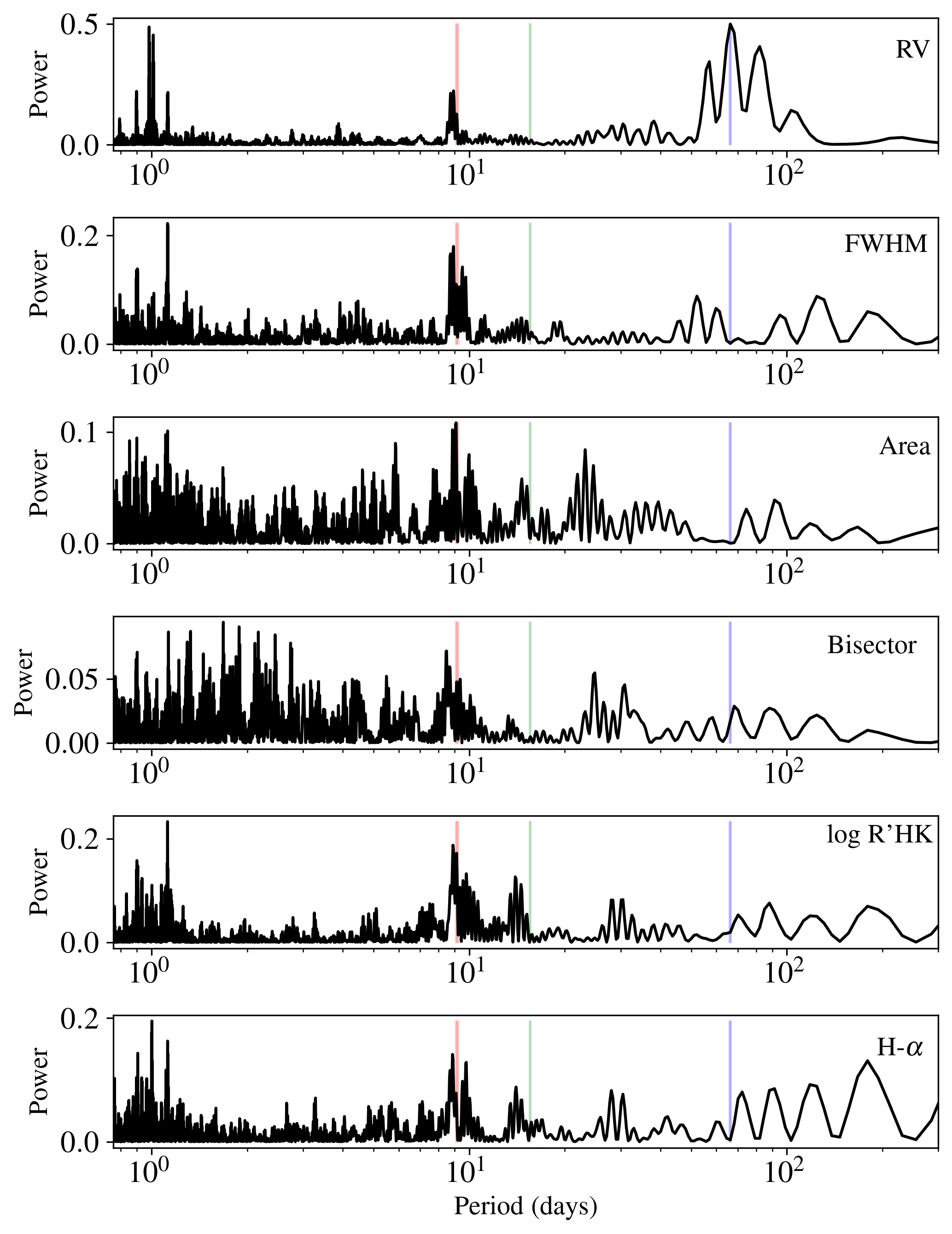}}
\caption{Generalized Lomb-Scargle periodograms for RV and activity indicators. GLSP are calculated on the basis of binned SOPHIE data, where we have eliminated long-term variation using a two-degree polynomial fit. From top to bottom, we show the GLSP for the RV, FWHM, the Area of the Gaussian fit to the CCF, Bisector, $\rm log R'_{\rm HK}$, and H$\alpha$ lines. The red, green, and blue vertical lines indicate the stellar rotation period, the orbital period of the transiting planet (b), and the orbital period of the candidate planet (c), respectively. }
\label{fig_indicators}  
\end{figure}

As a preliminary study based on the SOPHIE time series, we followed the approach of observing the star three times a night and grouping these data points together within each night to mitigate short timescale stellar variability \citepads{2011A&A...525A.140D}. The resulting binned dataset consists of $N=137$ nights of observations, spanning a time coverage of $485$ days across two SOPHIE campaigns.
%The RV RMS is $18.3$ m/s, which is $17$ times larger than the typical RV uncertainties (see Table~\ref{tab_data}). 

In Fig.~\ref{fig_indicators}, we show the GLSP of these observations. The GLSP for RV (top panel) reveals a prominent peak at $\sim66$ days. 
This period does not align with a harmonic of the stellar rotation period, which is identified as the second most dominant peak at around $9$ days (excluding the structure of peaks around one day induced by the time sampling). Importantly, none of the five activity indicators exhibit the presence of a signal with a period $\sim66$ day, as illustrated in the bottom panels.  However, they do exhibit strong periodicities at intermediate periods, specifically in the period range of $3-4$ $P_\mathrm{rot}$. This is discussed in Appendix \ref{App_C1}.
These findings, coupled with supplementary analyses detailed in Appendix \ref{App_C2}, strongly support the existence of a nontransiting planetary candidate\footnote{The reasons for not attempting to report false alarm probability (FAP) levels in Figs.~\ref{fig_indicators} and \ref{fig_RVperiodos} are also given in Appendix \ref{App_C2}.}, called planet c below, with period $P_\mathrm{c} \sim66$ days.
No significant signal is observed in the RV periodogram at the period of the transiting exoplanet ($P_\mathrm{b} \sim 15.6$ days), despite the planet being a mini-Neptune orbiting at close distance to the star. This is a direct consequence of the high level of stellar activity masking the planet's signature. Detailed modeling of the stellar signal is required to detect a peak at this period (see text below).

% PASTIS analysis of SOPHIE binned data: description (1GP, 2 pl) ==============
We then ran the \texttt{PASTIS} software \citepads{2014MNRAS.441..983D} on this dataset to extract the minimum mass estimates of the two planets. We constrained the ephemeris for the transiting planet based on the photometric data analyses (see Sect.~\ref{sec41}), incorporating Gaussian priors on $P_\mathrm{b}$ and $T_{0,\mathrm{b}}$, along with truncated Gaussian priors on the eccentricity $e_\mathrm{b}$. Uniform priors were applied to the argument of periapsis $\omega_\mathrm{b}$, RV semiamplitude $K_\mathrm{b}$, and the five Keplerian parameters of planet c ($K_\mathrm{c}, P_\mathrm{c}, T_\mathrm{p,c}, e_\mathrm{c}, \omega_\mathrm{c}$).
We employed a quasi-periodic (QP) GP model to capture the stellar variability induced by rotational modulation and evolution of the magnetic regions on the stellar surface (\citeads{2014MNRAS.443.2517H}; \citeads{2012MNRAS.419.3147A}; \citeads{2023A&A...674A.108S}). This kernel is defined as:
\begin{equation}
k_\mathrm{QP}(\tau; \Phi_\mathrm{QP}) = \alpha_\mathrm{QP}^2 ~ \rm{exp}~ \Big( -\frac{\tau^2}{2 ~\lambda_{1,\mathrm{QP}}^2} - \frac{2}{\lambda_{2,\mathrm{QP}}^2} \sin^2{\Big[\frac{\pi\tau}{P_\mathrm{rot}}\Big]}\Big),
\label{eq_QP}
\end{equation}
where $\Phi_\mathrm{QP} = \{\alpha_\mathrm{QP}, \lambda_{1,\mathrm{QP}}, \lambda_{2,\mathrm{QP}}, P_\mathrm{rot}\}$ represents the set of hyperparameters corresponding to the characteristic amplitude, decoherence timescales, harmonic complexity (or roughness of the signal), and rotation period. 
The stellar rotation period was constrained by a Gaussian prior based on results from Sect.~\ref{sec31}. Summary of all the priors used in this work is provided in Table~\ref{tab_transit_RV}. We used $40$ Markov chains of $500 000$ samples.

\begin{figure*}[t!]
\centering
\includegraphics[width=\textwidth]{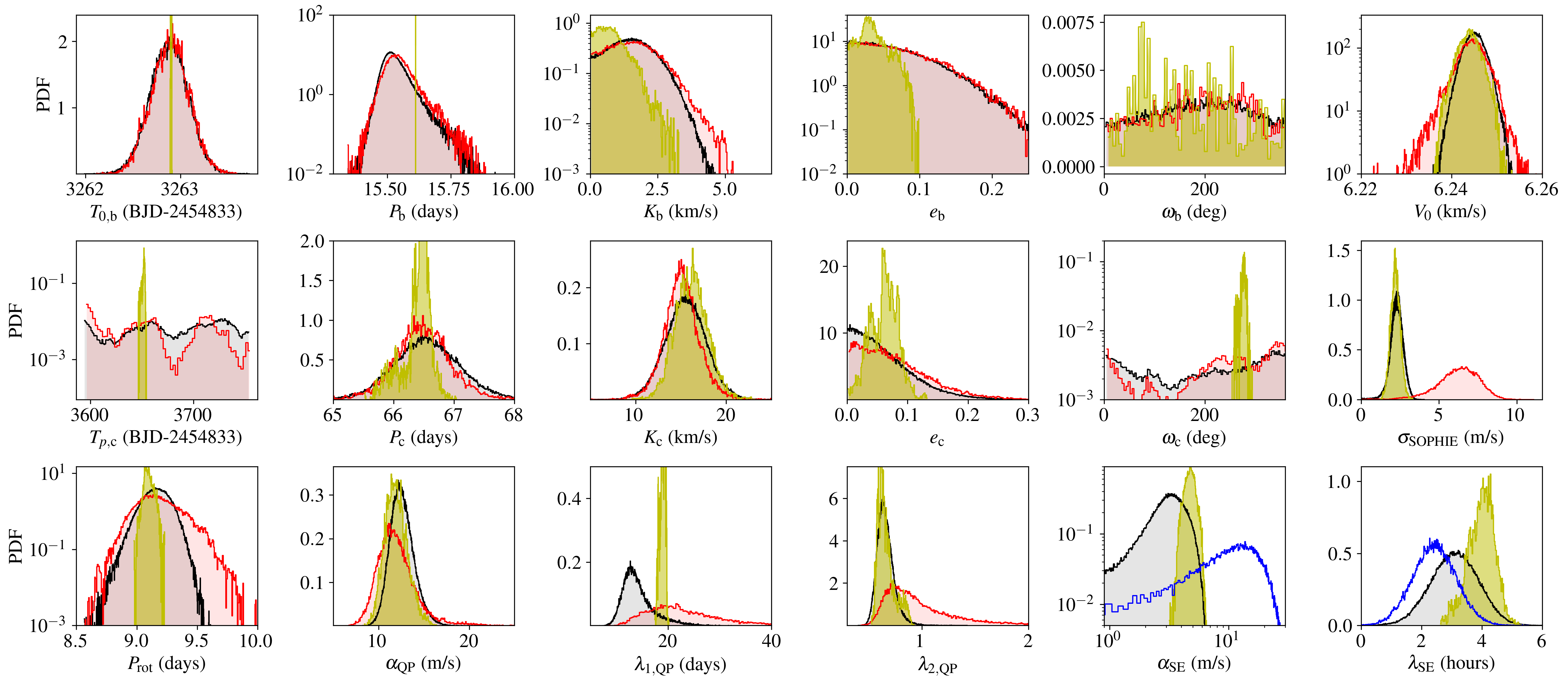}
\caption{Normalized posterior distribution of the parameters fitted to the RV data. The distributions resulting from the analyses of the SOPHIE binned and unbinned observations are shown in red and black, respectively. The  distributions resulting from the joint analysis combining the photometric (K2+TESS) and RV (SOPHIE+HIRES) observations are shown in yellow. \textit{Top}: Five Keplerian parameters of planet b ($T_{0,\mathrm{b}}$, $P_\mathrm{b}$, $K_\mathrm{b}$, $e_\mathrm{b}$, $\omega_\mathrm{b}$) and $V_0$. \textit{Middle}:  Five Keplerian parameters of the candidate planet ($T_{0,\mathrm{c}}$, $P_\mathrm{c}$, $K_\mathrm{c}$, $e_\mathrm{c}$, $\omega_\mathrm{c}$), and the RV jitter ($\sigma_\mathrm{SOPHIE}$). \textit{Bottom}:  GP hyperparameters of the stellar magnetic activity  model ($P_\mathrm{rot}$, $\alpha_\mathrm{QP}$, $\lambda_{1,\mathrm{QP}}$, $\lambda_{2,\mathrm{QP}}$), and the short-term stellar noise model ($\alpha_\mathrm{SE}$, $\lambda_\mathrm{SE}$). For the latter, the values resulting from the analysis of the two SOPHIE nights of observation (Sect.~\ref{sec_gr}) are shown for comparison (blue).}
\label{fig_posteriors}  
\end{figure*}

The posteriors of the best fitting parameters are shown in Fig.~\ref{fig_posteriors}.
For planet b, we found  $K_\mathrm{b}= 1.5 \pm 0.9 $ m/s, $P_\mathrm{b}=15.537 \pm 0.049$ days, and $e_\mathrm{b}=0.058 \pm 0.05$ among the inferred set of parameters. 
For planet c, we found $K_\mathrm{c}= 15.0 \pm 1.8$ m/s, $P_\mathrm{c}=66.46 \pm 0.44$ days, and $e_\mathrm{c}=0.07\pm 0.06$. The RV jitter is estimated to be $\sigma_\mathrm{SOPHIE} = 6.3 \pm 1.3$ m/s, far above the initial RV uncertainties (see Table~\ref{tab_data}). The GP model converged towards a period of $P_\mathrm{rot} =  9.16 \pm 0.17$ days, in agreement with Sect.~\ref{sec_star2}. It converged towards a characteristic amplitude $\alpha_\mathrm{QP} = 11.6 \pm 2$ m/s, and a decoherence timescale $\lambda_{1,\mathrm{QP}} = 22.1^{+11.6}_{-5.9}$ days. 
We note a very wide tail of the posterior distributions of the stellar activity model parameters. This indicates they are not well constrained by RV observations taken at the rate of one point per night.

The left panels of Fig.~\ref{fig_RVperiodos} show the GLSP of RV observations (top), iteratively subtracted (from top to bottom) by the best-fitting models for planet c, stellar activity, and planet b. The final RMS\footnote{RMS is calculated without weighting by RV uncertainties. We have not propagated the GP data correction into the residuals, which is why the RMS is $<\sigma_\mathrm{SOPHIE}$.} of the RV residuals is around $4.4$ m/s. 
In the GLSP of these residuals, we observe strong peaks at short periods. Without correcting the RV time series from these short-term noises, the peak at $P_\mathrm{b}$ is not prominent\footnote{It is worth noting that, in the absence of robust priors on the planet b's ephemeris ($P_\mathrm{b}$, $T_{0,\mathrm{b}}$), which are known from transit photometry, the peak at $P_\mathrm{b}$ completely disappears.} (see third row). 
This stellar signal remains significant over periods longer than a day, affects the RV characterization of the transiting planet, and needs to be corrected for \citepads{2015A&A...583A.118M}.\\

% PASTIS analysis of SOPHIE unbinned data: description (2GP, 2 pl) ==============
To this end, we analyzed the SOPHIE observations without grouping the $3$ data points by night. Based on the detailed analysis of the two full nights taken at high cadence rates (see Sect.~\ref{sec_gr}), we modeled short-timescale stellar variability with an SE covariance function (see Eq.~\eqref{eq_SE}). We parameterized Gaussian priors on these two hyperparameters with the values deduced in Sect.~\ref{sec_gr}. All other priors were kept identical to the analysis of the binned SOPHIE observation.

First, we find that the inferred Keplerian parameters for the two planets are consistent with the analysis of the binned RV data (see Fig.~\ref{fig_posteriors}). This means that adding a second GP to model stellar activity did not degrade the inferred planetary signals. We also observed narrower posterior distribution of the stellar activity parameters, indicating that the model is more well constrained than previously.

Second, when we compare the GLSPs of the unbinned (right panels of Fig.~\ref{fig_RVperiodos}) and binned (left panels) SOPHIE RV dataset, we see the planet b has now the largest peak (third row).  
Moreover, we see no strong residual periodic component in the GLSP of the RV residuals. This leads us to conclude that the main contribution of short-term correlated variability has been well constrained by the second GP noise model. 

Third, we note an RMS of the data residuals of $2.1$ m/s. This RMS corresponds now to the inferred RV jitter $\sigma_\mathrm{SOPHIE} = 2.2 \pm 0.4$ m/s, and is also in agreement with the initial RV uncertainties (see Table~\ref{tab_data}). 

\begin{figure*}[t!]
\centering
\includegraphics[width=\textwidth]{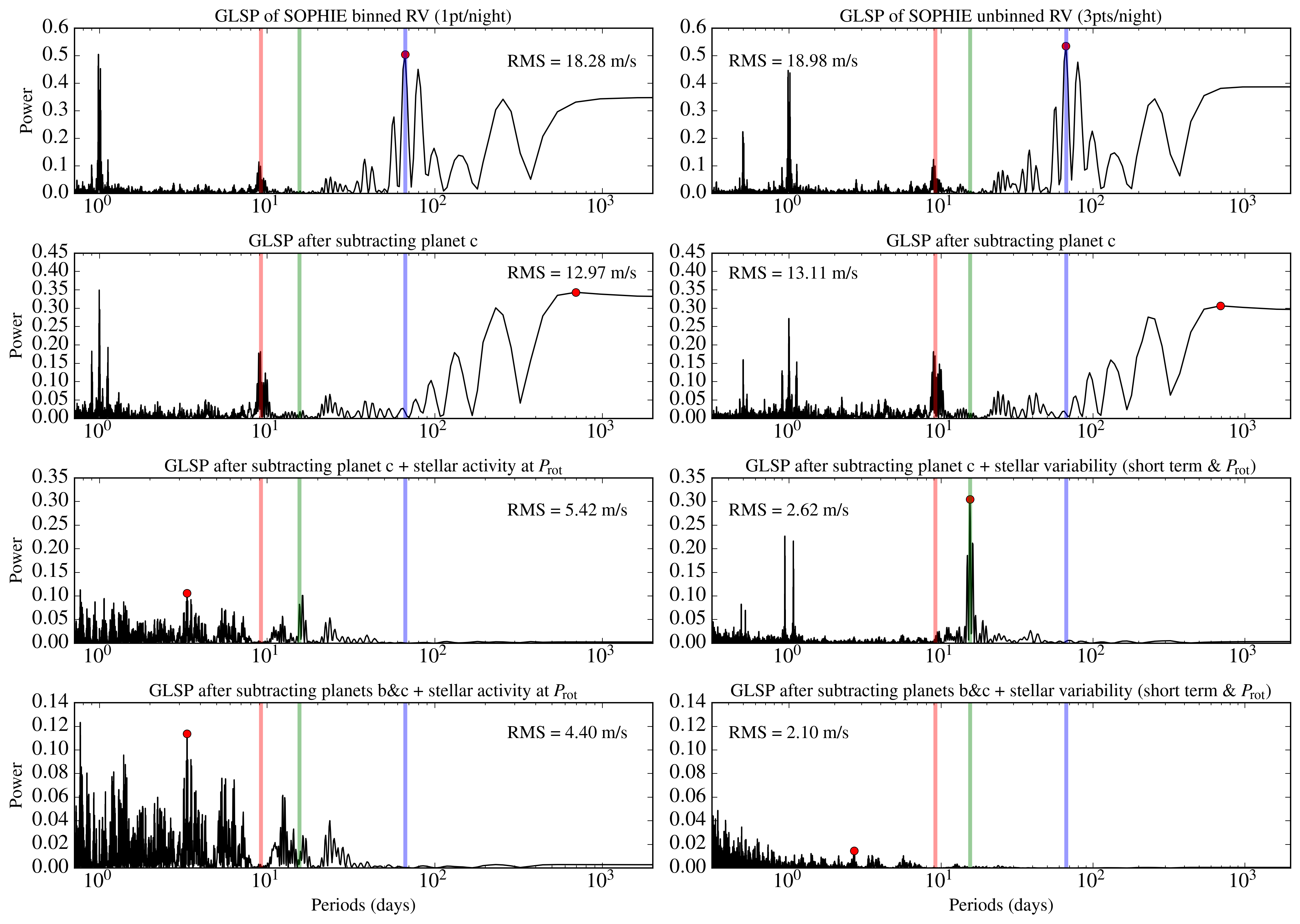}
\caption{Generalized Lomb-Scargle periodograms of the SOPHIE binned (left) and unbinned (right) data. From top to bottom: Raw data, then subtraction of the best-fitting Keplerian model for the planet candidate, the stellar activity model(s), and planet b. For the stellar activity model(s), we considered the mean of the predictive GP distribution resulting from the best fits.
The stellar rotation period, and the orbital period of planets b and c are indicated by the red, green and blue vertical lines, respectively. In each panel, the red dots indicate the highest periodogram peak taken over periods $>1$ day. We note that the planetary and activity signals are jointly estimated from the RV data, and then iteratively removed in the successive panels.
}
\label{fig_RVperiodos}  
\end{figure*}

%-------------------------------------- 
\subsection{Combined photometry and RV analyses}
\label{sec_combined}

Despite the robustness of the inferred results coming from the separate analyses of the K2$+$TESS light curve and the SOPHIE RV, we performed a joint orbital analysis of the photometric and spectroscopic dataset, along with stellar evolution tracks to refine the parameters and derive self-consistent uncertainties in the model parameters (taking into account the underlying correlations between some of them). We also added to this combined analysis the HIRES RV. Both the SOPHIE and HIRES observations are taken at the rate of three data points per night {(the raw HIRES data have been binned into three data points per night, as described in Sect.~\ref{sec_hires})}

We used again the \texttt{PASTIS} software, with the same setting as in Sects.~\ref{sec41} and \ref{sec42}: planet b in transit, planet c not transiting, two GPs to model stellar activity signals in the RV, and nine distinct GPs models for the photometric and instrumental variability observed around the transits of planet b. 
For the stellar parameters ($T_\mathrm{eff}$ , ${\rm log}g$, $[Fe/H]$), we used normal priors centered on the values derived from spectral analysis (see Table~\ref{tab_params}). For distance to Earth (d), we used a normal prior centered on the \textit{Gaia} DR3 value (see Table~\ref{tab_params}), and for stellar extinction E(B-V) we used a uniform prior. 
Priors on each parameter are listed in Table~\ref{tab_transit_RV}. In total the fitting procedure involved $75$ free parameters. The main inferred planet (Keplerian), stellar (GPs), and instrumental parameters are reported in Table~\ref{tab_transit_RV}.  

Overall, the results are in agreement with those derived with the individual analyses of the photometric and RV data.  \\

Concerning the stellar fundamental parameters, the results of this joint analysis are compatible with the results from spectral analysis (Sect. \ref{sec_combined}) withing the $1\sigma$ errorbars. 
Based on isochrones, \texttt{PASTIS} derived an age for the host star of $1.3\pm0.3$ Gyrs, which is not in exact agreement with the age derived from chemical clock relationships (see Sect.~\ref{sec31}). 
This could also be compared with the age estimated with gyrochronological relationships, as proposed by \citetads{2008ApJ...687.1264M}; \citetads{2019JOSS....4.1469A}; or \citetads{2023ApJ...952..131M}.  A visual examination of Figs. 1 and 5 in \citetads{2023ApJ...952..131M} (which relate $P_\mathrm{rot}$ and $S_\mathrm{ph}$ to stellar age) confirms that the age of HD 73344 should be between 1 and 2 Gyrs. The age-activity relationship (using $\rm log R'_{\rm HK}$) described in Eq.(3) of \citetads{2008ApJ...687.1264M} gives an age of $\sim 1.15$ Gyrs. 
Despite the lack of consensus on the stellar age derived by these different techniques (and in the absence of a proper asteroseismology study), all these age estimates nevertheless indicate that HD 73344 is certainly a young star, which is consistent with the high level of activity discussed in Sect.~\ref{sec3}.
 
The orbital period of planet b is refined for the RV analysis thanks to the photometric data, and it gives an RV semiamplitude for the planet smaller than with the analysis based on the RV data alone\footnote{We checked that when we fixed the planet period and analyzed the RV data alone, we found RV semiamplitude in total agreement with the values obtained from the present joint analysis.}.  We find a marginal RV signal for the planet b ($K_\mathrm{b} = 0.667^{+0.559}_{-0.426}$ m/s), despite the fact it is a mini-Neptune ($R_\mathrm{b} = 2.884^{+0.082}_{-0.072}$ $R_\oplus$) planet at short orbit ($P_\mathrm{b}=15.611\pm{0.00003}$). 
The marginal detection of planet b indicates the need of injection tests to secure the estimate of the planet mass, in presence of strong stellar variability noise \citepads{2023A&A...676A..82M}. 
If we consider the $3\sigma$ uncertainties on $K_\mathrm{b}$, we find a signal of $<2.34$ m/s, which is compatible with the RV jitter attributed to both SOPHIE and HIRES instruments. This corresponds to a planet with a mass $M_\mathrm{b} = 2.983^{+2.50}_{-1.90}$ $M_\oplus$ (or $M_\mathrm{b} <10.48$ $M_\oplus$ at $3\sigma$).  
However, this leaves us with a minor constraint on the bulk planet density of $\rho_\mathrm{b}<2.45$ g/cm$^3$ at $3\sigma$. According to the planet distance to its host star, we estimate the equilibrium temperature of the planet to be around $T_\mathrm{eq,b} = 910^{+9}_{-7}$ K (assuming zero albedo) and $T_\mathrm{lock,b} =1066^{+15}_{-12}$~K if the planet is tidally locked to its star (assuming homogeneous redistribution of heat in the atmosphere; \citeads{2011ApJ...726...82C}).

For planet c, we find an RV signal with an amplitude of $K_\mathrm{c}=16.1 \pm 1.8$ m/s. The planet is found on a nearly circular orbit ($e_\mathrm{c} = 0.061 \pm 0.02$). This corresponds to a planet with a minimum mass of $M_\mathrm{c}\sin i_\mathrm{c} = 116.3^{+12.8}_{-13.0}$ $M_\oplus$ $\sim 0.37\pm 0.04$  $M_\mathrm{J}$.

For the stellar activity parameters, we find values fully consistent with the previous analyses (see Fig.~\ref{fig_posteriors}). The stellar rotation period is $P_\mathrm{rot} = 9.09 \pm 0.04 $ days. The signatures of the stellar activity sources that are modulated with the stellar rotation have an amplitude around $11.8$ m/s, which is large compared to the signal of the transiting planet. It evolved over long timescales ($\lambda_{1,\mathrm{QP}} \sim 19$ days), close to the $\sim15$ days orbital period of the transiting planet.
The short-term stellar variability signals is correlated over longer timescales than solar-like stars ($\lambda_\mathrm{SE}=4.0\pm0.4$ hours), and generate a significant RV noise ($\alpha_\mathrm{SE}=4.8\pm0.5$ m/s), also above the RV signal of the transiting planet.  As we anticipated in Sect.~\ref{sec_gr}, the amplitude of the granulation signal is smaller than the one derived from the two nights of SOPHIE observations. This is explained by the longer exposure time used during the long observation campaign.

\begin{figure*}[t!]
\centering
\includegraphics[width=\textwidth]{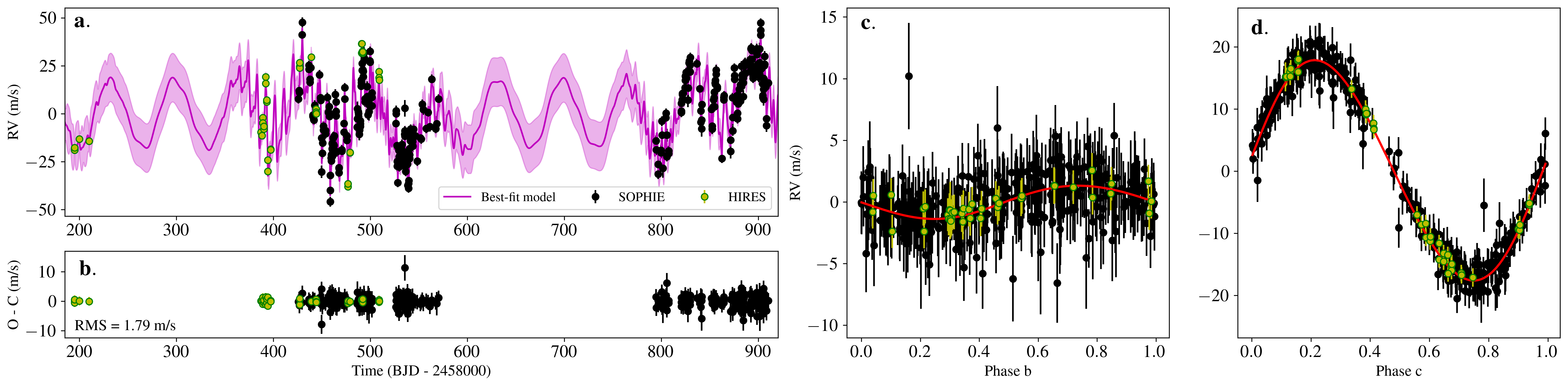}
\caption{Radial velocity of HD 73344 resulting from the joint analysis of the photometric and RV data. {Panel (a):} RV of SOPHIE (black) and HIRES (yellow) observations and best fitting model (purple). {Panel (b):} RV residuals. {Panel (c):} RV phased at the period of planet b (planet c and activity models subtracted). {Panel (d):} RV phased at the period of planet c (planet b and activity models subtracted).  Best-fitting models for planets b and c are shown in red.}
\label{fig_RV}  
\end{figure*}

\begin{figure}[h!]
\centering
\resizebox{\hsize}{!}{\includegraphics{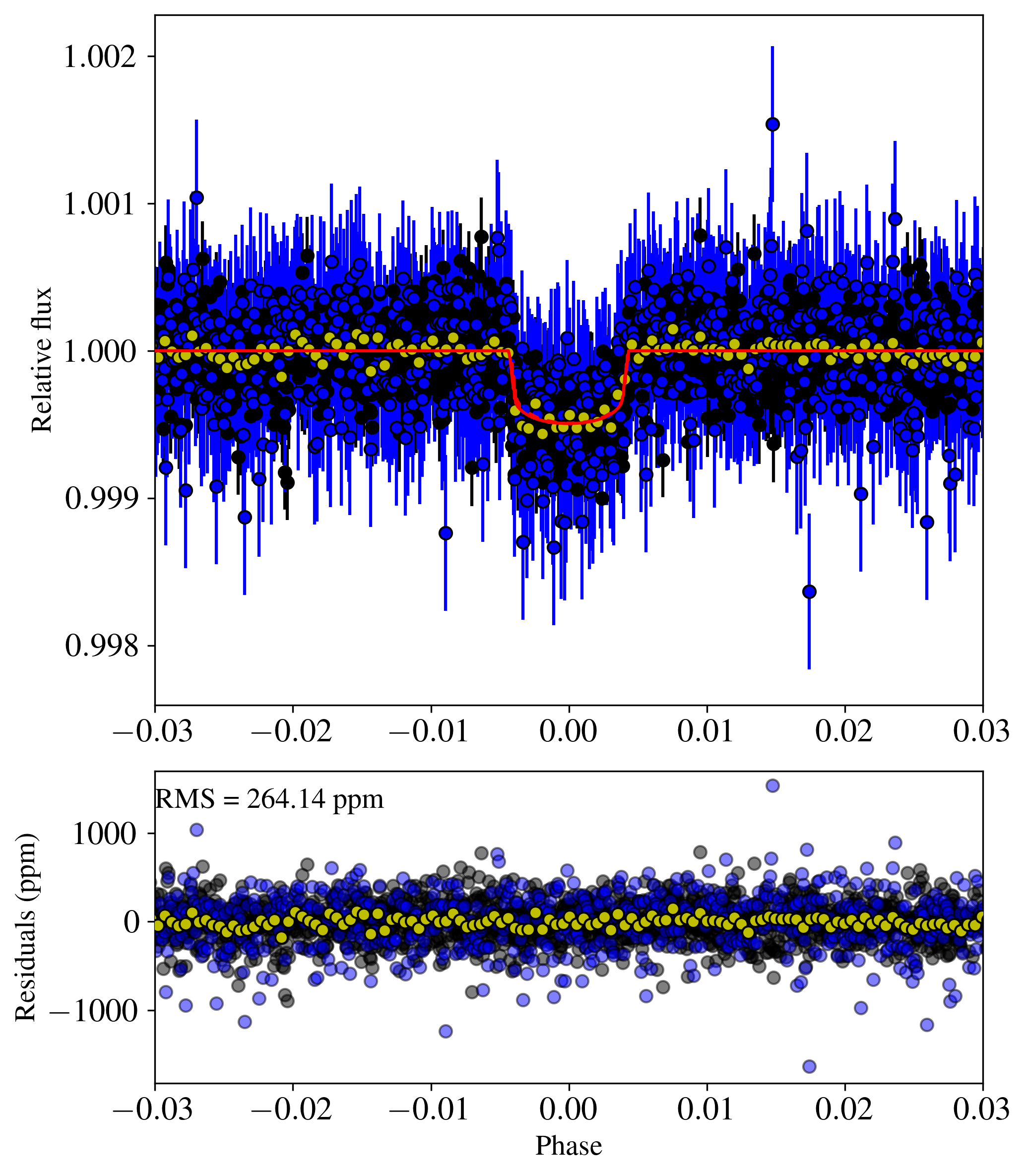}}
\caption{Phased light curve of HD 73344b  resulting from the joint analysis of the photometric and RV data. K2 data are shown in black, TESS data in blue. These light curves have been corrected by the best-fitting noise models. The best-fitting transit model is shown in red. Yellow dots represent the $120$min binned light curves (K2 and TESS combined).}
\label{fig_transits}  
\end{figure}

The final RV observations are shown in Fig.~\ref{fig_RV}, and the phased folded transit light curves in Fig.~\ref{fig_transits}. The RMS of their residuals are $1.79$ m/s and $264$ ppm, respectively.

%\clearpage
\subsection{Spitzer data analysis}
\label{sec_spitzer}

\begin{figure}[t!]
\centering
\resizebox{\hsize}{!}{\includegraphics{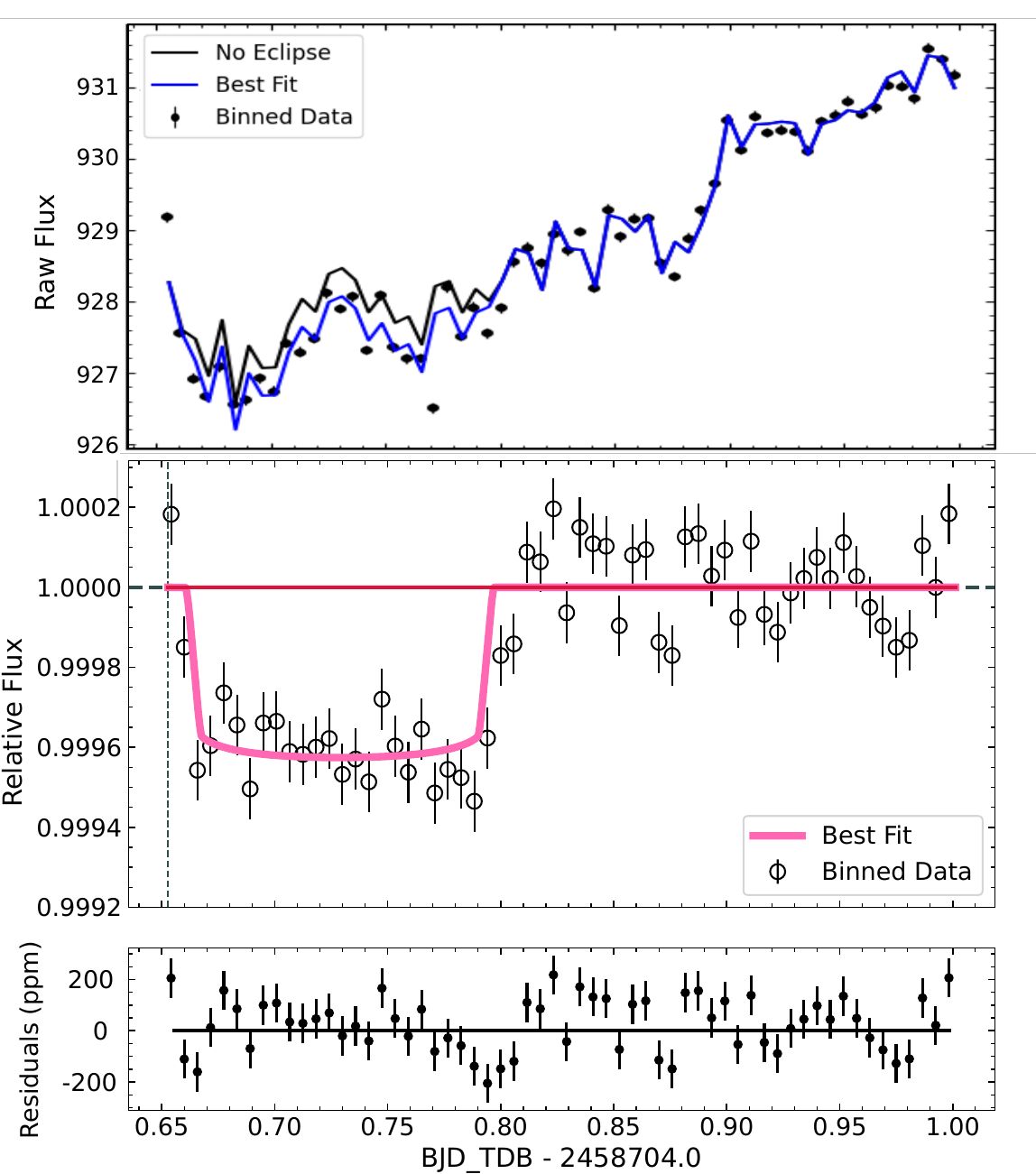}}
\caption{
{\em Spitzer} observations of HD 73344. {\em Top}: Raw {\em Spitzer} light curve (black points) of HD 73344b and best-fit models with (blue) and without (black) a transit fit. {\em Middle:} Detrended  {\em Spitzer} data and best-fit transit light curve.   {\em Bottom:} Residuals to the fit.
}
\label{fig_spitzer}  
\end{figure}

We analyzed the {\em Spitzer} photometry of HD~73344 using the \texttt{POET}\footnote{\url{https://github.com/kevin218/POET}} ({Photometry for Orbits, Eclipses, and
Transits}) code
\citep{stevenson:2012a,cubillos:2013}, following the same analysis approach as used by \cite{crossfield:2020}.  We identified a 2.5~pix aperture with 0.01~pix resolution on the pixel map  as giving the optimal precision. Because the available {\em K2} and TESS light curves give tighter constraints, we held all transit parameters fixed except the time of transit and $R_\mathrm{b}/R_\star$.  The quadratic limb-darkening coefficients were constrained by Gaussian priors, set to the mean and standard deviation of all the values tabulated by \cite{claret:2013} for model grid parameters closest to the stellar parameters of HD~73344 --- that is, $u_a=0.0106\pm0.0670$ and $u_b=0.229\pm0.142$. The final detrended light curve and best-fit transit model are shown in Fig.~\ref{fig_spitzer}; the median and standard deviation on the derived parameters are $T_\mathrm{0,b; Spitzer}=2458704.7289\pm0.0014$ BJD and $R_\mathrm{b; Spitzer}/R_\star=0.0203\pm0.0017$. The transit depth observed by \textit{Spitzer} in the NIR being fully compatible with the optical one measured by K2 and TESS, it confirms the achromatic property of this planetary transit \citep{2012ApJ...745...81F, 2015ApJ...804...59D} which nature is now considered as validated.% (see Appendix \ref{App_FP}).

%\newpage
%-------------------------------------- 
\section{Discussion}
\label{dis}

In this section, we first discuss the stability of the two-planet system (Sect. \ref{sec:stability}). We then study how we can constrain the inclination of the candidate planet from the transit probability of planet b (Sect. \ref{sec:proba}). We conclude with a general discussion of the composition of the transiting planet's interior, bearing in mind that our knowledge of this composition is drastically limited by the impact of stellar activity on the measurement of the planet's mass (Sect. \ref{sec:compo}).

% =================================
\subsection{Stability of the two-planet system}
\label{sec:stability}

We investigated whether it is possible to refine the orbital parameters of planets HD\,73344\,b and c using constraints on their orbital stability (see e.g., \citeads{Stalport-etal_2022}), and perhaps constrain the true mass of planet~c. To this end, we used a similar approach to what has been done to other two-planet systems by \citetads{Correia-etal_2005}, \citetads{Correia-etal_2009}, \citetads{Laskar-Correia_2009}, or \citetads{Couetdic-etal_2010}.

The stability of a dynamical system can be quantified by running a frequency analysis on the output of a numerical integration (\citeads{Laskar_1988}; \citeads{Laskar_1993}; \citeads{Laskar_2005}). In our case, we are interested in the short-term stability of the planets\footnote{Many planetary systems, including our Solar System, are not stable in the long term. Long-term stability can therefore not be used as a constraint to refine orbital parameters (see \citealp{Laskar-Petit_2017}).}, and so it is enough to analyze their mean longitude $\lambda$, which is related to their orbital motion. For a given planet, we define the stability coefficient
\begin{equation}
   \delta = \frac{|n_2 - n_1|}{n_0}\,,
\end{equation}
where $n_1$ and $n_2$ are the `mean' mean motions (i.e., the linear part of $\lambda$) obtained by running a frequency analysis on the first and second half of a numerical integration, and $n_0$ is a reference value, taken here to be the mean motion of the planet for its best-fit parameters. For a stable system, $\delta$ should be $0$ up to the numerical accuracy of the analysis, while values close to $1$ or above denote strongly unstable systems. As planet~b is much less massive than planet~c, it is much more sensitive to chaos, so we focus here on its stability coefficient $\delta_\mathrm{b}$.

We integrated the system with the numerical scheme SABA(10,6,4) of \citetads{Blanes-etal_2013} implemented in the \texttt{REBOUND} package \citepads{Rein-etal_2019}, and performed the frequency analysis with the dedicated function of \textsc{TRIP} \citepads{Gastineau-Laskar_2011}. We set the duration of our integrations to $200$~years, which represents about $4700$ orbits of planet~b and $1100$ orbits of planet~c. On such a short duration, tidal dissipation and general relativistic precession can be neglected. We checked that the oblateness of the star (expected to be $J_2\approx 10^{-7}$; see \citealp{Batygin-Adams_2013,Spalding-Millholland_2020}) also produces negligible orbital perturbations compared to planet-planet interactions. We first mapped the parameter space around the best-fit solution, by varying the semimajor axis and eccentricity of planet~c on a regular grid while all other parameters were set to their nominal values. As planet~c is only detected in radial velocity, there is a degeneracy between its mass $M_\mathrm{c}$ and inclination $i_\mathrm{c}$ with respect to the sky plane; hence, we repeated the same experiment for several values of $i_\mathrm{c}$ and modified $M_\mathrm{c}$ accordingly. In these simulations, we chose equal longitudes of node in the sky plane for the two planets, such that their mutual orbital inclination is simply $i_\mathrm{b}-i_\mathrm{c}$, where $i_\mathrm{b}\approx 88^\circ$ (see Table~\ref{tab_transit_RV}).

\begin{figure}
   \includegraphics[width=\columnwidth]{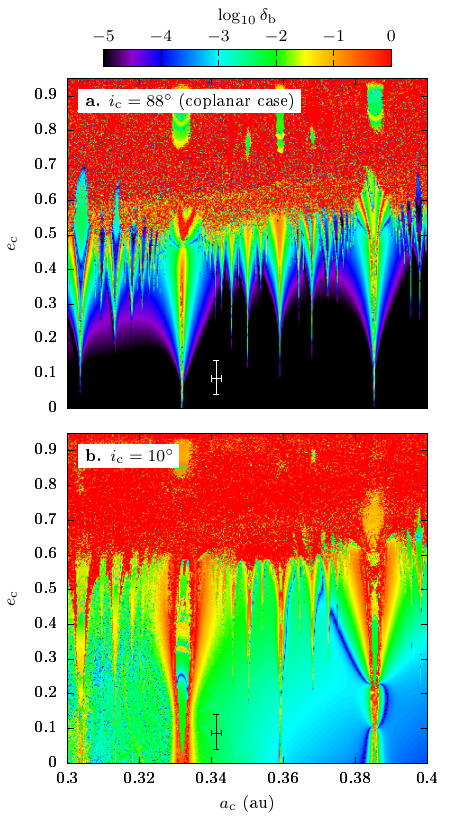}
   \caption{Stability of the HD\,73344 system as a function of the semimajor axis and eccentricity of planet~c. The cross shows the best-fit values and their $1$-$\sigma$ uncertainty interval. The colour scale is chosen so that dark blue denotes stable orbits and red is highly unstable. The inclination chosen for planet~c is labelled; its corresponding mass is $M_\mathrm{c}=0.36$~$M_\mathrm{J}$ for panel~{a} and $M_\mathrm{c}=2.05$~$M_\mathrm{J}$ for panel~{b}.}
   \label{fig:stabilitymap}
\end{figure}

Figure~\ref{fig:stabilitymap} shows the result obtained for $i_\mathrm{c}=88^\circ$ and $10^\circ$. We checked that integrating over a longer duration (e.g., $2000$~years) does not substantially alter our maps; this shows that $200$~years is long enough here for the frequency analysis to give a pertinent result. We rule out a strong mean-motion resonance between the two planets. The closest large resonance visible in Fig.~\ref{fig:stabilitymap} is the 4:1 mean-motion resonance (vertical structure on the left of the best-fit location), but it is more than $3$-$\sigma$ away from the most probable parameters of the planets. For the coplanar configuration, for which planet~c has its minimum mass, the best-fit solution lies in a very stable region (Fig.~\ref{fig:stabilitymap}a). If the sky-plane inclination of planet~c is small, however, its corresponding large mass and large mutual inclination with respect to planet~b are a source of instability (Fig.~\ref{fig:stabilitymap}b). We must therefore quantify the emergence of this instability.

Using the posterior distribution of the system's parameters obtained from the joint analysis in Sect.~\ref{sec_combined}, we computed the histogram of $\log_{10}\delta_\mathrm{b}$ for various inclinations of planet~c from $i_\mathrm{c}=88^\circ$ (coplanar case) to $i_\mathrm{c}=1^\circ$ (almost perpendicular case). Examples of the histograms obtained can be found in Appendix~\ref{asec:stability}. For inclination values $i_\mathrm{c}$ larger than about $30^\circ$, the posterior distribution of $\log_{10}\delta_\mathrm{b}$ has a single peak located below $-4$; this means that the whole sample is stable (similarly to the black and dark blue regions in Fig.~\ref{fig:stabilitymap}). For $5^\circ\lesssim i_\mathrm{c}\lesssim 30^\circ$, an unstable subsample appears as a second peak located above $-4$. As we decrease $i_\mathrm{c}$, and therefore increase the mass of planet~c, this unstable subsample grows. For $i_\mathrm{c}\lesssim 5^\circ$, there is again a single peak in the distribution, but located above $-2$; this means that the whole sample is now unstable (similarly to the yellow and red regions in Fig.~\ref{fig:stabilitymap}).

From this analysis, we deduce that the dynamical stability of the system would be able to constrain the planets' parameters only if $i_\mathrm{c}\lesssim 30^\circ$. However, such a small inclination $i_\mathrm{c}$ would correspond to a very large mutual inclination between the two planets, which we consider unlikely. First, the statistical distribution of multiplanetary systems shows that planets having small eccentricities tend to have small mutual inclinations, and vice versa \citepads{Xie-etal_2016}. This can be understood by the statistical equipartition of angular momentum deficit as a result of chaotic diffusion (see \citeads{Laskar-Petit_2017}). Second, a large mutual inclination would result in a fast precession of the orbital plane of planet~b in and out of transiting configuration, which would reduce its transit probability (see e.g., \citeads{Becker-Adams_2016}). In the next section, we use this last property to put more stringent constraints on the unknown parameters.

\subsection{Transit probability of planet~b}
\label{sec:proba}

\begin{figure*}
   \centering
   \includegraphics[width=\textwidth]{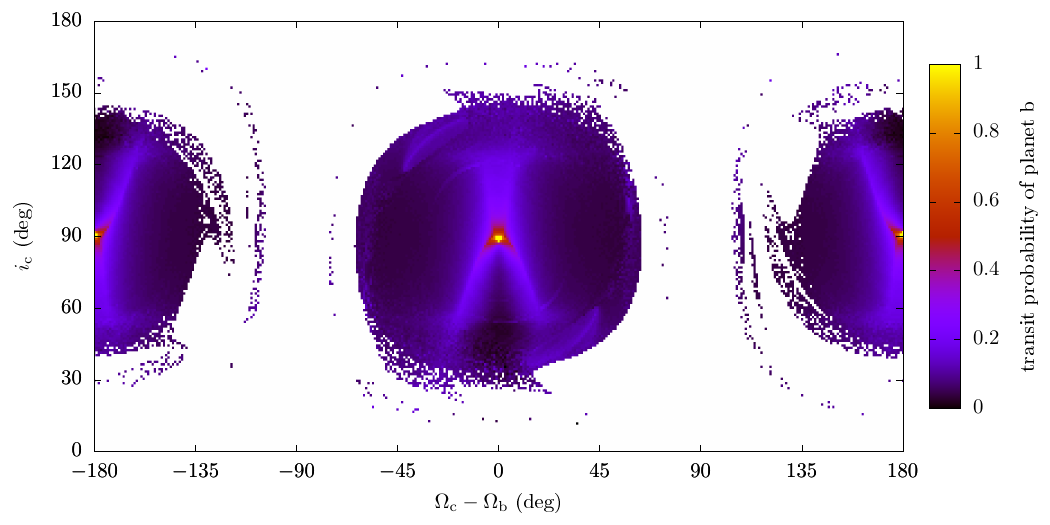}
   \caption{Transit probability of planet HD\,73344\,b computed as the fraction of time its impact parameter is smaller than $1$. The map is drawn as a function of the unknown longitude of node and inclination of planet~c with respect to the sky plane. Other parameters are set to their best-fit values (see Table~\ref{tab_transit_RV}). The transit probability (colour scale) is obtained from a $50$~kyr numerical integration (see text). Points are coloured white if a planet is ejected before the end of the simulation.}
   \label{fig:transitmap}
\end{figure*}

\begin{figure*}
   \centering
   \includegraphics[width=\textwidth]{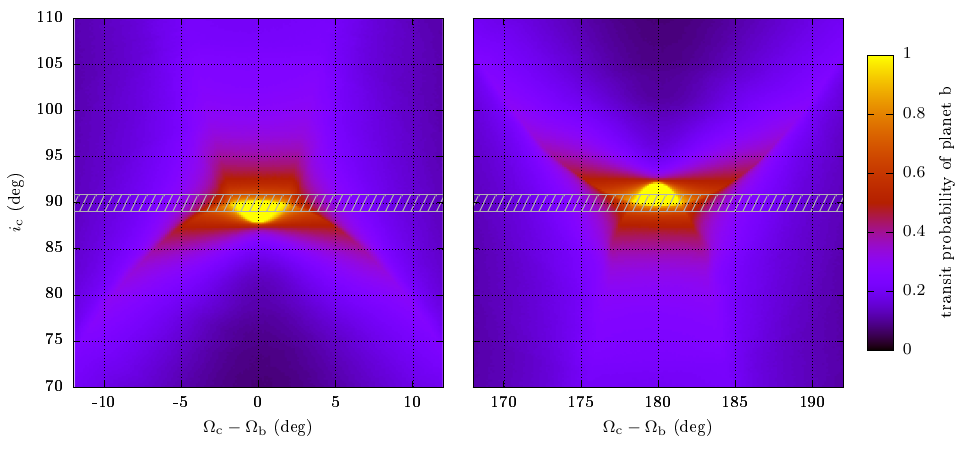}
   \caption{Same as Fig.~\ref{fig:transitmap}, but showing enlarged views. In the hatched band, planet~c would be observed to transit the star; this region is therefore excluded.}
   \label{fig:transitmap_zoom}
\end{figure*}

\begin{figure*}[th!]
   \centering
   \includegraphics[width=0.96\textwidth]{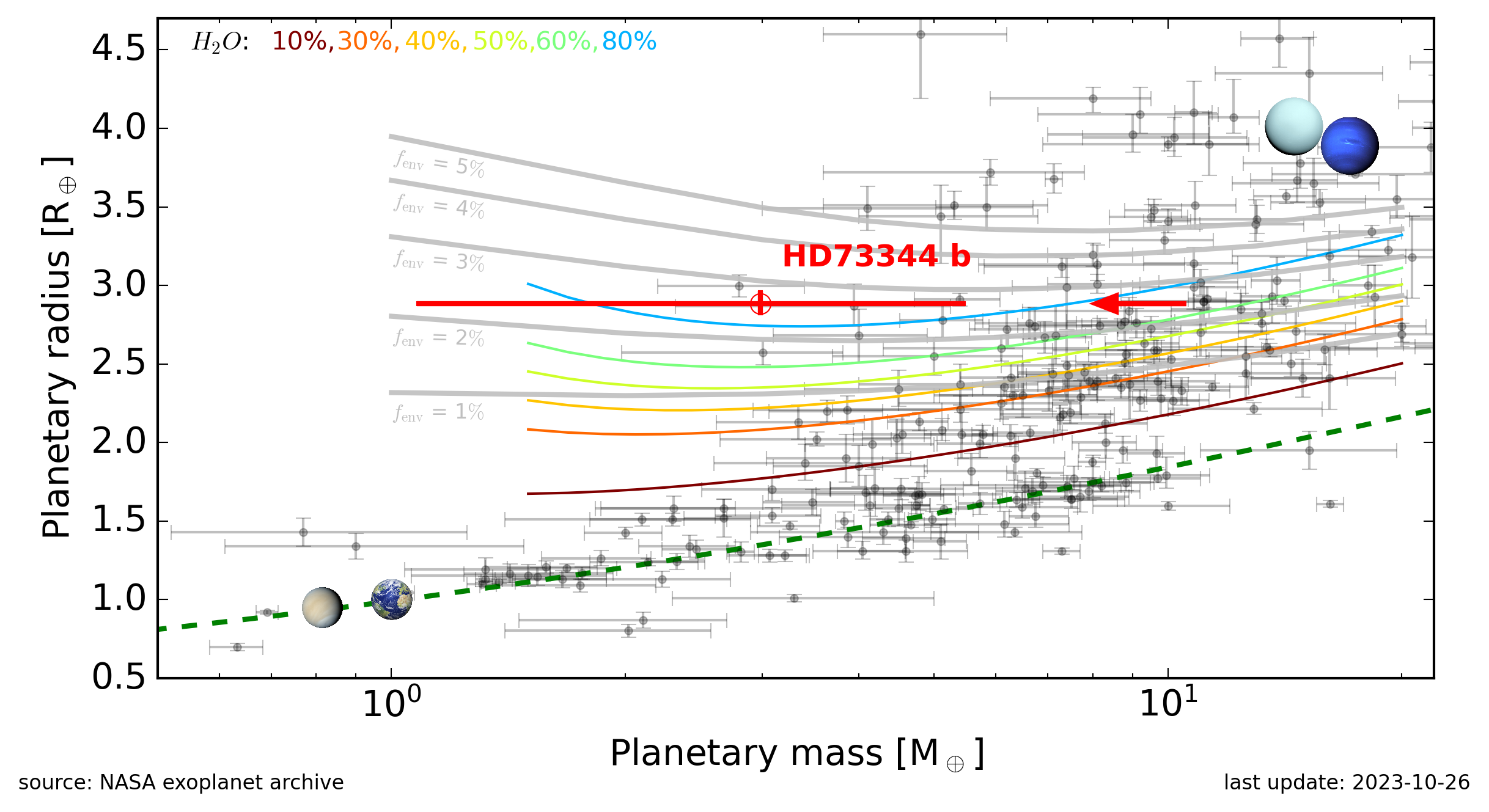}
   \caption{
   Mass--radius diagram of confirmed exoplanets with mass and radius precision better than $3\sigma$ and $10\sigma$, respectively. 
   Planetary parameters (black dots) are taken from the NASA Exoplanet Archive (\url{https://exoplanetarchive.ipac.caltech.edu}), and updated to October 2023. HD 73344b is represented by the red dot. The red arrow represents the upper $10\sigma$ limit on its mass. Composition models calculated from \citetads{2021ApJ...914...84A} with a core mass fraction of $30\%$ and various H$_2$O mass fraction (from $10\%$ to $80\%$) are represented by coloured lines (see text). Composition models calculated from \citetads{Lopez2014}
   for a $1$ Gyr-old system, $T_\mathrm{eq,b} \sim 900$ K, and an envelope mass fraction $f_\mathrm{env} = [1, 2, 3, 4, 5]~\%$ are shown in gray. 
   For comparison, the relation for an Earth-like composition ($32.5\%$ core,  $67.5\%$ mantle) from \citetads{2017ApJ...850...93B} is shown by the green dashed curve, and some solar system planets (Venus, the Earth, Uranus and Neptune) are displayed. 
   }
   \label{fig_compo}
\end{figure*}

During the orbital evolution of the system due to mutual planetary perturbations, the fraction of time the orbit of planet~b passes in front of the star (as observed today) gives an indication of the likelihood for the considered parameters. Assuming that the system does not contain additional unseen planets, the only parameters that are unconstrained by transit and RV data are the inclination of planet~c (linked to its mass through $M_\mathrm{c}\sin i_\mathrm{c}$) and the longitudes of node of the two planets in the sky plane. As the choice of origin for measuring the longitudes is arbitrary, only the difference $\Omega_\mathrm{c} - \Omega_\mathrm{b}$ actually matters, which reduces the unknown parameters to only two.

Figure~\ref{fig:transitmap} shows the transit probability of planet~b as a function of the two unknown parameters. For each pixel of the figure, a numerical integration is performed over $50$~kyr and the fraction of time steps the orbit of planet~b passes in front of the star is recorded\footnote{We checked that increasing the integration duration beyond $50$~kyr does not affect the probability values in a visible way; the integration duration is always more than ten times the period of the inclination precession cycles of planet~b.}. We used the integration scheme SABA(10,6,4) of \citetads{Blanes-etal_2013}, with the inclusion of the general relativistic precession implemented in the same way as \citetads{Saha-Tremaine_1994}.

The circular features in Fig.~\ref{fig:transitmap} roughly correspond to curves of constant mutual inclination between the two orbits. As obtained in Sect.~\ref{sec:stability}, a large mutual inclination is ruled out because it would make the system unstable. Figure~\ref{fig:transitmap} shows that the transit probability of planet~b sharply peaks at $100\%$ in two very small regions. These regions correspond to near coplanarity between the orbits of the two planets, which are either prograde ($\Omega_\mathrm{c}-\Omega_\mathrm{b}\approx 0^\circ$) or retrograde ($\Omega_\mathrm{c}-\Omega_\mathrm{b}\approx 180^\circ$) between each other. We point out that observational data cannot tell whether the inclination value of planet~b is $i_\mathrm{b}$ or $180^\circ - i_\mathrm{b}$. In the latter case, the corresponding transit probability map is obtained from Fig.~\ref{fig:transitmap} by the transformation $i_\mathrm{c}\rightarrow 180^\circ-i_\mathrm{c}$ and  $\Omega_\mathrm{c}\rightarrow 180^\circ+\Omega_\mathrm{c}$.

A zoom-in view of the regions of highest transit probability can be seen in Fig.~\ref{fig:transitmap_zoom}. As planet~c is not observed to transit the star, its inclination today is necessarily smaller than about $89.1^\circ$ (or larger than $180^\circ$ minus this value). Then, if we require the transit probability of planet~b to be $\mathcal{P}>0.9$, Fig.~\ref{fig:transitmap_zoom} gives that the inclination of planet~c must be $i_\mathrm{c}\gtrsim 87.9^\circ$ (or smaller than $180^\circ$ minus this value). Likewise, we obtain $i_\mathrm{c}\gtrsim 87.4^\circ$ for $\mathcal{P}>0.5$, and $i_\mathrm{c}\gtrsim 80.0^\circ$ for $\mathcal{P}>0.3$. The constraint obtained here is therefore much more stringent than the mere dynamical stability of the system (see Sect.~\ref{sec:stability}).

To estimate the uncertainty on the probability values depicted in Fig.~\ref{fig:transitmap_zoom}, we generated a similar map where, for each pixel, we propagated the full posterior distribution of the MCMC fit to the data (see Sect.~\ref{sec_combined}) instead of just the best-fit solution. This was made possible by reducing the resolution of the map to $15\times 15$ and by propagating the trajectories using the Lagrange-Laplace theory (see e.g., \citeads{Murray-Dermott_1999}). The statistics obtained for each pixel show that the $1$-$\sigma$ uncertainty on the probability value is everywhere smaller than $0.01$. This uncertainty translates into a $2$-$\sigma$ range of less than $0.1^\circ$ on the values of $i_\mathrm{c}$ cited above for $\mathcal{P}>0.9$ and $\mathcal{P}>0.5$, and a $2$-$\sigma$ range of about $2^\circ$ on the value cited for $\mathcal{P}>0.3$. These very small uncertainties mainly come from the tight constraint that we have on the inclination of planet~b (see Table~\ref{tab_transit_RV}).

The above analysis translates into a likelihood estimate for the inclination and mass of planet~c; however, we still do not have a direct measurement of their values. A way to break this degeneracy would be to detect the nodal precession of planet~b through a variation of its impact parameter $b_\mathrm{b}$ over time (see e.g., \citealp{Judkovsky-etal_2022}). We examined the behavior $b_\mathrm{b}$ in our simulations, and conclude that for an uncertainty of about $0.02$ on the measurement of $b_\mathrm{b}$ (see Table~\ref{tab_transit_RV}), we would need observations spanning at least ten years in order to detect a substantial noncoplanarity of several degrees between the orbits of the two planets. Conversely, no variation in $b_\mathrm{b}$ should be detectable if the system lies in the most likely region of Fig.~\ref{fig:transitmap_zoom}, which corresponds to near coplanarity between the two orbits.

% =================================
\subsection{Internal composition of planet b}
\label{sec:compo}

 This section proposes a preliminary investigation of the internal composition of planet b. However, it is crucial to bring to the reader's attention the significant degeneracy observed among the various parameters (e.g., core mass fraction and envelope size) of planet internal composition models. This degeneracy is particularly critical when dealing with planets with large uncertainties on their fundamental parameters (mass, radius), as it is the case here with the large uncertainty in the planet mass.
 
In the mass-radius diagram, planet b stands out the rocky super-Earth population and appears as a likely member of the sub-Neptune population (see Fig.~\ref{fig_compo}). 
At first, we compare the position of planet b in the mass-radius diagram with simple mass-radius relationships.  Given the mean density of the planet and its strong irradiation ($T_\mathrm{eq,b} = 910 \pm 7$ K assuming a zero albedo, see Sect. \ref{sec_combined}), we considered the mass-radius relations inferred from water-rich composition models (\citeads{Mousis2020}; \citeads{2021ApJ...914...84A}; \citeads{Acuna2022}). These models assume a water-dominated atmosphere either in vapour or supercritical state, on top of a high-pressure water layer or a mantle. 
For interior analysis, we computed the mass-radius relationships by employing Eq. (29) from \citetads{2021ApJ...914...84A} for different H$_2$O mass fractions at an equilibrium temperature of $T_\mathrm{eq} = 900$ K. We assumed a core mass fraction of 30\%, which aligns with the characteristics of an Earth-like interior.
Results are shown with the colored lines in Fig.~\ref{fig_compo}. This preliminary comparison shows that HD 73344b is compatible with a very high water-mass fraction of at least $80\%$. A more quantitative analysis also confirms that the water content is at least $75\%$ (see Appendix \ref{asec:interior}).

 However, such a high value is well above the most water-rich bodies in the solar system, such as the icy moons. This suggests that a water-dominated atmosphere is not inflated enough to account for the low density of the planet. We performed then a similar analysis but assuming that the envelope is made of gas of solar composition (i.e., dominated by \ce{H2} and He), on top of a core of Earth-like composition. To do so, we used the tabulated mass-radius relations from the interior structure model of \citetads{Lopez2014}. Intermediate mass-radius relations were obtained by linear interpolation on the grid provided in \citetads{Lopez2014}. In this case, we found that HD 73344b properties are compatible with an envelope that accounts for $2-3\%$ of the mass of the planet (see gray lines in Fig.~\ref{fig_compo} and Appendix \ref{asec:interior}). 
 
If we consider the scenario described above as realistic, it implies that HD 73344b lies in a parameter space where extreme atmospheric escape of hydrogen is expected (\citeads{Fossati2017}; \citeads{Zahnle2017}; \citeads{Rogers2023}). The restricted Jeans escape parameter can be computed as $\Lambda = G M_{\mathrm{p}} m_{\mathrm{H}} / (k_{\mathrm{B}} T_{\mathrm{eq}} R_{\mathrm{p}})$ \citep{Fossati2017}, where $m_{\mathrm{H}}$ is the mass of the proton. Studies of \cite{Owen2016,Cubillos2017,Fossati2017} conclude that planets with $\Lambda < 10-15$ are hydrodynamically unstable and their atmospheres must experience extreme atmospheric escape (e.g., Parker wind or boil-off). The analysis of \cite{Vivien2022} showed that pure water atmospheres can remain stable down to values of $\Lambda$ as small as $\sim 0.5$. For HD 73344b $\Lambda = 8.5^{+6.0}_{-5.6}$, meaning that a hydrogen dominated atmosphere is unlikely. To account for the planet's unusually large radius, the atmosphere and envelope would have to be a mixture of hydrogen and heavier volatile elements. In this case, an atmosphere with a higher mean molecular weight would result in a reduced escape rate, potentially explaining how this planet has retained its volatile envelope.
This underlines the need for spectroscopic characterization to break the degeneracy on the planet's composition. It also shows that HD 73344b is a promising target for testing the escape of \ce{H2} and He.
We calculated HD~73344b's Transmission Spectroscopy Metric (TSM; \citeads{2018PASP..130k4401K}) while propagating all parameter uncertainties and found TSM=$260^{+640}_{-30}$. Despite the large upper error bar (corresponding to lower planet masses), this metric suggests that HD~73344b could be a particularly promising target for transmission spectroscopy. However, transmission spectra depends on the scale height of the atmosphere, which depends on the planet mass. %Improving the mass measurement of this planet is then required (see discussions in \citeads{2022A&A...667L..11A} for similar conclusions) {to achieve, at least, the $50\%$ mass precision level necessary to produce reliable atmopsheric retrievals }.
{It is therefore necessary to improve the measurement of the planet mass (see discussions in \citeads{2022A&A...667L..11A} for similar conclusions) to reach, at the very least, the $50\%$ level of accuracy on the mass required to produce reliable atmospheric retrievals \citepads{2023A&A...669A.150D}.}
Moreover, we expect the transmission spectra to be contaminated by stellar activity signatures, while this may complicate the interpretation of such observations it could also bring very useful information on the chromaticity of the stellar signal. 
%\footnote{We also calculated HD~73344b's Emission spectroscopy metric (ESM) and found ESM=$9.9\pm 0.4$.} 

 % =================================
 
\section{Conclusions}
\label{ccl}

We observed the bright star HD 73344 ($V_\mathrm{mag}=6.9$) with SOPHIE and HIRES in order to confirm the transiting planet with a period of $P_\mathrm{b} \sim 15.6$ days, which was initially a candidate in the K2 data \citepads{2018AJ....156...22Y}. This planet was also confirmed by contemporary TESS (TOI 5140.01) and \textit{Spitzer} observations. Our main results are listed below.

\begin{itemize}

    \item Analysis of the spectroscopic SOPHIE and HIRES spectra made it possible to refine the parameters of the host star and measure the stellar abundances.

    \item In the RV data, a candidate planet with $P_\mathrm{c} = 66.456^{+0.100}_{-0.250}$ days and a minimum mass of $M_\mathrm{c}\sin i_\mathrm{c}  = 0.37\pm 0.04$~$M_\mathrm{J}$ is detected. A two-planet system in this configuration is dynamically stable if $i_\mathrm{c}\gtrsim 30^\circ$ (which translates into $M_\mathrm{c}\lesssim 0.7$~$M_\mathrm{J}$). Moreover, imposing that planet~b transits the star (as observed) more than $50\%$ of the time requires a very near coplanarity between the two planetary orbits, such that $87.4^\circ\lesssim i_\mathrm{c}\lesssim 89.1^\circ$. This inclination interval translates into a very tight range for the true mass $M_\mathrm{c}$ ---much tighter than the observational uncertainty on $M_\mathrm{c}\sin i_\mathrm{c}$.

    \item In the RV data, the variability of the host star completely masks the RV signal of the transiting planet. Both spectroscopic and photometric data show that the star is indeed particularly active. The rotation period ($P_\mathrm{rot} \sim 9$ days) is close to the orbital period of the transiting planet. The coherence of the activity signal extends to $3-4$ $P_\mathrm{rot}$ {and is particularly strong in the area of the Gaussian fit to the CCF indicator (see Appendix~\ref{appC}). According to \citetads{10.1093/mnras/stab1183}, this may indicate that HD 73344 is faculae-dominated.} %, which is the typical lifetime of {faculae} for solar-type stars. This is also consistent with the analysis of the activity indicators, particularly when we look at the Area of the Gaussian fit to the CCF, known as a good marker of stellar plages.

    \item The prevailing approach in RV observations today involves averaging a limited number of data points gathered over a single night to mitigate the influence of short-term stellar variability, such as p-modes, granulation, and supergranulation. However, this strategy falls short when the goal is to detect planetary signals of just a few meters per second around evolved stars. In our efforts to better understand and characterize the short-term variability exhibited by the star HD 73344, we conducted observations spanning two entire nights using SOPHIE. In doing so, we identified a signal with an amplitude of $12.8 \pm 6.0$ m/s, and we observed variation in this signal over a coherence time of $2.4 \pm 0.7$ hours. These findings provide us with robust constraints for modeling these sources of noise when employing nonbinned RV data.
    
    \item Tests based on SOPHIE RV data reveal that not binning the data (1) provides a more accurate estimation of stellar activity, and (2) yields planetary parameters that are consistent with the binned case, indicating that binning does not enhance precision in planetary parameter determination.
    The planetary signal is clearly evident in the periodograms when using unbinned data (unlike the binned case), as the high-frequency noise has been effectively modeled. We note, however, that without the use of priors on $P_\mathrm{b}$ (coming from photometry), the planet RV  signal at $P_\mathrm{b}$ is not detected. 
    
    \item The joint analysis of photometric and spectroscopic data using a model featuring two planets and two Gaussian processes (one for capturing the effects of rotationally modulated magnetic activity and another for the variability occurring on short timescales) allows a more comprehensive understanding of both the planetary system and the overall activity of the host star. For the transiting planet, we infer a radius of $R_\mathrm{b} = 2.884^{+0.082}_{-0.072}$ $R_\oplus$ and a mass of $M_\mathrm{b} = 2.983^{+2.500}_{-1.905}$ $M_\oplus$ (marginal detection). This gives an average density of $\rho_\mathrm{b}=0.681^{+0.590}_{-0.438}$ g/cm$^3$, which is consistent with the density expected for gaseous planets. %{We note that, despite our effort to mitigate the stellar activity noise, the RV signal of the transiting planet is marginally detected. Long-term RV monitoring of this system is still needed. 

    \item Our initial assessment suggests the presence of an atmosphere enriched in volatile gases, such as hydrogen and helium. However, due to the significant uncertainty regarding the mass of planet b, we refrained from conducting an in depth analysis of its internal structure, including a detailed examination of its atmospheric composition (e.g., employing models from \citeads{Acuna2022}).  This also underscores the critical importance of obtaining a precise estimate of the planet's mass in order to reveal its true nature. 

\end{itemize}

As a perspective of this study, we note that alternative data analysis techniques based on multidimensional Gaussian processes could improve the RV detection of the transiting planet (see e.g., \citeads{10.1093/mnras/stv1428}; \citeads{10.1093/mnras/stab2889}; \citeads{2023arXiv230408489H}). Not yet implemented in the \texttt{PASTIS} software, such analyses were beyond the scope of the present study.
Another perspective is to carry out injection tests to secure the uncertainty on the inferred planet mass (see \citeads{2023A&A...676A..82M}), and to optimize the observational strategy for future observations of this system.

Finally, HD 73344 is a very bright star, and planet b is a sub-Neptune planet with an ideal orbital period for future observations by JWST and/or ARIEL (according to the TSM metric). However, the high activity level of the host star may complicate the interpretation of transmission spectra (\citeads{2023RASTI...2..148R}; \citeads{2023arXiv230315418R}), and refining the planet mass should be considered first. On the other hand, if observed with ARIEL, this target could serve as a benchmark for testing stellar activity diagnostic tools and correction methods for transmission spectra (\citeads{2021MNRAS.501.1733C}; \citeads{2024ApJ...960..107T}).
 
% =================================

\begin{acknowledgements}

{The authors thank the anonymous referee for her/his helpful
comments that improved the quality of the paper.}
We warmly thank the OHP and Keck staff for their support on the observations. In particular, we thank Dr. L. Bouma,  Dr. E. Petigura, B. Famaey, and Dr. J.-B. Salomon for the time they devoted to the observations of HD 73344 with the HIRES and SOPHIE spectrographs.
M.~S. thanks Aur{\'e}lie Astoul and Yubo Su for valuable discussions about the stellar structure and star-planet interactions.
We also thank Dr. F. Thevenin for the helpful discussion about \textit{Gaia}'s observations of this system. 
This paper includes data collected by the K2 and TESS missions, which are publicly available from the Mikulski Archive for Space Telescopes (MAST). Funding for these missions is provided by NASA’s Science Mission directorate.  We acknowledge the use of public TESS data from pipelines at the TESS Science Office and at the TESS Science Processing Operations Center.  
This work is based in part on observations made with the \textit{Spitzer} Space Telescope, which was operated by the Jet Propulsion Laboratory, California Institute of Technology under a contract with NASA.  
This research has made use of the VizieR catalog access tool, CDS, Strasbourg, France (DOI : 10.26093/cds/vizier). 
This research has made use of the NASA Exoplanet Archive, which is operated by the California Institute of Technology, under contract with the National Aeronautics and Space Administration under the Exoplanet Exploration Program.  
S.S., N.M., and D.M acknowledge support from the Programme National de Planétologie (PNP), and the Programme National de Physique Stellaire (PNPS) of CNRS-INSU. 
N.C.S, S.S., E.D.M and V.A. acknowledge funding by the European Union (ERC, FIERCE, 101052347). Views and opinions expressed are however those of the author(s) only and do not necessarily reflect those of the European Union or the European Research Council. Neither the European Union nor the granting authority can be held responsible for them. This work was supported by FCT - - Funda\c{c}\~ao para a Ci\^encia e Tecnologia through national funds and by FEDER through COMPETE2020 - Programa Operacional Competitividade e Internacionalização by these grants: UIDB/04434/2020; UIDP/04434/2020. 
E.D.M. acknowledges the support from FCT through Stimulus FCT contract 2021.01294. 
V.A. was supported by FCT through national funds and by FEDER through COMPETE2020 - Programa Operacional Competitividade e Internacionaliza\c{c}\~ao by these grants: UIDB/04434/2020; UIDP/04434/2020; 2022.06962.PTDC. 
This material is based upon work supported by NASA’S Interdisciplinary Consortia for Astrobiology Research (NNH19ZDA001N-ICAR) under award number 19-ICAR19\_2-0041.
 S.M.\ acknowledges support by the Spanish Ministry of Science and Innovation with the Ramon y Cajal fellowship number RYC-2015-17697, the grant number PID2019-107187GB-I00, the grant no. PID2019-107061GB-C66, and through AEI under the Severo Ochoa Centres of Excellence Programme 2020--2023 (CEX2019-000920-S).
  J.C. acknowledges funding from the European Research Council (ERC) under the European Union’s Horizon 2020 research and innovation programme (grant agreement No. 757561).
 S.B. gratefully acknowledges support from MIT International Science and Technology Initiatives (MISTI) and the MIT-France program.
 S.D. is funded by the UK Science and Technology Facilities Council (grant number ST/V004735/1).
 The project leading to this publication has received funding from the Excellence Initiative of Aix-Marseille Universit\'e--A*Midex, a French ``Investissements d’Avenir program'' AMX-21-IET-018. This research holds as part of the project FACOM (ANR-22-CE49-0005-01\_ACT) and has benefited from a funding provided by l'Agence Nationale de la Recherche (ANR) under the Generic Call for Proposals 2022.
We acknowledge funding from the French ANR under contract number ANR18CE310019 (SPlaSH). This work is supported in part by the French National Research Agency in the framework of the Investissements d’Avenir program (ANR-15-IDEX-02), through the funding of the ”Origin of Life” project of the Grenoble-Alpes University.
O.V. acknowledges funding from the ANR project `EXACT' (ANR-21-CE49-0008-01), from the Programme de Plan\'etologie (PNP) and from the Centre National d'\'{E}tudes Spatiales (CNES).
E.W. acknowledges support from the ERC Consolidator Grant funding scheme (project ASTEROCHRONOMETRY, G.A. no. 772293 http://www.asterochronometry.eu).

\end{acknowledgements}

%%%%%%%%%%%%%%%%%%%%%%%%%%%%%%%%%%%%%%%%
% BIBLIOGRAPHY
%%%%%%%%%%%%%%%%%%%%%%%%%%%%%%%%%%%%%%%%
\newpage

\bibliographystyle{aa} 
\bibliography{bibfile} 

\newcommand{\noop}[1]{}
\begin{thebibliography}{132}
\expandafter\ifx\csname natexlab\endcsname\relax\def\natexlab#1{#1}\fi

\bibitem[{{Acu{\~n}a} {et~al.}(2022){Acu{\~n}a}, {Lopez}, {Morel}, {Deleuil},
  {Mousis}, {Aguichine}, {Marcq}, \& {Santerne}}]{Acuna2022}
{Acu{\~n}a}, L., {Lopez}, T.~A., {Morel}, T., {et~al.} 2022, \aap, 660, A102

\bibitem[{{Adibekyan} {et~al.}(2015){Adibekyan}, {Figueira}, {Santos}, {Sousa},
  {Faria}, {Delgado-Mena}, {Oshagh}, {Tsantaki}, {Hakobyan}, {Gonz{\'a}lez
  Hern{\'a}ndez}, {Su{\'a}rez-Andr{\'e}s}, \& {Israelian}}]{Adibekyan-15}
{Adibekyan}, V., {Figueira}, P., {Santos}, N.~C., {et~al.} 2015, \aap, 583, A94

\bibitem[{{Adibekyan} {et~al.}(2012){Adibekyan}, {Sousa}, {Santos}, {Delgado
  Mena}, {Gonz{\'a}lez Hern{\'a}ndez}, {Israelian}, {Mayor}, \&
  {Khachatryan}}]{Adibekyan-12}
{Adibekyan}, V.~Z., {Sousa}, S.~G., {Santos}, N.~C., {et~al.} 2012, \aap, 545,
  A32

\bibitem[{{Aguichine} {et~al.}(2021){Aguichine}, {Mousis}, {Deleuil}, \&
  {Marcq}}]{2021ApJ...914...84A}
{Aguichine}, A., {Mousis}, O., {Deleuil}, M., \& {Marcq}, E. 2021, \apj, 914,
  84

\bibitem[{{Aigrain} {et~al.}(2004){Aigrain}, {Favata}, \&
  {Gilmore}}]{2004A&A...414.1139A}
{Aigrain}, S., {Favata}, F., \& {Gilmore}, G. 2004, \aap, 414, 1139

\bibitem[{{Aigrain} {et~al.}(2012){Aigrain}, {Pont}, \&
  {Zucker}}]{2012MNRAS.419.3147A}
{Aigrain}, S., {Pont}, F., \& {Zucker}, S. 2012, \mnras, 419, 3147

\bibitem[{{Allard} {et~al.}(2012){Allard}, {Homeier}, \&
  {Freytag}}]{2012RSPTA.370.2765A}
{Allard}, F., {Homeier}, D., \& {Freytag}, B. 2012, Philosophical Transactions
  of the Royal Society of London Series A, 370, 2765

\bibitem[{{Almenara} {et~al.}(2022){Almenara}, {Bonfils}, {Forveille},
  {Astudillo-Defru}, {Ciardi}, {Schwarz}, {Collins}, {Cointepas}, {Lund},
  {Bouchy}, {Charbonneau}, {D{\'\i}az}, {Delfosse}, {Kidwell}, {Kunimoto},
  {Latham}, {Lissauer}, {Murgas}, {Ricker}, {Seager}, {Vezie}, \&
  {Watanabe}}]{2022A&A...667L..11A}
{Almenara}, J.~M., {Bonfils}, X., {Forveille}, T., {et~al.} 2022, \aap, 667,
  L11

\bibitem[{{Angus} {et~al.}(2019){Angus}, {Morton}, \&
  {Foreman-Mackey}}]{2019JOSS....4.1469A}
{Angus}, R., {Morton}, T., \& {Foreman-Mackey}, D. 2019, The Journal of Open
  Source Software, 4, 1469

\bibitem[{{Baluev}(2008)}]{2008MNRAS.385.1279B}
{Baluev}, R.~V. 2008, \mnras, 385, 1279

\bibitem[{Barragán {et~al.}(2021)Barragán, Aigrain, Rajpaul, \&
  Zicher}]{10.1093/mnras/stab2889}
Barragán, O., Aigrain, S., Rajpaul, V.~M., \& Zicher, N. 2021, Monthly Notices
  of the Royal Astronomical Society, 509, 866

\bibitem[{{Batygin} \& {Adams}(2013)}]{Batygin-Adams_2013}
{Batygin}, K. \& {Adams}, F.~C. 2013, \apj, 778, 169

\bibitem[{{Becker} \& {Adams}(2016)}]{Becker-Adams_2016}
{Becker}, J.~C. \& {Adams}, F.~C. 2016, \mnras, 455, 2980

\bibitem[{{Bertran de Lis} {et~al.}(2015){Bertran de Lis}, {Delgado Mena},
  {Adibekyan}, {Santos}, \& {Sousa}}]{Bertrandelis-15}
{Bertran de Lis}, S., {Delgado Mena}, E., {Adibekyan}, V.~Z., {Santos}, N.~C.,
  \& {Sousa}, S.~G. 2015, \aap, 576, A89

\bibitem[{{Blanes} {et~al.}(2013){Blanes}, {Casas}, {Farr\'es}, {Laskar},
  {Makazaga}, \& {Murua}}]{Blanes-etal_2013}
{Blanes}, S., {Casas}, F., {Farr\'es}, A., {et~al.} 2013, Applied Numerical
  Mathematics, 68, 58

\bibitem[{{Boisse} {et~al.}(2011){Boisse}, {Bouchy}, {H{\'e}brard}, {Bonfils},
  {Santos}, \& {Vauclair}}]{2011A&A...528A...4B}
{Boisse}, I., {Bouchy}, F., {H{\'e}brard}, G., {et~al.} 2011, \aap, 528, A4

\bibitem[{{Boisse} {et~al.}(2010){Boisse}, {Eggenberger}, {Santos}, {Lovis},
  {Bouchy}, {H{\'e}brard}, {Arnold}, {Bonfils}, {Delfosse}, {Desort},
  {D{\'\i}az}, {Ehrenreich}, {Forveille}, {Gallenne}, {Lagrange}, {Moutou},
  {Udry}, {Pepe}, {Perrier}, {Perruchot}, {Pont}, {Queloz}, {Santerne},
  {S{\'e}gransan}, \& {Vidal-Madjar}}]{2010A&A...523A..88B}
{Boisse}, I., {Eggenberger}, A., {Santos}, N.~C., {et~al.} 2010, \aap, 523, A88

\bibitem[{Bouchy {et~al.}(2013)Bouchy, D{\'\i}az, H{\'e}brard, Arnold, Boisse,
  Delfosse, Perruchot, \& Santerne}]{bouchy2013sophie+}
Bouchy, F., D{\'\i}az, R., H{\'e}brard, G., {et~al.} 2013, Astronomy \&
  Astrophysics, 549, A49

\bibitem[{Bouchy {et~al.}(2009)Bouchy, H{\'e}brard, Udry, Delfosse, Boisse,
  Desort, Bonfils, Eggenberger, Ehrenreich, Forveille,
  {et~al.}}]{bouchy2009sophie}
Bouchy, F., H{\'e}brard, G., Udry, S., {et~al.} 2009, Astronomy \&
  Astrophysics, 505, 853

\bibitem[{{Brandenburg} {et~al.}(2017){Brandenburg}, {Mathur}, \&
  {Metcalfe}}]{2017ApJ...845...79B}
{Brandenburg}, A., {Mathur}, S., \& {Metcalfe}, T.~S. 2017, \apj, 845, 79

\bibitem[{Brewer {et~al.}(2016)Brewer, Fischer, Valenti, \&
  Piskunov}]{Brewer2016}
Brewer, J.~M., Fischer, D.~A., Valenti, J.~A., \& Piskunov, N. 2016, The
  Astrophysical Journal Supplement Series, 225, 32

\bibitem[{{Brugger} {et~al.}(2017){Brugger}, {Mousis}, {Deleuil}, \&
  {Deschamps}}]{2017ApJ...850...93B}
{Brugger}, B., {Mousis}, O., {Deleuil}, M., \& {Deschamps}, F. 2017, \apj, 850,
  93

\bibitem[{{Chaplin} {et~al.}(2011){Chaplin}, {Bedding}, {Bonanno}, {Broomhall},
  {Garc{\'\i}a}, {Hekker}, {Huber}, {Verner}, {Basu}, {Elsworth}, {Houdek},
  {Mathur}, {Mosser}, {New}, {Stevens}, {Appourchaux}, {Karoff}, {Metcalfe},
  {Molenda-{\.Z}akowicz}, {Monteiro}, {Thompson}, {Christensen-Dalsgaard},
  {Gilliland}, {Kawaler}, {Kjeldsen}, {Ballot}, {Benomar}, {Corsaro},
  {Campante}, {Gaulme}, {Hale}, {Handberg}, {Jarvis}, {R{\'e}gulo}, {Roxburgh},
  {Salabert}, {Stello}, {Mullally}, {Li}, \& {Wohler}}]{2011ApJ...732L...5C}
{Chaplin}, W.~J., {Bedding}, T.~R., {Bonanno}, A., {et~al.} 2011, \apjl, 732,
  L5

\bibitem[{{Chaplin} {et~al.}(2019){Chaplin}, {Cegla}, {Watson}, {Davies}, \&
  {Ball}}]{2019AJ....157..163C}
{Chaplin}, W.~J., {Cegla}, H.~M., {Watson}, C.~A., {Davies}, G.~R., \& {Ball},
  W.~H. 2019, \aj, 157, 163

\bibitem[{{Christiansen}(2022)}]{2022NatAs...6..516C}
{Christiansen}, J.~L. 2022, Nature Astronomy, 6, 516

\bibitem[{{Claret} \& {Bloemen}(2011)}]{2011A&A...529A..75C}
{Claret}, A. \& {Bloemen}, S. 2011, \aap, 529, A75

\bibitem[{{Claret} {et~al.}(2013){Claret}, {Hauschildt}, \&
  {Witte}}]{claret:2013}
{Claret}, A., {Hauschildt}, P.~H., \& {Witte}, S. 2013, \aap, 552, A16

\bibitem[{{Collier Cameron} {et~al.}(2019){Collier Cameron}, {Mortier},
  {Phillips}, {Dumusque}, {Haywood}, {Langellier}, {Watson}, {Cegla}, {Costes},
  {Charbonneau}, {Coffinet}, {Latham}, {Lopez-Morales}, {Malavolta},
  {Maldonado}, {Micela}, {Milbourne}, {Molinari}, {Saar}, {Thompson},
  {Buchschacher}, {Cecconi}, {Cosentino}, {Ghedina}, {Glenday}, {Gonzalez},
  {Li}, {Lodi}, {Lovis}, {Pepe}, {Poretti}, {Rice}, {Sasselov}, {Sozzetti},
  {Szentgyorgyi}, {Udry}, \& {Walsworth}}]{2019MNRAS.487.1082C}
{Collier Cameron}, A., {Mortier}, A., {Phillips}, D., {et~al.} 2019, \mnras,
  487, 1082

\bibitem[{{Correia} {et~al.}(2009){Correia}, {Udry}, {Mayor}, {Benz},
  {Bertaux}, {Bouchy}, {Laskar}, {Lovis}, {Mordasini}, {Pepe}, \&
  {Queloz}}]{Correia-etal_2009}
{Correia}, A.~C.~M., {Udry}, S., {Mayor}, M., {et~al.} 2009, \aap, 496, 521

\bibitem[{{Correia} {et~al.}(2005){Correia}, {Udry}, {Mayor}, {Laskar}, {Naef},
  {Pepe}, {Queloz}, \& {Santos}}]{Correia-etal_2005}
{Correia}, A.~C.~M., {Udry}, S., {Mayor}, M., {et~al.} 2005, \aap, 440, 751

\bibitem[{Costes {et~al.}(2021)Costes, Watson, de~Mooij, Saar, Dumusque,
  Cameron, Phillips, Günther, Jenkins, Mortier, \&
  Thompson}]{10.1093/mnras/stab1183}
Costes, J.~C., Watson, C.~A., de~Mooij, E., {et~al.} 2021, MNRAS, 505, 830

\bibitem[{{Couetdic} {et~al.}(2010){Couetdic}, {Laskar}, {Correia}, {Mayor}, \&
  {Udry}}]{Couetdic-etal_2010}
{Couetdic}, J., {Laskar}, J., {Correia}, A.~C.~M., {Mayor}, M., \& {Udry}, S.
  2010, \aap, 519, A10

\bibitem[{Courcol {et~al.}(2015)Courcol, Bouchy, Pepe, Santerne, Delfosse,
  Arnold, Astudillo-Defru, Boisse, Bonfils, Borgniet,
  {et~al.}}]{courcol2015sophie}
Courcol, B., Bouchy, F., Pepe, F., {et~al.} 2015, Astronomy \& Astrophysics,
  581, A38

\bibitem[{{Cowan} \& {Agol}(2011)}]{2011ApJ...726...82C}
{Cowan}, N.~B. \& {Agol}, E. 2011, \apj, 726, 82

\bibitem[{{Cracchiolo} {et~al.}(2021){Cracchiolo}, {Micela}, \&
  {Peres}}]{2021MNRAS.501.1733C}
{Cracchiolo}, G., {Micela}, G., \& {Peres}, G. 2021, \mnras, 501, 1733

\bibitem[{{Crossfield} {et~al.}(2019){Crossfield}, {Gorjian}, \&
  {Benneke}}]{crossfield:2019spitz}
{Crossfield}, I., {Gorjian}, V., \& {Benneke}, B. 2019, {K2's Greatest Hits:
  Spitzer Transits of 2 Exceptional Worlds}, Spitzer Proposal ID \#14292

\bibitem[{{Crossfield} {et~al.}(2020){Crossfield}, {Dragomir}, {Cowan},
  {Daylan}, {Wong}, {Kataria}, {Deming}, {Kreidberg}, {Mikal-Evans}, {Gorjian},
  {Jenkins}, {Benneke}, {Collins}, {Burke}, {Henze}, {McDermott}, {Mireles},
  {Watanabe}, {Wohler}, {Ricker}, {Vand erspek}, {Seager}, \&
  {Jenkins}}]{crossfield:2020}
{Crossfield}, I. J.~M., {Dragomir}, D., {Cowan}, N.~B., {et~al.} 2020, \apjl,
  903, L7

\bibitem[{{Cubillos} {et~al.}(2017){Cubillos}, {Erkaev}, {Juvan}, {Fossati},
  {Johnstone}, {Lammer}, {Lendl}, {Odert}, \& {Kislyakova}}]{Cubillos2017}
{Cubillos}, P., {Erkaev}, N.~V., {Juvan}, I., {et~al.} 2017, \mnras, 466, 1868

\bibitem[{{Cubillos} {et~al.}(2013){Cubillos}, {Harrington}, {Madhusudhan},
  {Stevenson}, {Hardy}, {Blecic}, {Anderson}, {Hardin}, \&
  {Campo}}]{cubillos:2013}
{Cubillos}, P., {Harrington}, J., {Madhusudhan}, N., {et~al.} 2013, \apj, 768,
  42

\bibitem[{{Delgado Mena} {et~al.}(2021){Delgado Mena}, {Adibekyan}, {Santos},
  {Tsantaki}, {Gonz{\'a}lez Hern{\'a}ndez}, {Sousa}, \& {Bertr{\'a}n de
  Lis}}]{Delgado-21}
{Delgado Mena}, E., {Adibekyan}, V., {Santos}, N.~C., {et~al.} 2021, \aap, 655,
  A99

\bibitem[{{Delgado Mena} {et~al.}(2019){Delgado Mena}, {Moya}, {Adibekyan},
  {Tsantaki}, {Gonz{\'a}lez Hern{\'a}ndez}, {Israelian}, {Davies}, {Chaplin},
  {Sousa}, {Ferreira}, \& {Santos}}]{2019A&A...624A..78D}
{Delgado Mena}, E., {Moya}, A., {Adibekyan}, V., {et~al.} 2019, \aap, 624, A78

\bibitem[{{Delgado Mena} {et~al.}(2017){Delgado Mena}, {Tsantaki}, {Adibekyan},
  {Sousa}, {Santos}, {Gonz{\'a}lez Hern{\'a}ndez}, \& {Israelian}}]{Delgado-17}
{Delgado Mena}, E., {Tsantaki}, M., {Adibekyan}, V.~Z., {et~al.} 2017, \aap,
  606, A94

\bibitem[{{D{\'e}sert} {et~al.}(2015){D{\'e}sert}, {Charbonneau}, {Torres},
  {Fressin}, {Ballard}, {Bryson}, {Knutson}, {Batalha}, {Borucki}, {Brown},
  {Deming}, {Ford}, {Fortney}, {Gilliland}, {Latham}, \&
  {Seager}}]{2015ApJ...804...59D}
{D{\'e}sert}, J.-M., {Charbonneau}, D., {Torres}, G., {et~al.} 2015, \apj, 804,
  59

\bibitem[{{Di Maio} {et~al.}(2023){Di Maio}, {Changeat}, {Benatti}, \&
  {Micela}}]{2023A&A...669A.150D}
{Di Maio}, C., {Changeat}, Q., {Benatti}, S., \& {Micela}, G. 2023, \aap, 669,
  A150

\bibitem[{{D{\'\i}az} {et~al.}(2014){D{\'\i}az}, {Almenara}, {Santerne},
  {Moutou}, {Lethuillier}, \& {Deleuil}}]{2014MNRAS.441..983D}
{D{\'\i}az}, R.~F., {Almenara}, J.~M., {Santerne}, A., {et~al.} 2014, \mnras,
  441, 983

\bibitem[{{Dotter} {et~al.}(2008){Dotter}, {Chaboyer}, {Jevremovi{\'c}},
  {Kostov}, {Baron}, \& {Ferguson}}]{2008ApJS..178...89D}
{Dotter}, A., {Chaboyer}, B., {Jevremovi{\'c}}, D., {et~al.} 2008, \apjs, 178,
  89

\bibitem[{{Doyle} {et~al.}(2014){Doyle}, {Davies}, {Smalley}, {Chaplin}, \&
  {Elsworth}}]{Doyle-14}
{Doyle}, A.~P., {Davies}, G.~R., {Smalley}, B., {Chaplin}, W.~J., \&
  {Elsworth}, Y. 2014, \mnras, 444, 3592

\bibitem[{{Dumusque} {et~al.}(2011){Dumusque}, {Udry}, {Lovis}, {Santos}, \&
  {Monteiro}}]{2011A&A...525A.140D}
{Dumusque}, X., {Udry}, S., {Lovis}, C., {Santos}, N.~C., \& {Monteiro},
  M.~J.~P.~F.~G. 2011, \aap, 525, A140

\bibitem[{{Fazio} {et~al.}(2004){Fazio}, {Hora}, {Allen}, {Ashby}, {Barmby},
  {Deutsch}, {Huang}, {Kleiner}, {Marengo}, {Megeath}, {Melnick}, {Pahre},
  {Patten}, {Polizotti}, {Smith}, {Taylor}, {Wang}, {Willner}, {Hoffmann},
  {Pipher}, {Forrest}, {McMurty}, {McCreight}, {McKelvey}, {McMurray}, {Koch},
  {Moseley}, {Arendt}, {Mentzell}, {Marx}, {Losch}, {Mayman}, {Eichhorn},
  {Krebs}, {Jhabvala}, {Gezari}, {Fixsen}, {Flores}, {Shakoorzadeh}, {Jungo},
  {Hakun}, {Workman}, {Karpati}, {Kichak}, {Whitley}, {Mann}, {Tollestrup},
  {Eisenhardt}, {Stern}, {Gorjian}, {Bhattacharya}, {Carey}, {Nelson},
  {Glaccum}, {Lacy}, {Lowrance}, {Laine}, {Reach}, {Stauffer}, {Surace},
  {Wilson}, {Wright}, {Hoffman}, {Domingo}, \& {Cohen}}]{fazio:2005}
{Fazio}, G.~G., {Hora}, J.~L., {Allen}, L.~E., {et~al.} 2004, \apjs, 154, 10

\bibitem[{{Fossati} {et~al.}(2017){Fossati}, {Erkaev}, {Lammer}, {Cubillos},
  {Odert}, {Juvan}, {Kislyakova}, {Lendl}, {Kubyshkina}, \&
  {Bauer}}]{Fossati2017}
{Fossati}, L., {Erkaev}, N.~V., {Lammer}, H., {et~al.} 2017, \aap, 598, A90

\bibitem[{{Fressin} {et~al.}(2012){Fressin}, {Torres}, {Pont}, {Knutson},
  {Charbonneau}, {Mazeh}, {Aigrain}, {Fridlund}, {Henze}, {Guillot}, \&
  {Rauer}}]{2012ApJ...745...81F}
{Fressin}, F., {Torres}, G., {Pont}, F., {et~al.} 2012, \apj, 745, 81

\bibitem[{{Garc{\'\i}a} {et~al.}(2010){Garc{\'\i}a}, {Mathur}, {Salabert},
  {Ballot}, {R{\'e}gulo}, {Metcalfe}, \& {Baglin}}]{2010Sci...329.1032G}
{Garc{\'\i}a}, R.~A., {Mathur}, S., {Salabert}, D., {et~al.} 2010, Science,
  329, 1032

\bibitem[{{Gardner} {et~al.}(2006){Gardner}, {Mather}, {Clampin}, {Doyon},
  {Greenhouse}, {Hammel}, {Hutchings}, {Jakobsen}, {Lilly}, {Long}, {Lunine},
  {McCaughrean}, {Mountain}, {Nella}, {Rieke}, {Rieke}, {Rix}, {Smith},
  {Sonneborn}, {Stiavelli}, {Stockman}, {Windhorst}, \&
  {Wright}}]{2006SSRv..123..485G}
{Gardner}, J.~P., {Mather}, J.~C., {Clampin}, M., {et~al.} 2006, \ssr, 123, 485

\bibitem[{Gastineau \& Laskar(2011)}]{Gastineau-Laskar_2011}
Gastineau, M. \& Laskar, J. 2011, ACM Commun. Comput. Algebra, 44, 194

\bibitem[{{Gupta} {et~al.}(2022){Gupta}, {Luhn}, {Wright}, {Mahadevan}, {Ford},
  {Stef{\'a}nsson}, {Bender}, {Blake}, {Halverson}, {Hearty}, {Kanodia},
  {Logsdon}, {McElwain}, {Ninan}, {Robertson}, {Roy}, {Schwab}, \&
  {Terrien}}]{2022AJ....164..254G}
{Gupta}, A.~F., {Luhn}, J., {Wright}, J.~T., {et~al.} 2022, \aj, 164, 254

\bibitem[{{Hara} \& {Delisle}(2023)}]{2023arXiv230408489H}
{Hara}, N.~C. \& {Delisle}, J.-B. 2023, arXiv e-prints, arXiv:2304.08489

\bibitem[{{Harvey}(1988)}]{1988IAUS..123..497H}
{Harvey}, J.~W. 1988, in IAU Symposium, Vol. 123, Advances in Helio- and
  Asteroseismology, ed. J.~{Christensen-Dalsgaard} \& S.~{Frandsen}, 497

\bibitem[{{Haywood} {et~al.}(2014){Haywood}, {Collier Cameron}, {Queloz},
  {Barros}, {Deleuil}, {Fares}, {Gillon}, {Lanza}, {Lovis}, {Moutou}, {Pepe},
  {Pollacco}, {Santerne}, {S{\'e}gransan}, \& {Unruh}}]{2014MNRAS.443.2517H}
{Haywood}, R.~D., {Collier Cameron}, A., {Queloz}, D., {et~al.} 2014, \mnras,
  443, 2517

\bibitem[{Heidari(2022)}]{heidari:tel-04043297}
Heidari, N. 2022, Theses, {Univ. C{\^o}te d'Azur ; Shahid Beheshti Univ.
  (Tehran)}

\bibitem[{{Heidari} {et~al.}(2024){Heidari}, {Boisse}, {Hara}, {Wilson},
  {Kiefer}, {H{\'e}brard}, {Philipot}, {Hoyer}, {Stassun}, {Henry}, {Santos},
  {Acu{\~n}a}, {Almasian}, {Arnold}, {Astudillo-Defru}, {Attia}, {Bonfils},
  {Bouchy}, {Bourrier}, {Collet}, {Cort{\'e}s-Zuleta}, {Carmona}, {Delfosse},
  {Dalal}, {Deleuil}, {Demangeon}, {D{\'\i}az}, {Dumusque}, {Ehrenreich},
  {Forveille}, {Hobson}, {Jenkins}, {Jenkins}, {Lagrange}, {Latham}, {Larue},
  {Liu}, {Moutou}, {Mignon}, {Osborn}, {Pepe}, {Rapetti}, {Rodrigues},
  {Santerne}, {Segransan}, {Shporer}, {Sulis}, {Torres}, {Udry}, {Vakili},
  {Vanderburg}, {Venot}, {Vivien}, \& {Vines}}]{2024A&A...681A..55H}
{Heidari}, N., {Boisse}, I., {Hara}, N.~C., {et~al.} 2024, \aap, 681, A55

\bibitem[{{Howard} {et~al.}(2010{\natexlab{a}}){Howard}, {Johnson}, {Marcy},
  {Fischer}, {Wright}, {Bernat}, {Henry}, {Peek}, {Isaacson}, {Apps}, {Endl},
  {Cochran}, {Valenti}, {Anderson}, \& {Piskunov}}]{2010ApJ...721.1467H}
{Howard}, A.~W., {Johnson}, J.~A., {Marcy}, G.~W., {et~al.} 2010{\natexlab{a}},
  \apj, 721, 1467

\bibitem[{{Howard} {et~al.}(2010{\natexlab{b}}){Howard}, {Marcy}, {Johnson},
  {Fischer}, {Wright}, {Isaacson}, {Valenti}, {Anderson}, {Lin}, \&
  {Ida}}]{2010Sci...330..653H}
{Howard}, A.~W., {Marcy}, G.~W., {Johnson}, J.~A., {et~al.} 2010{\natexlab{b}},
  Science, 330, 653

\bibitem[{{Howell} {et~al.}(2014){Howell}, {Sobeck}, {Haas}, {Still},
  {Barclay}, {Mullally}, {Troeltzsch}, {Aigrain}, {Bryson}, {Caldwell},
  {Chaplin}, {Cochran}, {Huber}, {Marcy}, {Miglio}, {Najita}, {Smith},
  {Twicken}, \& {Fortney}}]{2014PASP..126..398H}
{Howell}, S.~B., {Sobeck}, C., {Haas}, M., {et~al.} 2014, \pasp, 126, 398

\bibitem[{{Huber} {et~al.}(2013){Huber}, {Carter}, {Barbieri}, {Miglio},
  {Deck}, {Fabrycky}, {Montet}, {Buchhave}, {Chaplin}, {Hekker},
  {Montalb{\'a}n}, {Sanchis-Ojeda}, {Basu}, {Bedding}, {Campante},
  {Christensen-Dalsgaard}, {Elsworth}, {Stello}, {Arentoft}, {Ford},
  {Gilliland}, {Handberg}, {Howard}, {Isaacson}, {Johnson}, {Karoff},
  {Kawaler}, {Kjeldsen}, {Latham}, {Lund}, {Lundkvist}, {Marcy}, {Metcalfe},
  {Silva Aguirre}, \& {Winn}}]{2013Sci...342..331H}
{Huber}, D., {Carter}, J.~A., {Barbieri}, M., {et~al.} 2013, Science, 342, 331

\bibitem[{{Huber} {et~al.}(2022){Huber}, {White}, {Metcalfe}, {Chontos},
  {Fausnaugh}, {Ho}, {Van Eylen}, {Ball}, {Basu}, {Bedding}, {Benomar},
  {Bossini}, {Breton}, {Buzasi}, {Campante}, {Chaplin},
  {Christensen-Dalsgaard}, {Cunha}, {Deal}, {Garc{\'\i}a}, {Garc{\'\i}a
  Mu{\~n}oz}, {Gehan}, {Gonz{\'a}lez-Cuesta}, {Jiang}, {Kayhan}, {Kjeldsen},
  {Lundkvist}, {Mathis}, {Mathur}, {Monteiro}, {Nsamba}, {Ong},
  {Pak{\v{s}}tien{\.{e}}}, {Serenelli}, {Silva Aguirre}, {Stassun}, {Stello},
  {Norgaard Stilling}, {Lykke Winther}, {Wu}, {Barclay}, {Daylan},
  {G{\"u}nther}, {Hermes}, {Jenkins}, {Latham}, {Levine}, {Ricker}, {Seager},
  {Shporer}, {Twicken}, {Vanderspek}, \& {Winn}}]{2022AJ....163...79H}
{Huber}, D., {White}, T.~R., {Metcalfe}, T.~S., {et~al.} 2022, \aj, 163, 79

\bibitem[{Jenkins {et~al.}(2010)Jenkins, Caldwell, Chandrasekaran, Twicken,
  Bryson, Quintana, Clarke, Li, Allen, Tenenbaum, Wu, Klaus, Cleve, Dotson,
  Haas, Gilliland, Koch, \& Borucki}]{Jenkins_2010}
Jenkins, J.~M., Caldwell, D.~A., Chandrasekaran, H., {et~al.} 2010, The
  Astrophysical Journal Letters, 713, L120

\bibitem[{{Judkovsky} {et~al.}(2022){Judkovsky}, {Ofir}, \&
  {Aharonson}}]{Judkovsky-etal_2022}
{Judkovsky}, Y., {Ofir}, A., \& {Aharonson}, O. 2022, \aj, 163, 91

\bibitem[{{Kallinger} {et~al.}(2014){Kallinger}, {De Ridder}, {Hekker},
  {Mathur}, {Mosser}, {Gruberbauer}, {Garc{\'\i}a}, {Karoff}, \&
  {Ballot}}]{2014A&A...570A..41K}
{Kallinger}, T., {De Ridder}, J., {Hekker}, S., {et~al.} 2014, \aap, 570, A41

\bibitem[{{Kempton} {et~al.}(2018){Kempton}, {Bean}, {Louie}, {Deming}, {Koll},
  {Mansfield}, {Christiansen}, {L{\'o}pez-Morales}, {Swain}, {Zellem},
  {Ballard}, {Barclay}, {Barstow}, {Batalha}, {Beatty}, {Berta-Thompson},
  {Birkby}, {Buchhave}, {Charbonneau}, {Cowan}, {Crossfield}, {de Val-Borro},
  {Doyon}, {Dragomir}, {Gaidos}, {Heng}, {Hu}, {Kane}, {Kreidberg}, {Mallonn},
  {Morley}, {Narita}, {Nascimbeni}, {Pall{\'e}}, {Quintana}, {Rauscher},
  {Seager}, {Shkolnik}, {Sing}, {Sozzetti}, {Stassun}, {Valenti}, \& {von
  Essen}}]{2018PASP..130k4401K}
{Kempton}, E. M.~R., {Bean}, J.~L., {Louie}, D.~R., {et~al.} 2018, \pasp, 130,
  114401

\bibitem[{{Kjeldsen} \& {Bedding}(1995)}]{1995A&A...293...87K}
{Kjeldsen}, H. \& {Bedding}, T.~R. 1995, \aap, 293, 87

\bibitem[{{Kov{\'a}cs} {et~al.}(2016){Kov{\'a}cs}, {Zucker}, \&
  {Mazeh}}]{2016ascl.soft07008K}
{Kov{\'a}cs}, G., {Zucker}, S., \& {Mazeh}, T. 2016, {BLS: Box-fitting Least
  Squares}, Astrophysics Source Code Library, record ascl:1607.008

\bibitem[{{Kurucz}(1993)}]{Kurucz-93}
{Kurucz}, R.~L. 1993, {SYNTHE spectrum synthesis programs and line data}

\bibitem[{{Laskar}(1988)}]{Laskar_1988}
{Laskar}, J. 1988, \aap, 198, 341

\bibitem[{{Laskar}(1993)}]{Laskar_1993}
{Laskar}, J. 1993, Physica D Nonlinear Phenomena, 67, 257

\bibitem[{{Laskar}(2005)}]{Laskar_2005}
{Laskar}, J. 2005, Frequency map analysis and quasiperiodic decompositions
  (Cambridge Scientific Pub), 99--134

\bibitem[{{Laskar} \& {Correia}(2009)}]{Laskar-Correia_2009}
{Laskar}, J. \& {Correia}, A.~C.~M. 2009, \aap, 496, L5

\bibitem[{{Laskar} \& {Petit}(2017)}]{Laskar-Petit_2017}
{Laskar}, J. \& {Petit}, A.~C. 2017, \aap, 605, A72

\bibitem[{{Lopez} \& {Fortney}(2014)}]{Lopez2014}
{Lopez}, E.~D. \& {Fortney}, J.~J. 2014, \apj, 792, 1

\bibitem[{{Luger} {et~al.}(2016){Luger}, {Agol}, {Kruse}, {Barnes}, {Becker},
  {Foreman-Mackey}, \& {Deming}}]{2016AJ....152..100L}
{Luger}, R., {Agol}, E., {Kruse}, E., {et~al.} 2016, \aj, 152, 100

\bibitem[{{Luger} {et~al.}(2018){Luger}, {Kruse}, {Foreman-Mackey}, {Agol}, \&
  {Saunders}}]{2018AJ....156...99L}
{Luger}, R., {Kruse}, E., {Foreman-Mackey}, D., {Agol}, E., \& {Saunders}, N.
  2018, \aj, 156, 99

\bibitem[{{Mamajek} \& {Hillenbrand}(2008)}]{2008ApJ...687.1264M}
{Mamajek}, E.~E. \& {Hillenbrand}, L.~A. 2008, \apj, 687, 1264

\bibitem[{{Mathur} {et~al.}(2023){Mathur}, {Claytor}, {Santos}, {Garc{\'\i}a},
  {Amard}, {Bugnet}, {Corsaro}, {Bonanno}, {Breton}, {Godoy-Rivera},
  {Pinsonneault}, \& {van Saders}}]{2023ApJ...952..131M}
{Mathur}, S., {Claytor}, Z.~R., {Santos}, {\^A}. R.~G., {et~al.} 2023, \apj,
  952, 131

\bibitem[{{Mathur} {et~al.}(2014{\natexlab{a}}){Mathur}, {Garc{\'\i}a},
  {Ballot}, {Ceillier}, {Salabert}, {Metcalfe}, {R{\'e}gulo}, {Jim{\'e}nez}, \&
  {Bloemen}}]{2014A&A...562A.124M}
{Mathur}, S., {Garc{\'\i}a}, R.~A., {Ballot}, J., {et~al.} 2014{\natexlab{a}},
  \aap, 562, A124

\bibitem[{{Mathur} {et~al.}(2019){Mathur}, {Garc{\'\i}a}, {Bugnet}, {Santos},
  {Santiago}, \& {Beck}}]{2019FrASS...6...46M}
{Mathur}, S., {Garc{\'\i}a}, R.~A., {Bugnet}, L., {et~al.} 2019, Frontiers in
  Astronomy and Space Sciences, 6, 46

\bibitem[{{Mathur} {et~al.}(2014{\natexlab{b}}){Mathur}, {Salabert},
  {Garc{\'\i}a}, \& {Ceillier}}]{2014JSWSC...4A..15M}
{Mathur}, S., {Salabert}, D., {Garc{\'\i}a}, R.~A., \& {Ceillier}, T.
  2014{\natexlab{b}}, Journal of Space Weather and Space Climate, 4, A15

\bibitem[{{Meunier} {et~al.}(2015){Meunier}, {Lagrange}, {Borgniet}, \&
  {Rieutord}}]{2015A&A...583A.118M}
{Meunier}, N., {Lagrange}, A.~M., {Borgniet}, S., \& {Rieutord}, M. 2015, \aap,
  583, A118

\bibitem[{{Meunier} {et~al.}(2023){Meunier}, {Pous}, {Sulis}, {Mary}, \&
  {Lagrange}}]{2023A&A...676A..82M}
{Meunier}, N., {Pous}, R., {Sulis}, S., {Mary}, D., \& {Lagrange}, A.~M. 2023,
  \aap, 676, A82

\bibitem[{Modigliani {et~al.}(2019)Modigliani, Sownsowska, \&
  Lovis}]{modigliani2019espresso}
Modigliani, A., Sownsowska, D., \& Lovis, C. 2019, ESPRESSO Pipeline User
  Manual

\bibitem[{{Mousis} {et~al.}(2020){Mousis}, {Deleuil}, {Aguichine}, {Marcq},
  {Naar}, {Aguirre}, {Brugger}, \& {Gon{\c{c}}alves}}]{Mousis2020}
{Mousis}, O., {Deleuil}, M., {Aguichine}, A., {et~al.} 2020, \apjl, 896, L22

\bibitem[{{Murray} \& {Dermott}(1999)}]{Murray-Dermott_1999}
{Murray}, C.~D. \& {Dermott}, S.~F. 1999, {Solar system dynamics} (Cambridge
  University Press)

\bibitem[{{Ness}(2018)}]{Ness2018}
{Ness}, M. 2018, \pasa, 35, e003

\bibitem[{{Noyes} {et~al.}(1984){Noyes}, {Hartmann}, {Baliunas}, {Duncan}, \&
  {Vaughan}}]{1984ApJ...279..763N}
{Noyes}, R.~W., {Hartmann}, L.~W., {Baliunas}, S.~L., {Duncan}, D.~K., \&
  {Vaughan}, A.~H. 1984, \apj, 279, 763

\bibitem[{{Owen} \& {Wu}(2016)}]{Owen2016}
{Owen}, J.~E. \& {Wu}, Y. 2016, \apj, 817, 107

\bibitem[{{Perruchot} {et~al.}(2008){Perruchot}, {Kohler}, {Bouchy}, {Richaud},
  {Richaud}, {Moreaux}, {Merzougui}, {Sottile}, {Hill}, {Knispel}, {Regal},
  {Meunier}, {Ilovaisky}, {Le Coroller}, {Gillet}, {Schmitt}, {Pepe}, {Fleury},
  {Sosnowska}, {Vors}, {M{\'e}gevand}, {Blanc}, {Carol}, {Point}, {Laloge}, \&
  {Brunel}}]{2008SPIE.7014E..0JP}
{Perruchot}, S., {Kohler}, D., {Bouchy}, F., {et~al.} 2008, in Society of
  Photo-Optical Instrumentation Engineers (SPIE) Conference Series, Vol. 7014,
  Ground-based and Airborne Instrumentation for Astronomy II, ed. I.~S.
  {McLean} \& M.~M. {Casali}, 70140J

\bibitem[{{Piaulet} {et~al.}(2021){Piaulet}, {Benneke}, {Rubenzahl}, {Howard},
  {Lee}, {Thorngren}, {Angus}, {Peterson}, {Schlieder}, {Werner}, {Kreidberg},
  {Jaouni}, {Crossfield}, {Ciardi}, {Petigura}, {Livingston}, {Dressing},
  {Fulton}, {Beichman}, {Christiansen}, {Gorjian}, {Hardegree-Ullman}, {Krick},
  \& {Sinukoff}}]{Piaulet2021}
{Piaulet}, C., {Benneke}, B., {Rubenzahl}, R.~A., {et~al.} 2021, \aj, 161, 70

\bibitem[{{Polanski} {et~al.}(2022){Polanski}, {Crossfield}, {Howard},
  {Isaacson}, \& {Rice}}]{2022RNAAS...6..155P}
{Polanski}, A.~S., {Crossfield}, I. J.~M., {Howard}, A.~W., {Isaacson}, H., \&
  {Rice}, M. 2022, Research Notes of the American Astronomical Society, 6, 155

\bibitem[{{Queloz} {et~al.}(2001){Queloz}, {Henry}, {Sivan}, {Baliunas},
  {Beuzit}, {Donahue}, {Mayor}, {Naef}, {Perrier}, \&
  {Udry}}]{2001A&A...379..279Q}
{Queloz}, D., {Henry}, G.~W., {Sivan}, J.~P., {et~al.} 2001, \aap, 379, 279

\bibitem[{{Rackham} \& {de Wit}(2023)}]{2023arXiv230315418R}
{Rackham}, B.~V. \& {de Wit}, J. 2023, arXiv e-prints, arXiv:2303.15418

\bibitem[{{Rackham} {et~al.}(2023){Rackham}, {Espinoza}, {Berdyugina},
  {Korhonen}, {MacDonald}, {Montet}, {Morris}, {Oshagh}, {Shapiro}, {Unruh},
  {Quintana}, {Zellem}, {Apai}, {Barclay}, {Barstow}, {Bruno}, {Carone},
  {Casewell}, {Cegla}, {Criscuoli}, {Fischer}, {Fournier}, {Giampapa}, {Giles},
  {Iyer}, {Kopp}, {Kostogryz}, {Krivova}, {Mallonn}, {McGruder},
  {Molaverdikhani}, {Newton}, {Panja}, {Peacock}, {Reardon}, {Roettenbacher},
  {Scandariato}, {Solanki}, {Stassun}, {Steiner}, {Stevenson}, {Tregloan-Reed},
  {Valio}, {Wedemeyer}, {Welbanks}, {Yu}, {Alam}, {Davenport}, {Deming},
  {Dong}, {Ducrot}, {Fisher}, {Gilbert}, {Kostov}, {L{\'o}pez-Morales}, {Line},
  {Mo{\v{c}}nik}, {Mullally}, {Paudel}, {Ribas}, \&
  {Valenti}}]{2023RASTI...2..148R}
{Rackham}, B.~V., {Espinoza}, N., {Berdyugina}, S.~V., {et~al.} 2023, RAS
  Techniques and Instruments, 2, 148

\bibitem[{{Radick} {et~al.}(2018){Radick}, {Lockwood}, {Henry}, {Hall}, \&
  {Pevtsov}}]{2018ApJ...855...75R}
{Radick}, R.~R., {Lockwood}, G.~W., {Henry}, G.~W., {Hall}, J.~C., \&
  {Pevtsov}, A.~A. 2018, \apj, 855, 75

\bibitem[{Rajpaul {et~al.}(2015)Rajpaul, Aigrain, Osborne, Reece, \&
  Roberts}]{10.1093/mnras/stv1428}
Rajpaul, V., Aigrain, S., Osborne, M.~A., Reece, S., \& Roberts, S. 2015,
  Monthly Notices of the Royal Astronomical Society, 452, 2269

\bibitem[{{Rein} {et~al.}(2019){Rein}, {Tamayo}, \& {Brown}}]{Rein-etal_2019}
{Rein}, H., {Tamayo}, D., \& {Brown}, G. 2019, \mnras, 489, 4632

\bibitem[{{Rice} \& {Brewer}(2020)}]{Rice2020}
{Rice}, M. \& {Brewer}, J.~M. 2020, \apj, 898, 119

\bibitem[{{Ricker} {et~al.}(2015){Ricker}, {Winn}, {Vanderspek}, {Latham},
  {Bakos}, {Bean}, {Berta-Thompson}, {Brown}, {Buchhave}, {Butler}, {Butler},
  {Chaplin}, {Charbonneau}, {Christensen-Dalsgaard}, {Clampin}, {Deming},
  {Doty}, {De Lee}, {Dressing}, {Dunham}, {Endl}, {Fressin}, {Ge}, {Henning},
  {Holman}, {Howard}, {Ida}, {Jenkins}, {Jernigan}, {Johnson}, {Kaltenegger},
  {Kawai}, {Kjeldsen}, {Laughlin}, {Levine}, {Lin}, {Lissauer}, {MacQueen},
  {Marcy}, {McCullough}, {Morton}, {Narita}, {Paegert}, {Palle}, {Pepe},
  {Pepper}, {Quirrenbach}, {Rinehart}, {Sasselov}, {Sato}, {Seager},
  {Sozzetti}, {Stassun}, {Sullivan}, {Szentgyorgyi}, {Torres}, {Udry}, \&
  {Villasenor}}]{2015JATIS...1a4003R}
{Ricker}, G.~R., {Winn}, J.~N., {Vanderspek}, R., {et~al.} 2015, Journal of
  Astronomical Telescopes, Instruments, and Systems, 1, 014003

\bibitem[{{Rogers} {et~al.}(2023){Rogers}, {Schlichting}, \&
  {Owen}}]{Rogers2023}
{Rogers}, J.~G., {Schlichting}, H.~E., \& {Owen}, J.~E. 2023, \apjl, 947, L19

\bibitem[{{Saha} \& {Tremaine}(1994)}]{Saha-Tremaine_1994}
{Saha}, P. \& {Tremaine}, S. 1994, \aj, 108, 1962

\bibitem[{{Santos} {et~al.}(2021){Santos}, {Breton}, {Mathur}, \&
  {Garc{\'\i}a}}]{2021ApJS..255...17S}
{Santos}, A.~R.~G., {Breton}, S.~N., {Mathur}, S., \& {Garc{\'\i}a}, R.~A.
  2021, \apjs, 255, 17

\bibitem[{{Santos} {et~al.}(2019){Santos}, {Garc{\'\i}a}, {Mathur}, {Bugnet},
  {van Saders}, {Metcalfe}, {Simonian}, \&
  {Pinsonneault}}]{2019ApJS..244...21S}
{Santos}, A.~R.~G., {Garc{\'\i}a}, R.~A., {Mathur}, S., {et~al.} 2019, \apjs,
  244, 21

\bibitem[{{Santos} {et~al.}(2013){Santos}, {Sousa}, {Mortier}, {Neves},
  {Adibekyan}, {Tsantaki}, {Delgado Mena}, {Bonfils}, {Israelian}, {Mayor}, \&
  {Udry}}]{2013A&A...556A.150S}
{Santos}, N.~C., {Sousa}, S.~G., {Mortier}, A., {et~al.} 2013, \aap, 556, A150

\bibitem[{{Sneden}(1973)}]{Sneden-73}
{Sneden}, C.~A. 1973, PhD thesis, THE UNIVERSITY OF TEXAS AT AUSTIN.

\bibitem[{{Sousa}(2014)}]{2014dapb.book..297S}
{Sousa}, S.~G. 2014, in Determination of Atmospheric Parameters of B, 297--310

\bibitem[{{Sousa} {et~al.}(2021){Sousa}, {Adibekyan}, {Delgado-Mena}, {Santos},
  {Rojas-Ayala}, {Soares}, {Legoinha}, {Ulmer-Moll}, {Camacho}, {Barros},
  {Demangeon}, {Hoyer}, {Israelian}, {Mortier}, {Tsantaki}, \&
  {Monteiro}}]{2021A&A...656A..53S}
{Sousa}, S.~G., {Adibekyan}, V., {Delgado-Mena}, E., {et~al.} 2021, \aap, 656,
  A53

\bibitem[{{Sousa} {et~al.}(2015){Sousa}, {Santos}, {Adibekyan}, {Delgado-Mena},
  \& {Israelian}}]{Sousa-15}
{Sousa}, S.~G., {Santos}, N.~C., {Adibekyan}, V., {Delgado-Mena}, E., \&
  {Israelian}, G. 2015, \aap, 577, A67

\bibitem[{{Sousa} {et~al.}(2007){Sousa}, {Santos}, {Israelian}, {Mayor}, \&
  {Monteiro}}]{Sousa-07}
{Sousa}, S.~G., {Santos}, N.~C., {Israelian}, G., {Mayor}, M., \& {Monteiro},
  M.~J.~P.~F.~G. 2007, A\&A, 469, 783

\bibitem[{{Sousa} {et~al.}(2008){Sousa}, {Santos}, {Mayor}, {Udry},
  {Casagrande}, {Israelian}, {Pepe}, {Queloz}, \& {Monteiro}}]{Sousa-08}
{Sousa}, S.~G., {Santos}, N.~C., {Mayor}, M., {et~al.} 2008, \aap, 487, 373

\bibitem[{{Spalding} \& {Millholland}(2020)}]{Spalding-Millholland_2020}
{Spalding}, C. \& {Millholland}, S.~C. 2020, \aj, 160, 105

\bibitem[{{Stalport} {et~al.}(2022){Stalport}, {Delisle}, {Udry}, {Matthews},
  {Bourrier}, \& {Leleu}}]{Stalport-etal_2022}
{Stalport}, M., {Delisle}, J.~B., {Udry}, S., {et~al.} 2022, \aap, 664, A53

\bibitem[{{Stevenson} {et~al.}(2012){Stevenson}, {Harrington}, {Fortney},
  {Loredo}, {Hardy}, {Nymeyer}, {Bowman}, {Cubillos}, {Bowman}, \&
  {Hardin}}]{stevenson:2012a}
{Stevenson}, K.~B., {Harrington}, J., {Fortney}, J.~J., {et~al.} 2012, \apj,
  754, 136

\bibitem[{{Stock} {et~al.}(2023){Stock}, {Kemmer}, {Kossakowski}, {Sabotta},
  {Reffert}, \& {Quirrenbach}}]{2023A&A...674A.108S}
{Stock}, S., {Kemmer}, J., {Kossakowski}, D., {et~al.} 2023, \aap, 674, A108

\bibitem[{{Sulis} {et~al.}(2023){Sulis}, {Lendl}, {Cegla}, {Rodr{\'\i}guez
  D{\'\i}az}, {Bigot}, {Van Grootel}, {Bekkelien}, {Cameron}, {Maxted},
  {Simon}, {Lovis}, {Scandariato}, {Bruno}, {Nardiello}, {Bonfanti},
  {Fridlund}, {Persson}, {Salmon}, {Sousa}, {Wilson}, {Krenn}, {Hoyer},
  {Santerne}, {Ehrenreich}, {Alibert}, {Alonso}, {Anglada}, {B{\'a}rczy},
  {Barrado y Navascues}, {Barros}, {Baumjohann}, {Beck}, {Beck}, {Benz},
  {Billot}, {Bonfils}, {Borsato}, {Brandeker}, {Broeg}, {Cabrera}, {Charnoz},
  {Corral van Damme}, {Csizmadia}, {Davies}, {Deleuil}, {Deline}, {Delrez},
  {Demangeon}, {Demory}, {Erikson}, {Fortier}, {Fossati}, {Gandolfi}, {Gillon},
  {G{\"u}del}, {Heng}, {Isaak}, {Kiss}, {Laskar}, {Lecavelier des Etangs},
  {Magrin}, {Munari}, {Nascimbeni}, {Olofsson}, {Ottensamer}, {Pagano},
  {Pall{\'e}}, {Peter}, {Piotto}, {Pollacco}, {Queloz}, {Ragazzoni}, {Rando},
  {Rauer}, {Ribas}, {Rieder}, {Santos}, {S{\'e}gransan}, {Smith},
  {Steinberger}, {Steller}, {Szab{\'o}}, {Thomas}, {Udry}, {Walton}, \&
  {Wolter}}]{2023A&A...670A..24S}
{Sulis}, S., {Lendl}, M., {Cegla}, H.~M., {et~al.} 2023, \aap, 670, A24

\bibitem[{{Sulis} {et~al.}(2020){Sulis}, {Lendl}, {Hofmeister}, {Veronig},
  {Fossati}, {Cubillos}, \& {Van Grootel}}]{2020A&A...636A..70S}
{Sulis}, S., {Lendl}, M., {Hofmeister}, S., {et~al.} 2020, \aap, 636, A70

\bibitem[{{Sulis} {et~al.}(2022){Sulis}, {Mary}, {Bigot}, \&
  {Deleuil}}]{2022A&A...667A.104S}
{Sulis}, S., {Mary}, D., {Bigot}, L., \& {Deleuil}, M. 2022, \aap, 667, A104

\bibitem[{{Thompson} {et~al.}(2024){Thompson}, {Biagini}, {Cracchiolo},
  {Petralia}, {Changeat}, {Saba}, {Morello}, {Morvan}, {Micela}, \&
  {Tinetti}}]{2024ApJ...960..107T}
{Thompson}, A., {Biagini}, A., {Cracchiolo}, G., {et~al.} 2024, \apj, 960, 107

\bibitem[{{Tinetti} {et~al.}(2018){Tinetti}, {Drossart}, {Eccleston},
  {Hartogh}, {Heske}, {Leconte}, {Micela}, {Ollivier}, {Pilbratt}, {Puig},
  {Turrini}, {Vandenbussche}, {Wolkenberg}, {Beaulieu}, {Buchave}, {Ferus},
  {Griffin}, {Guedel}, {Justtanont}, {Lagage}, {Machado}, {Malaguti}, {Min},
  {N{\o}rgaard-Nielsen}, {Rataj}, {Ray}, {Ribas}, {Swain}, {Szabo}, {Werner},
  {Barstow}, {Burleigh}, {Cho}, {Coud{\'e} du Foresto}, {Coustenis}, {Decin},
  {Encrenaz}, {Galand}, {Gillon}, {Helled}, {Morales}, {Garc{\'\i}a Mu{\~n}oz},
  {Moneti}, {Pagano}, {Pascale}, {Piccioni}, {Pinfield}, {Sarkar}, {Selsis},
  {Tennyson}, {Triaud}, {Venot}, {Waldmann}, {Waltham}, {Wright}, {Amiaux},
  {Augu{\`e}res}, {Berth{\'e}}, {Bezawada}, {Bishop}, {Bowles}, {Coffey},
  {Colom{\'e}}, {Crook}, {Crouzet}, {Da Peppo}, {Sanz}, {Focardi}, {Frericks},
  {Hunt}, {Kohley}, {Middleton}, {Morgante}, {Ottensamer}, {Pace}, {Pearson},
  {Stamper}, {Symonds}, {Rengel}, {Renotte}, {Ade}, {Affer}, {Alard}, {Allard},
  {Altieri}, {Andr{\'e}}, {Arena}, {Argyriou}, {Aylward}, {Baccani}, {Bakos},
  {Banaszkiewicz}, {Barlow}, {Batista}, {Bellucci}, {Benatti}, {Bernardi},
  {B{\'e}zard}, {Blecka}, {Bolmont}, {Bonfond}, {Bonito}, {Bonomo}, {Brucato},
  {Brun}, {Bryson}, {Bujwan}, {Casewell}, {Charnay}, {Pestellini}, {Chen},
  {Ciaravella}, {Claudi}, {Cl{\'e}dassou}, {Damasso}, {Damiano}, {Danielski},
  {Deroo}, {Di Giorgio}, {Dominik}, {Doublier}, {Doyle}, {Doyon}, {Drummond},
  {Duong}, {Eales}, {Edwards}, {Farina}, {Flaccomio}, {Fletcher}, {Forget},
  {Fossey}, {Fr{\"a}nz}, {Fujii}, {Garc{\'\i}a-Piquer}, {Gear}, {Geoffray},
  {G{\'e}rard}, {Gesa}, {Gomez}, {Graczyk}, {Griffith}, {Grodent}, {Guarcello},
  {Gustin}, {Hamano}, {Hargrave}, {Hello}, {Heng}, {Herrero}, {Hornstrup},
  {Hubert}, {Ida}, {Ikoma}, {Iro}, {Irwin}, {Jarchow}, {Jaubert}, {Jones},
  {Julien}, {Kameda}, {Kerschbaum}, {Kervella}, {Koskinen}, {Krijger}, {Krupp},
  {Lafarga}, {Landini}, {Lellouch}, {Leto}, {Luntzer}, {Rank-L{\"u}ftinger},
  {Maggio}, {Maldonado}, {Maillard}, {Mall}, {Marquette}, {Mathis}, {Maxted},
  {Matsuo}, {Medvedev}, {Miguel}, {Minier}, {Morello}, {Mura}, {Narita},
  {Nascimbeni}, {Nguyen Tong}, {Noce}, {Oliva}, {Palle}, {Palmer}, {Pancrazzi},
  {Papageorgiou}, {Parmentier}, {Perger}, {Petralia}, {Pezzuto},
  {Pierrehumbert}, {Pillitteri}, {Piotto}, {Pisano}, {Prisinzano}, {Radioti},
  {R{\'e}ess}, {Rezac}, {Rocchetto}, {Rosich}, {Sanna}, {Santerne}, {Savini},
  {Scandariato}, {Sicardy}, {Sierra}, {Sindoni}, {Skup}, {Snellen}, {Sobiecki},
  {Soret}, {Sozzetti}, {Stiepen}, {Strugarek}, {Taylor}, {Taylor}, {Terenzi},
  {Tessenyi}, {Tsiaras}, {Tucker}, {Valencia}, {Vasisht}, {Vazan}, {Vilardell},
  {Vinatier}, {Viti}, {Waters}, {Wawer}, {Wawrzaszek}, {Whitworth}, {Yung},
  {Yurchenko}, {Zapatero Osorio}, {Zellem}, {Zingales}, \&
  {Zwart}}]{2018ExA....46..135T}
{Tinetti}, G., {Drossart}, P., {Eccleston}, P., {et~al.} 2018, Experimental
  Astronomy, 46, 135

\bibitem[{{Torres} {et~al.}(2010){Torres}, {Andersen}, \&
  {Gim{\'e}nez}}]{Torres-2010}
{Torres}, G., {Andersen}, J., \& {Gim{\'e}nez}, A. 2010, \aapr, 18, 67

\bibitem[{{Van Eylen} {et~al.}(2019){Van Eylen}, {Albrecht}, {Huang},
  {MacDonald}, {Dawson}, {Cai}, {Foreman-Mackey}, {Lundkvist}, {Silva Aguirre},
  {Snellen}, \& {Winn}}]{2019AJ....157...61V}
{Van Eylen}, V., {Albrecht}, S., {Huang}, X., {et~al.} 2019, \aj, 157, 61

\bibitem[{{Vivien} {et~al.}(2022){Vivien}, {Aguichine}, {Mousis}, {Deleuil}, \&
  {Marcq}}]{Vivien2022}
{Vivien}, H.~G., {Aguichine}, A., {Mousis}, O., {Deleuil}, M., \& {Marcq}, E.
  2022, \apj, 931, 143

\bibitem[{{Vogt} {et~al.}(1994){Vogt}, {Allen}, {Bigelow}, {Bresee}, {Brown},
  {Cantrall}, {Conrad}, {Couture}, {Delaney}, {Epps}, {Hilyard}, {Hilyard},
  {Horn}, {Jern}, {Kanto}, {Keane}, {Kibrick}, {Lewis}, {Osborne},
  {Pardeilhan}, {Pfister}, {Ricketts}, {Robinson}, {Stover}, {Tucker}, {Ward},
  \& {Wei}}]{1994SPIE.2198..362V}
{Vogt}, S.~S., {Allen}, S.~L., {Bigelow}, B.~C., {et~al.} 1994, in Society of
  Photo-Optical Instrumentation Engineers (SPIE) Conference Series, Vol. 2198,
  Instrumentation in Astronomy VIII, ed. D.~L. {Crawford} \& E.~R. {Craine},
  362

\bibitem[{{Xie} {et~al.}(2016){Xie}, {Dong}, {Zhu}, {Huber}, {Zheng}, {De Cat},
  {Fu}, {Liu}, {Luo}, {Wu}, {Zhang}, {Zhang}, {Zhou}, {Cao}, {Hou}, {Wang}, \&
  {Zhang}}]{Xie-etal_2016}
{Xie}, J.-W., {Dong}, S., {Zhu}, Z., {et~al.} 2016, Proceedings of the National
  Academy of Science, 113, 11431

\bibitem[{{Yu} {et~al.}(2018){Yu}, {Crossfield}, {Schlieder}, {Kosiarek},
  {Feinstein}, {Livingston}, {Howard}, {Benneke}, {Petigura}, {Bristow},
  {Christiansen}, {Ciardi}, {Crepp}, {Dressing}, {Fulton}, {Gonzales},
  {Hardegree-Ullman}, {Henning}, {Isaacson}, {L{\'e}pine}, {Martinez},
  {Morales}, \& {Sinukoff}}]{2018AJ....156...22Y}
{Yu}, L., {Crossfield}, I. J.~M., {Schlieder}, J.~E., {et~al.} 2018, \aj, 156,
  22

\bibitem[{{Zahnle} \& {Catling}(2017)}]{Zahnle2017}
{Zahnle}, K.~J. \& {Catling}, D.~C. 2017, \apj, 843, 122

\bibitem[{{Zechmeister} \& {K{\"u}rster}(2009)}]{2009A&A...496..577Z}
{Zechmeister}, M. \& {K{\"u}rster}, M. 2009, \aap, 496, 577

\end{thebibliography}

%%%%%%%%%%%%%%%%%%%%%%%%%%%%%%%%%%%%%%%%
% APPENDIX
%%%%%%%%%%%%%%%%%%%%%%%%%%%%%%%%%%%%%%%%

 \begin{appendix}
%-------------------------------------- 

\section{Individual transits of HD 73344b seen by K2 and TESS}
\label{app_transits}

%Figure \ref{fig_lcs} shows the nine individual transits of HD 73344b observed by K2 (first two rows) and TESS (last row). K2 observations are significantly affected by strong instrumental systematics that distort the shape of the transit. Although some of these systematics are poorly captured by the GP noise models described in Sec.~\ref{sec41} (especially for the second and fourth transits), the fit of the out-of-transit data are generally acceptable. The stellar variability signal is strong in K2 and TESS data and needs to be modeled jointly with planetary transit.

\begin{figure*}[h!]
\centering
\resizebox{0.8\hsize}{!}{\includegraphics{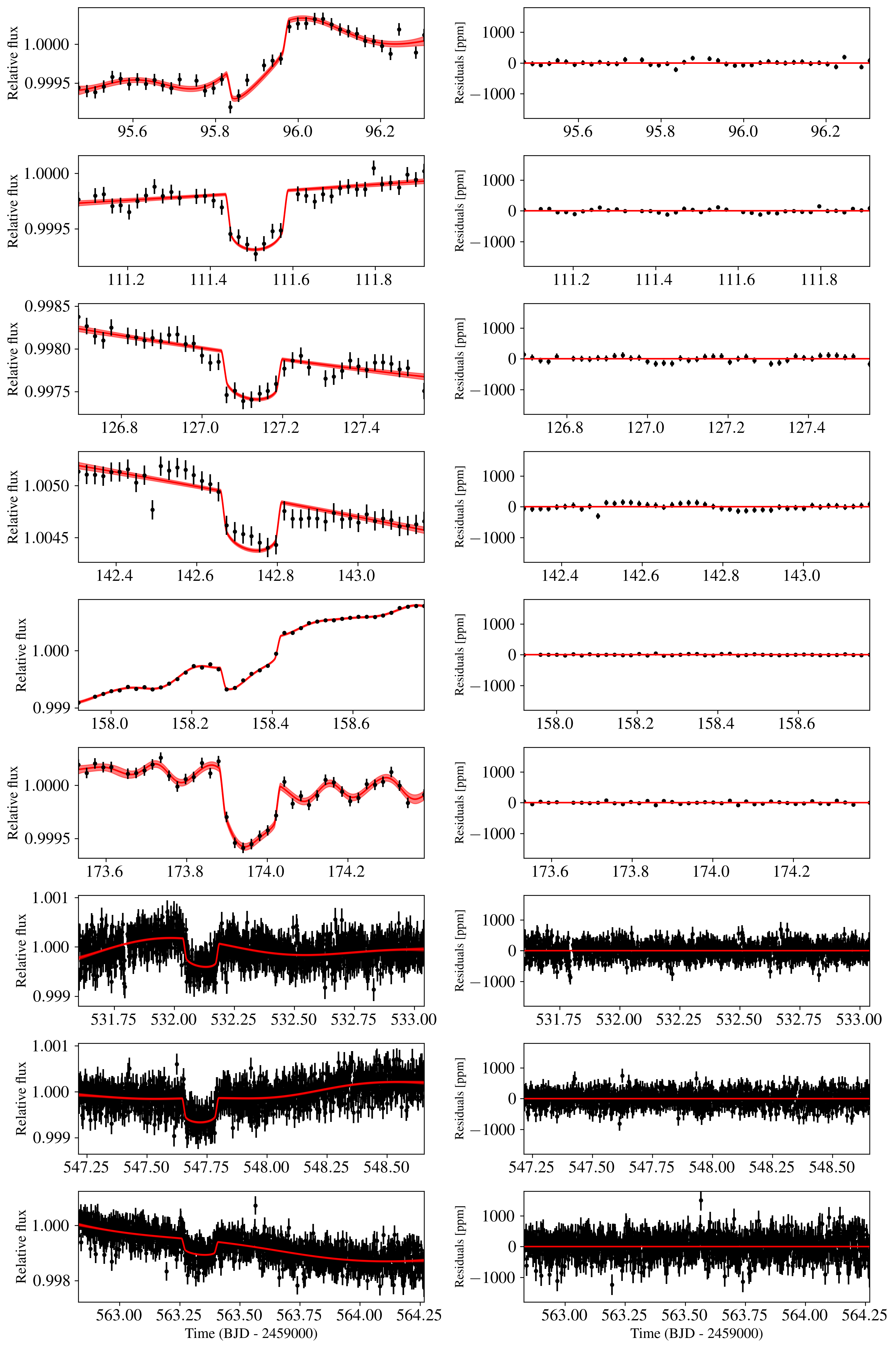}}
\caption{{Light curves of HD 73344 showing the individual transits of planet b as seen with K2 ($6$ transits; top left panels) and TESS ($3$ transits; bottom left panels). The best-fitting models are shown in red. The corresponding residuals are shown in the right panels.}
}
\label{fig_lcs}  
\end{figure*}

{
Figure \ref{fig_lcs} illustrates the individual transits of HD 73344b observed by K2 (six transits) and TESS (three transits). The corresponding residuals are displayed in the right panels. The RMS values of the residuals are (from top to bottom): [80, 62, 87, 90, 17, 40] ppm for K2 transits, and [229, 227, 341] ppm for TESS transits.
}

{
The K2 observations exhibit pronounced instrumental systematics that distort the transit shapes, particularly noticeable near the transit bottoms. The details of these systematics are poorly captured by the GP noise models described in Sect.~\ref{sec41}. However, these systematics remain of relatively low in amplitude (residuals <200 ppm if we zoom into the right panels), which is attributed to their averaging effect over the 30-min K2 integration time. In contrast, TESS observations offer high-quality data with precise transit events characterization, owing to its 120-s sampling rate. 
}

The K2 and TESS data present notable signals of stellar variability, underlining the importance of incorporating them into joint modeling with planetary transits. However, K2 observations are also particularly sensitive to instrumental systematics. This complicates the modeling that also needs to take into account the different bandwidths and integration times of K2/TESS observations. Given our aim to conduct a comprehensive analysis integrating photometric and RV data, we opted for a pragmatic approach in Sect.~\ref{sec_combined}: i.e., rather than using overly complex noise models for the photometric part, we used simpler GP noise models for each individual transits. These models effectively capture the global variability evolving around individual transits. Nonetheless, there is scope for further refinement in future analyses.

%-------------------------------------- 

\section{Stellar abundances from HIRES spectra}
\label{App_abun}

The stellar abundances of HD 73344, extracted from the SOPHIE and HIRES spectra (with \texttt{KeckSpec}), are reported in Table~\ref{tab_abundance}. They were extracted using two different codes, which are compared here. For the chemical elements extracted from the two spectra, all are in agreement at $1\sigma$, with the exception of oxygen. %This can be due to the different line indicators used in each work, but also to possible not noticed contamination around the oxygen forbidden line (6300$\AA$) in individual spectra (which depends on the observing conditions). Furthermore, we note that the two oxygen lines (6158$\AA$ and 6300$\AA$) available in the SOPHIE spectra are very weak and subject to large errors.
However, this is easily explained by the sensitivity of oxygen abundances to different line indicators.

\def\arraystretch{1.1}
\begin{table}[h]
\caption{Abundances of multiple chemical elements extracted from the SOPHIE and HIRES spectra of HD 73344. Last lines shows the Lithium abundance and the $\alpha$ element enhancement.}
\begin{center} \centering
\begin{tabular}{|l|c|c|}
\hline
Label                     & SOPHIE spectra & HIRES spectra \\
\hline
[C/H] & $0.159 \pm 0.049$ &  0.10$\pm$0.07 \\
\hline
[N/H] & -                 &  0.18$\pm$0.09 \\
\hline
[O/H] & $0.088 \pm 0.078$ &  0.24$\pm$0.09 \\
\hline
[Na/H] & $0.21 \pm 0.03$ & 0.13$\pm$0.07 \\
\hline
[Mg/H] & $0.13 \pm 0.04$ &  0.11$\pm$0.04 \\
\hline
[Al/H] & - &  0.05$\pm$0.08 \\
\hline
[Si/H] & $0.19 \pm 0.05$ &  0.14$\pm$0.06 \\
\hline
[Ca/H] & - &  0.18$\pm$0.07 \\
\hline
[Ti/H] & $0.15 \pm 0.04$ &  0.14$\pm$0.05 \\
\hline
[V/H] & -                 &  0.14$\pm$0.07 \\
\hline
[Cr/H] & -                &  0.17$\pm$0.05 \\
\hline
[Mn/H] & -                &  0.16$\pm$0.07 \\
\hline
[Fe/H] & $0.18 \pm 0.043$ &  0.17$\pm$0.06 \\
\hline
[Ni/H] & $0.17 \pm 0.02$ &  0.13$\pm$0.05 \\
\hline
[Cu/H] & $0.123 \pm 0.035$ & - \\
\hline
[Zn/H] & $0.120 \pm 0.030$ & - \\
\hline
[Sr/H] & $0.131 \pm 0.077$ & - \\
\hline
[Y/H] & $0.147 \pm 0.086$ &  0.23$\pm$0.09 \\
\hline
[Zr/H] & $0.070 \pm 0.065$ & - \\
\hline
[Ba/H] & $0.166 \pm 0.060$ &  - \\
\hline
[Ce/H] & $0.064 \pm 0.042$ & - \\
\hline
[Nd/H] & $-0.001 \pm 0.070$ & - \\
\hline
A(Li) & $2.81 \pm 0.05$ & - \\ 
\hline
[$\alpha$/Fe] & - & -0.03$\pm$0.06 \\
\hline
\end{tabular}
\label{tab_abundance}
\end{center}
\end{table}

From the stellar abundances derived from the SOPHIE spectra, we also estimated the stellar age from 3D chemical clock formulas based on $T_\mathrm{eff}$ and [Fe/H] (see Table 10 of \citeads{2019A&A...624A..78D}). The results are presented in Table~\ref{tab_age}. These ages are consistent with the value given in Table~\ref{tab_transit_RV}. The weighted average age of 2.0\,$\pm$\,0.2 Gyr is however twice the value extracted from isochrones analysis with \texttt{PASTIS} (see Sect.~\ref{sec_combined} and discussion therein).

\def\arraystretch{1.1}
\begin{table}[h]
\caption{Ages from chemical clocks 3D formulas derived in \citetads{2019A&A...624A..78D}.}
\begin{center} \centering
\begin{tabular}{|l|c|}
\hline
Element & Age [Gyr] \\
\hline
[Y/Zn] &          $2.17 \pm 1.47$ \\
\hline
[Y/Ti]  &        $1.53 \pm 1.70$ \\
\hline
[Y/Mg]   &       $1.59 \pm 1.42$\\
\hline
[Sr/Ti] &         $1.98 \pm 1.49$ \\
\hline
[Sr/Zn]  &       $2.20 \pm 1.21$ \\
\hline
[Sr/Mg]   &      $1.94 \pm 1.23$ \\
\hline
[Y/Si]     &     $2.11 \pm 1.68$ \\
\hline
[Sr/Si]     &    $2.08 \pm 1.45$ \\
\hline
\end{tabular}
\label{tab_age}
\end{center}
\end{table}

%-------------------------------------- 

\section{Temporal evolution of stellar activity and origin of the $\sim 66$ days signal}
\label{appC}

In this appendix, we first analyze the temporal evolution of stellar activity over the two years of observations with the SOPHIE spectrograph. Then, we discuss the origin of the $\sim 66$ days signal spotted in Sect.~\ref{sec42}.  
Finally, we look at the impact of stellar activity on the planet derived parameters.

\subsection{Temporal evolution of stellar activity}
\label{App_C1}
The temporal evolution of stellar activity in the RV observations is here compared with that of the chromospheric indicators discussed in Sect.~\ref{sec42}. 
We used the binned RV observations and the GLSP computed as in Eq.~\eqref{eq_GLS1}. These observations were obtained over two long campaigns of $144$ and $115$ days respectively, separated by a loss of observations during $224$ days. First campaign (C1) and second campaign (C2) of observations contain 78 and 61 individual nights, respectively.
 On each time series, we removed all signals generated by long-term changes in stellar activity with a second-order polynomial, following the recommendations given by \citetads{10.1093/mnras/stab1183}.

\begin{figure*}
\centering
\resizebox{\hsize}{!}{\includegraphics{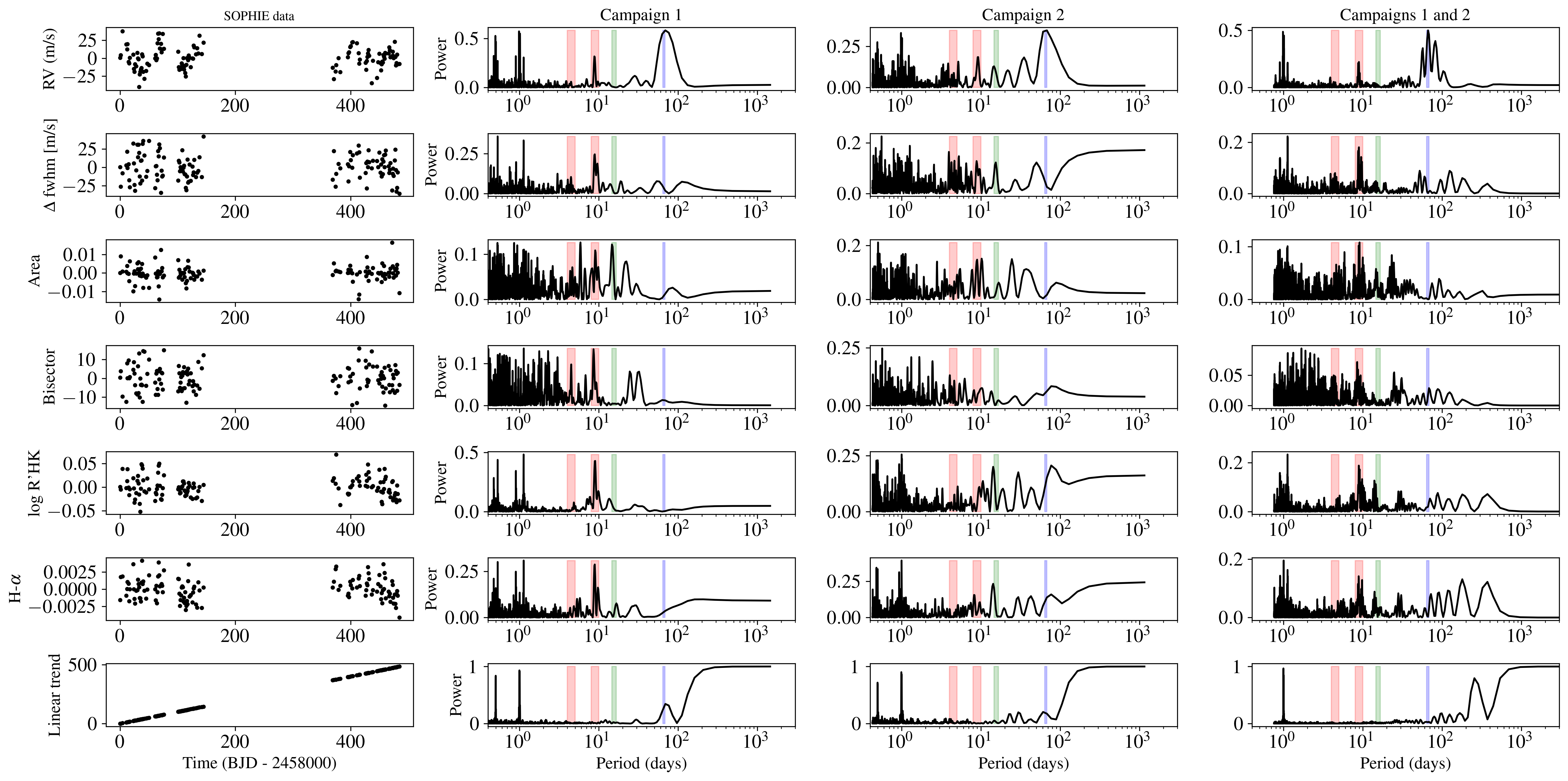}}
\caption{Temporal evolution of the stellar activity. \textit{From top to bottom:} RV data, FWHM, the Area of the Gaussian fit to the CCF, Bisector, $\rm log R'_{\rm HK}$, H$\alpha$ lines, and linear trend. \textit{From left to right:} Time series, GLSP for first campaign, second campaign and both campaigns. In all time series, we have eliminated long-term variation by a two-degree polynomial fit. The large red, green and blue vertical lines indicate the stellar rotation period (and half its period), the orbital period of the transiting planet (b), and the orbital period of the candidate planet (c), respectively. }
\label{app_indicators}  
\end{figure*}

In Fig.~\ref{app_indicators}, we show the RV data (first line) followed by the chromospheric indicator data (lines 2 to 7).  The last row represents a linear trend introduced to track the influence of gaps in the frequency domain. The last three columns show the periodograms of each series calculated for the two campaigns separately and jointly (Sect.~\ref{sec_combined}). The stellar rotation period, its first harmonic at half the rotation period, and the orbital periods of the planets are highlighted by large colored vertical bars to facilitate visual inspection of the results. 

During the first campaign, we observe significant peaks at the star's rotation period in all indicators. However, these peaks lose significance during the second campaign, while we see that other structures appear both at short and long periods (in $\log\,$R$'  _{\textrm{HK}}$ for instance). 
This suggests a strong temporal evolution of the star's activity between the two campaigns and potentially the presence of several active regions at different latitude of the stellar surface during the second campaign.

We also observe a non-negligible activity signal close to the transit period of the planet, particularly in the Area indicator (in both campaigns). During the second campaign, a prominent peak at $P_\mathrm{b}$ is also observed in the  $\rm log R'_{\rm HK}$ and H$\alpha$ activity indicators. For all activity indicators, however, the peak at $P_\mathrm{b}$ disappears when the two campaigns are analyzed together. This shows the temporal decoherence of stellar activity at this particular period, which may have helped to disentangle the planet's coherent signal at $P_\mathrm{b}$ from the activity signal in our previous analyses (Sect.~\ref{sec4}).  

Close to the candidate planet's period $P_\mathrm{c}$, weak periodicities also appear in some activity indicators and in the linear trend signal.
%. 
However, all such peaks are absent in the joint analysis of the two campaigns. 

In the joint analysis, we observe a set of peaks in some indicators at periods around $\sim 3$ to $4$ $P_\mathrm{rot}$. These peaks are particularly visible in the Area indicator. 
It is not easy to interpret the evolving signatures of these activity indicators in periodograms and associate them with a clearly identified stellar origin (spot/faculae).
Placing HD 73344 in \citetads{2018ApJ...855...75R} correlation plot between  luminosity and $\rm log R'_{\rm HK}$ (see their Fig. 14, with B-V $\sim 0.547$ for  HD 73344), we see that HD 73344 is located in the transition regime  between spot- and faculae-dominated stars. However, spots lifetime  rarely last more than two stellar rotations, in comparison with faculae that have a much longer lifetime (e.g., from a few months up to a year in the case of the Sun; \citeads{2019MNRAS.487.1082C}). In addition,  F-type stars are known to be predominantly faculae-dominated (see e.g., \citeads{10.1093/mnras/stab1183}) due to their shallow convective envelope. Thus, we suggest that the peaks seen around $\sim 3$ to $4$ $P_\mathrm{rot}$ in the periodogram  of some activity indicators may come from active faculae regions.

\subsection{Origin of the RV signal at $\sim 66$ days period}
\label{App_C2}

In Sect.~\ref{sec4} we spotted a periodicity at $\approx 66$ days.
Indeed, there are several aspects that point towards a sustained periodicity at $\approx 66$ days.\\
First, the peak appears in the GLSP of the SOPHIE RV data of each campaign, and jointly (see Fig.~\ref{app_indicators}).
The phase of the periodic signal is found to be consistent for the two individual campaigns ($73 ^{\circ} \pm 6  ^{\circ}$ for campaign 1 and  $55 ^{\circ} \pm 26  ^{\circ}$ for campaign 2\footnote{The stated uncertainty on the phase estimates is based on Monte Carlo simulations that used a jacknife (leave one out) resampling and accounted for an uncertainty of $\pm 0.6 $ d on the $66.46$ d period.}). This coherence is not found, for instance, for the $\approx 9$ days peak, which is present in the Area indicator for individual campaigns, but not jointly. \\
Second, we see that the linear trend signal presents a peak at $\approx 66$ days (at least in individual campaigns), so that the $\approx 66$ d signal in the RV could simply be the trace of a remaining trend in the data.  To check whether this is correct, we extended the GLSP to the case where the model includes not only a constant but also a trend (that is, the GLSP $P(\nu)$ now computes the reduction in the RSS when jointly fitting constant plus trend plus sinusoid at frequency $\nu$, with respect to the RSS when fitting only constant plus trend). As a result, the $\approx 66$ d signal remained\footnote{Regarding this point, we also note that the GLSP of the estimated activity signal (the two GP processes alone) does not show any peak at $\approx 66$ days.}. \\
Third, since the activity indicators capture (some of the) frequency contents about activity, we investigated whether these signals could be used to capture the $66$ d periodicity of the RV data. To do so, we computed another modified  (`activity aware') GLSP, where the model now includes all the activity indicators in addition to the constant and the linear trend of the previous case. Again, the result showed a dominant peak at $\approx 66$ days. \\

By construction, activity indicators can trace activity signals but not  planetary signals  (see e.g., \citeads{2001A&A...379..279Q}; \citeads{2011A&A...528A...4B}) and the results above support the fact that the $66$ d peak is not caused by activity. If this signal is indeed caused by stellar activity, we cannot explain why it is invisible in the activity indicators.
 
Another possibility, besides a planetary signature, is that this periodic signal is caused by an unmodeled instrumental noise. To our knowledge, no instrumental variation at this timescale is known, and a coherent signal with an amplitude of $K\ge \sim 16$ m/s coming from an instrumental systematics would have been easily spotted by the SOPHIE technical team. 
As an additional check, we analyzed the RV data taken with the HIRES spectrograph alone to exclude possible instrumental systematics from SOPHIE. We indeed observe that the $\approx 66$ d peak is also present in the GLSP of HIRES data (third largest peak, not shown here), but many other high peaks also exist, which prevents us from drawing a very clear conclusion on the interpretation of these data. We note that the RV from the telluric lines alone do not show any $\approx 66$ d periodicity. 

In summary, there is a periodic signal at $\approx 66$ d in the RV observations. We cannot exclude the hypothesis of a nonplanetary origin, but none of the dataset we analyzed favor this hypothesis. We assume then that the peak spotted in the GLSP at $P_\mathrm{c} \sim 66$ days period is the signature of a non transiting planet.

As a final note, we comment on the `significance' of this peak. Any statistical significance estimation algorithm relies on a noise model. In our case, the main  question is whether the $66$ d is caused by the stellar activity noise or not. If we test this hypothesis with a noise model free from this component, and investigate the probability that such a large peak occurs with this noise (that is, the p-value of this peak), then this peak is so large that it is declared highly significant by standard procedures. For instance, we obtain a p-value less than $10^{-4}$ for a classical bootstrap procedure based on the permutation of the residuals. This procedure is often used but provides reliable estimates only if the noise is white. The more elaborated \textsf{3SD} procedure \citepads{2022A&A...667A.104S} based on an activity  noise model composed with the two estimated GPs also leads to a low p-value ($8\times10^{-8}$). However, considering these p-values as a definitive support for the significance of this peak  would be somewhat adventurous for such an active star, as these values are strongly conditioned to the noise model, which is not well constrained.

\subsection{Impact of the temporal evolution of stellar activity on planet-derived parameters}

As a final test, we evaluated the impact of the temporal evolution of stellar activity on planet-derived parameters.
This assessment involved comparing outcomes derived from SOPHIE RV data collected during the first campaign (C1), the second campaign (C2), and the combined dataset (C1$+$C2; as detailed in Sect.~\ref{sec42}). Our benchmark was established through the joint analysis integrating both photometric (K2, TESS) and nonbinned RV data (SOPHIE, HIRES), as outlined in Sect.~\ref{sec_combined}.

\begin{figure*}[t!]
\centering
\includegraphics[width=\textwidth]{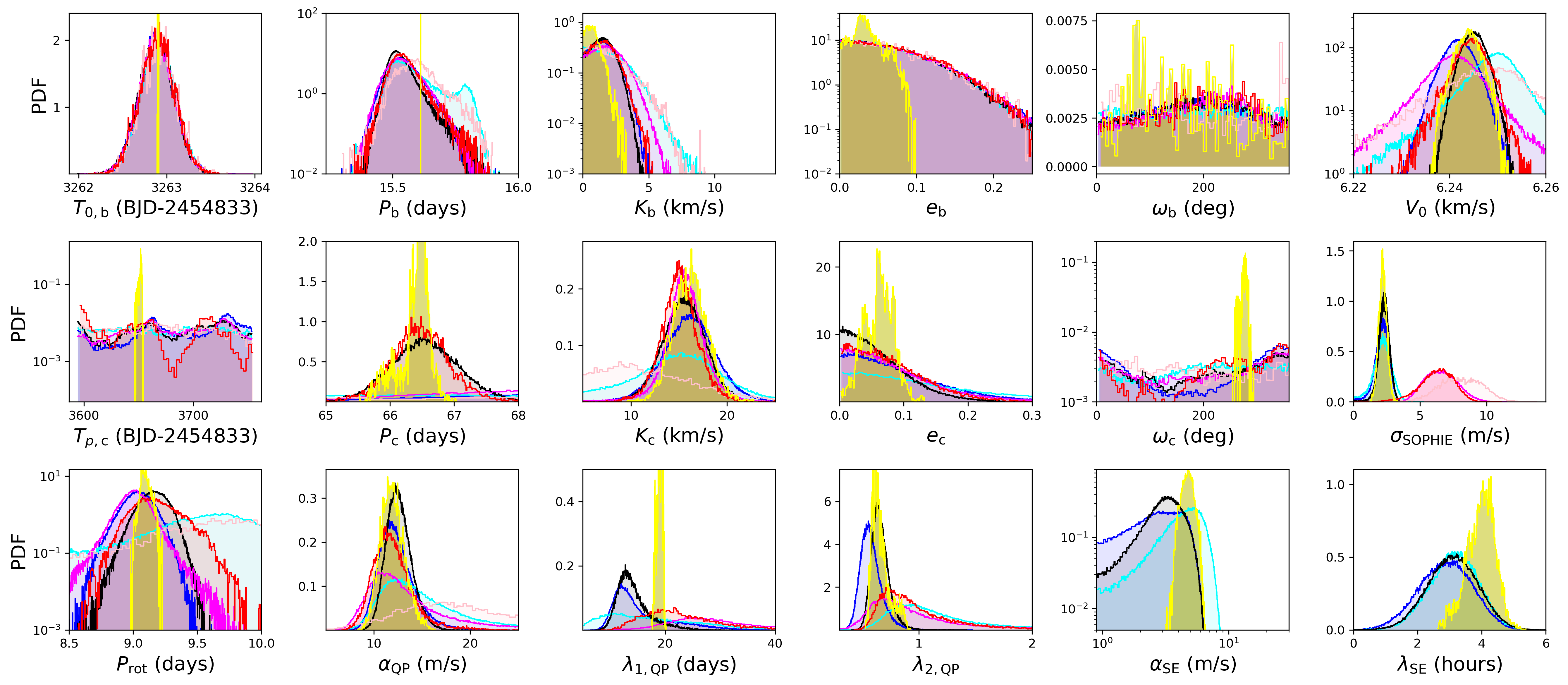}
\caption{Normalized posteriors distribution of the parameters fitted to the RV data. The distributions resulting from the analyses of the SOPHIE binned observations are shown in magenta (C1), pink (C2), and red (C1+C2). The distributions corresponding to the analyses of the unbinned data are shown in blue (C1), cyan (C2), and black (C1+C2).
The distributions resulting from the joint analysis combining the photometric (K2+TESS) and nonbinned RV (SOPHIE+HIRES) observations are shown in yellow (benchmark).
\textit{Top}: the five Keplerian parameters of planet b ($T_{0,\mathrm{b}}$, $P_\mathrm{b}$, $K_\mathrm{b}$, $e_\mathrm{b}$, $\omega_\mathrm{b}$), and $V_0$. \textit{Middle}:  the five Keplerian parameters of the candidate planet ($T_{0,\mathrm{c}}$, $P_\mathrm{c}$, $K_\mathrm{c}$, $e_\mathrm{c}$, $\omega_\mathrm{c}$), and the RV jitter ($\sigma_\mathrm{SOPHIE}$). \textit{Bottom}: the GP hyperparameters of the stellar magnetic activity  model ($P_\mathrm{rot}$, $\alpha_\mathrm{QP}$, $\lambda_{1,\mathrm{QP}}$, $\lambda_{2,\mathrm{QP}}$), and the short-term stellar noise model ($\alpha_\mathrm{SE}$, $\lambda_\mathrm{SE}$). }
\label{ann_posteriors}  
\end{figure*}

The distribution of pertinent posterior probabilities resulting from these diverse analyses is shown in Fig.~\ref{ann_posteriors}, incorporating both binned and unbinned RV data.

The RV signature of planet c is not well constrained when examining SOPHIE's individual RV campaigns, as their duration closely aligns with the planet orbital period $P_\mathrm{c}$.

We see that, during the second SOPHIE observing campaign (represented by cyan and pink posteriors), the stellar activity signal exhibited a notably higher amplitude (see GP amplitude and Sect.~\ref{App_C1}). As a result, this has been accompanied by larger uncertainties in the planetary parameters (see the enlarged posterior tails in Fig.~\ref{ann_posteriors}). We also note a smaller difference between the parameters derived during the C1 and C2 campaigns when using unbinned RV data (2 GP) compared to binned RV data (1 GP).

We see that a better parameter convergence is consistently observed in both campaign when utilizing the nonbinned dataset (comparing blue to magenta; or cyan to pink).

Ultimately, we present the RMS of the RV residuals associated with the various analyses in Table \ref{tab_RMS_res}. 
Notably, the application of the two-GP noise models on the unbinned RV dataset considerably improves the quality of our noise modeling and leads to similar residual RMS for both campaigns; while larger differences are observed on the RMS of the binned RV dataset between C1 and C2.

\def\arraystretch{1.1}
\begin{table}[h]
\caption{RMS values of the SOPHIE RV residuals in m/s.}
\begin{center} \centering
\begin{tabular}{|c|c|c|c|}
\hline
                       & C1  & C2 & C1$+$C2 \\
\hline
Binned RV       & 3.38          &  4.62      &  4.40 \\
Non-binned RV   & 1.82          &  1.63      &  2.10 \\
\hline
\end{tabular}
\vspace{0.2cm}\\
\footnotesize 
Notes. As a reference, the RMS of the data residuals found during the joint analysis in Sect.~\ref{sec_combined} was $1.79$ m/s. 
\vspace{-0.1cm}
\label{tab_RMS_res}
\end{center}
\end{table}

% =================================
\section{List of the main physical parameters of the HD 73344 planetary system}
\label{asec:main_table}

\begingroup
\renewcommand{\arraystretch}{1.25} % Default value: 1

\begin{table*}[t] 
\centering
\caption{Main physical and orbital parameters of the HD 73344 planetary system derived from the joint analysis of the photometric and spectroscopic data, and stellar evolution tracks.  The median values and $68.3\%$ credible interval are reported in the last column.}
% ==============
\begin{tabular}{l c c c}
\label{tab_transit_RV} 
Parameters & Units & Prior  & Posterior \\
\noalign{\smallskip}\hline\hline\noalign{\smallskip}
%%%%%%%%%%%% Fitted parameters
\textit{Fitted parameters}  &   &   &    \\
\hline
%%%%%%%%%%%% Host star
$T_\mathrm{eff}$          &  K &  $\mathcal{N}(6220,64)$ & $6252.602^{+1.869}_{-1.507}$  \\
${\rm log}g$              & cgs & $\mathcal{N}(4.39, 0.02)$  & $4.372^{+0.013}_{-0.014}$ \\
$\rho_\star$              & $\rho_\odot$  & $\mathcal{U}(0,10)$ & $0.719^{+0.033}_{-0.033}$  \\
$\rm [Fe/H]$              & dex  & $\mathcal{N}(0.18, 0.043)$ & $0.143^{+0.027}_{-0.015}$ \\
Age                       & Gyrs  & $\mathcal{U}(0,100)$ &  $1.150^{+0.300}_{-0.326}$ \\
d                         & pc   & $\mathcal{AN}(35.2093, 0.0361, 0.0718)$ & $35.193^{+0.023}_{-0.014}$ \\
E(B-V)                    & mag     & $\mathcal{U}(0,1)$    & $0.170^{+0.034}_{-0.097}$ \\
\hline
%%%%%%%%%%%% Planet b
$P_\mathrm{b}$                     &  days &  $\mathcal{N}(15.612,0.08)$ & $15.61100^{+0.00003}_{-0.00003}$ \\
$T_{0, \mathrm{b}}$                & BJD-2454833 & $\mathcal{N}(3262.854,0.2)$  & $3262.900^{+0.003}_{-0.003}$ \\
$R_\mathrm{b}/R_\star$           & $\%$ & $\mathcal{U}(0,0.1)$ &  $2.217^{+0.042}_{-0.046}$   \\
$i_\mathrm{b}$                     & deg  & $\mathrm{Sine}(80,90)$ & $88.082^{+0.051}_{-0.056}$  \\
$K_\mathrm{b}^\dagger$             & m/s  & $\mathcal{U}(0,100)$ &   $0.667^{+0.559}_{-0.426}$ ($<2.34$)\\
$e_\mathrm{b}$                     &  --    & $\mathcal{TN}(0,0.83)$ & $0.030^{+0.019}_{-0.013}$ \\
$\omega_\mathrm{b}$                &  deg   & $\mathcal{U}(0,360)$ & $153.891^{+119.020}_{-90.006}$ \\
\hline
%%%%%%%%%%%% 
$P_\mathrm{c}$                     & days  & $\mathcal{U}(50,90)$ &  $66.456^{+0.100}_{-0.250}$ \\
$T_\mathrm{p, c}$                & BJD-2454833 & $\mathcal{U}(3593,3753)$ & $3651.901^{+0.701}_{-1.906}$ \\
$K_\mathrm{c}$                     & m/s  & $\mathcal{U}(0,100)$ &  $16.070^{+1.775}_{-1.790}$ \\
$e_\mathrm{c}$                     &  -- & $\mathcal{U}(0,1.0)$ & $0.061^{+0.021}_{-0.026}$ \\
$\omega_\mathrm{c}$                &  deg  & $\mathcal{U}(0,360)$ & $276.633^{+4.891}_{-6.392}$  \\
\hline
%%%%%%%%%%%% Stellar activity 
$P_\mathrm{rot}$                 & days  & $\mathcal{N}(9,2)$ & $9.088^{+0.040}_{-0.024}$  \\
$\alpha_\mathrm{QP}$             & m/s  & $\mathcal{U}(0,100)$ & $11.802^{+1.294}_{-1.177}$ \\
$\lambda_{1,\mathrm{QP}}$          & days  & $\mathcal{U}(0,100)$ & $18.992^{+0.461}_{-0.603}$  \\
$\lambda_{2,\mathrm{QP}}$          &  --     & $\mathcal{U}(0,5)$ & $0.638^{+0.098}_{-0.049}$\\
$\alpha_\mathrm{SE}$             &  m/s  & $\mathcal{N}(12.8,6.0)$ & $4.773^{+0.495}_{-0.489}$  \\
$\lambda_\mathrm{SE}$            & hours  & $\mathcal{N}(2.4,0.7)$ & $3.997^{+0.383}_{-0.476}$  \\
\hline
%%%%%%%%%%%% Instrumental
$\sigma_\mathrm{SOPHIE}$         & m/s  & $\mathcal{U}(0,100)$ & $2.18 \pm 0.30$ \\
$\sigma_\mathrm{HIRES}$          & m/s  & $\mathcal{U}(0,100)$ & $0.85 \pm 0.45$  \\
$V_0$                       &  km/s   & $\mathcal{U}(0,360)$ & $6.244 \pm 0.002$  \\
\hline
\vspace{-0.1cm}\\
%%%%%%%%%%% Derived parameters
\textit{Derived parameters}  &   &   &\\
\hline
%%%%%%%%%%%% 
$u_\mathrm{a, K2}$                    &    --   &     & $0.329^{+0.0015}_{-0.0007}$ \\ 
$u_\mathrm{b, K2}$                    &     --  &     & $0.303^{+0.00009}_{-0.0001}$ \\
$u_\mathrm{a, TESS}$                    &    --   &     & $0.247^{+0.001}_{-0.0004}$ \\
$u_\mathrm{b, TESS}$                    &    --   &     & $0.306^{+0.0004}_{-0.0004}$ \\
\hline
%%%%%%%%%%%% 
$b_\mathrm{b}$                    &   --  &   & $0.783^{+0.017}_{-0.019}$ \\
$a_\mathrm{b}/R_\star$            &   --  &   & $23.556^{+0.359}_{-0.366}$ \\
$T_{dur, b}$             & hours & & $3.304^{+0.047}_{-0.048}$ \\
$a_\mathrm{b}$                    & AU  &  &$0.131^{+0.0003}_{-0.0002}$ \\
$R_\mathrm{b}$                    & $R_\oplus$  &  & $2.884^{+0.082}_{-0.072}$ \\
$M_\mathrm{b}^\dagger$            & $M_\oplus$  &  & $2.983^{+2.50}_{-1.90}$ ($<10.48$) \\
$\rho_\mathrm{b}^\dagger$         &g/cm$^3$  &  & $0.681^{+0.590}_{-0.438}$ ($<2.451$) \\
$T_\mathrm{eq,b}$               & K  &  & $911 \pm 7$\\
$T_\mathrm{lock,b}$               & K  &  & $1066^{+15}_{-12}$\\
\hline
%%%%%%%%%%%% 
$a_\mathrm{c}$                    & AU  &  & $0.343^{+0.0009}_{-0.0006}$ \\
$M_\mathrm{c} \sin(i_\mathrm{c})$             & $M_\oplus$  &  & $116.3^{+12.8}_{-13.0}$ \\
$T_\mathrm{eq,c}$               & K &  & $562 \pm 4$ \\
%%%%%%%%%%%\noalign{\smallskip}\hline\noalign{\smallskip}
\end{tabular}
\parbox{7in}{
}
\tablefoot{
We only list the parameters that are relevant to follow-up analyses. However, the joint analysis involves $75$ free parameters. The priors of all these additional parameters were taken as non informative. 
We assumed $R_\odot=695~508$ km, $M_\odot=1.98842 \times 10^{30}$ kg, $R_\oplus = 6378$ km, $M_\oplus=5.9736\times10^{24}$ kg, and $1$ AU $=149~597~870$ km. Temperature $T_\mathrm{eq}$ was derived assuming a null albedo, and $T_\mathrm{lock}$ assuming tidally synchronized rotation. Symbol $\dagger$ indicates that the $99\%$ confidence interval is also given into parentheses. \\
Notation: $\mathcal{N}(\mu,\sigma)$ refers to a Gaussian distribution with mean $\mu$ and standard deviation $\sigma$; $\mathcal{TN}(\mu,\sigma)$  to a truncated-normal distribution; $\mathcal{AN}(\mu,\sigma_-, \sigma_+)$  to an asymmetric normal distribution with asymmetric width $\sigma_-$/$\sigma_+$; $\mathcal{U}(a,b)$ to a uniform distribution between $[a,b]$; and $\mathrm{Sine}(a,b)$ to a sinusoidal distribution between $a$ and $b$.
}
\end{table*}
\endgroup

% =================================

%--------------------------------------

\section{Posterior distribution of the stability coefficient}
\label{asec:stability}

\begin{figure}
   \includegraphics[width=\columnwidth]{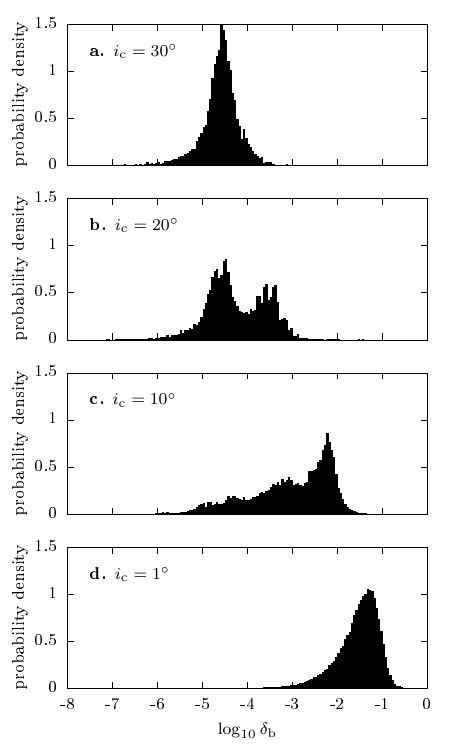}
   \caption{Posterior distribution of the stability coefficient $\delta_\mathrm{b}$ coming from the joint fit (radial velocity and transit data). Different values for the inclination of planet~c in the sky plane are assumed (see labels). The corresponding mass of planet~c ranges from $0.7$~$M_\mathrm{J}$ for panel~{\bf a} to $20$~$M_\mathrm{J}$ for panel~{\bf d}.}
   \label{fig:stabilityhistogram}
\end{figure}

In Sect.~\ref{sec:stability}, we quantify the dynamical stability of the HD\,73344 system by using the stability coefficient $\delta_\mathrm{b}$. As noted by \cite{Stalport-etal_2022}, an instructive insight of the data can be obtained by computing the distribution of $\delta_\mathrm{b}$ that corresponds to the posterior distribution of the parameters. To this aim, we use the MCMC posterior sample coming from a joint fit (radial velocity and transit data; see Sect.~\ref{sec_combined}), made of $10^5$ realizations of the system\footnote{We used here a more conservative version of the fit presented in Sect.~\ref{sec_combined} in which the stellar parameters have slightly larger uncertainties.}. The histogram of $\delta_\mathrm{b}$ drawn from these realizations is presented in Fig.~\ref{fig:stabilityhistogram} for different values of the inclination of planet~c. Histograms for values $30^\circ<i_\mathrm{c}<88^\circ$ are not shown; they all present a single peak at $\log_{10}\delta_\mathrm{b}\approx -5$ or below. When we decrease $i_\mathrm{c}$ below $30^\circ$, however, Fig.~\ref{fig:stabilityhistogram} shows that the distribution gradually transitions from a single stable population (for $i_\mathrm{c}\gtrsim 30^\circ$) to a single highly unstable population (for $i_\mathrm{c}\lesssim 5^\circ$). In between, the population contains both stable, metastable, and unstable subsamples, visible as the multiple bumps in the histogram.

\section{Inferring the bulk H$_2$O and H$_2$-He content using interior structure models} \label{asec:interior}

To infer the bulk composition of HD 73344b, we used the open source tool \texttt{SMINT}\footnote{\url{https://github.com/cpiaulet/smint}} \citepads{Piaulet2021}. This tool performs a MCMC retrieval on grids of interior structure models existing in the literature, to infer the composition of a planet based on its physical properties. We considered two possible compositions for the interior: i) an Earth-like core with a H\textsubscript{2}-He envelope of solar metallicity \citepads{Lopez2014}, and ii) a refractory core with a variable core mass fraction and a pure H\textsubscript{2}O envelope and atmosphere on top \citepads{2021ApJ...914...84A}. These compositions represent end-member cases between an envelope that would form with a Sun-like composition, and a H\textsubscript{2}-He free envelope. Our goal was to determine the range of possible bulk volatile contents in HD 73344b.

\begin{figure}
   \includegraphics[width=\columnwidth]{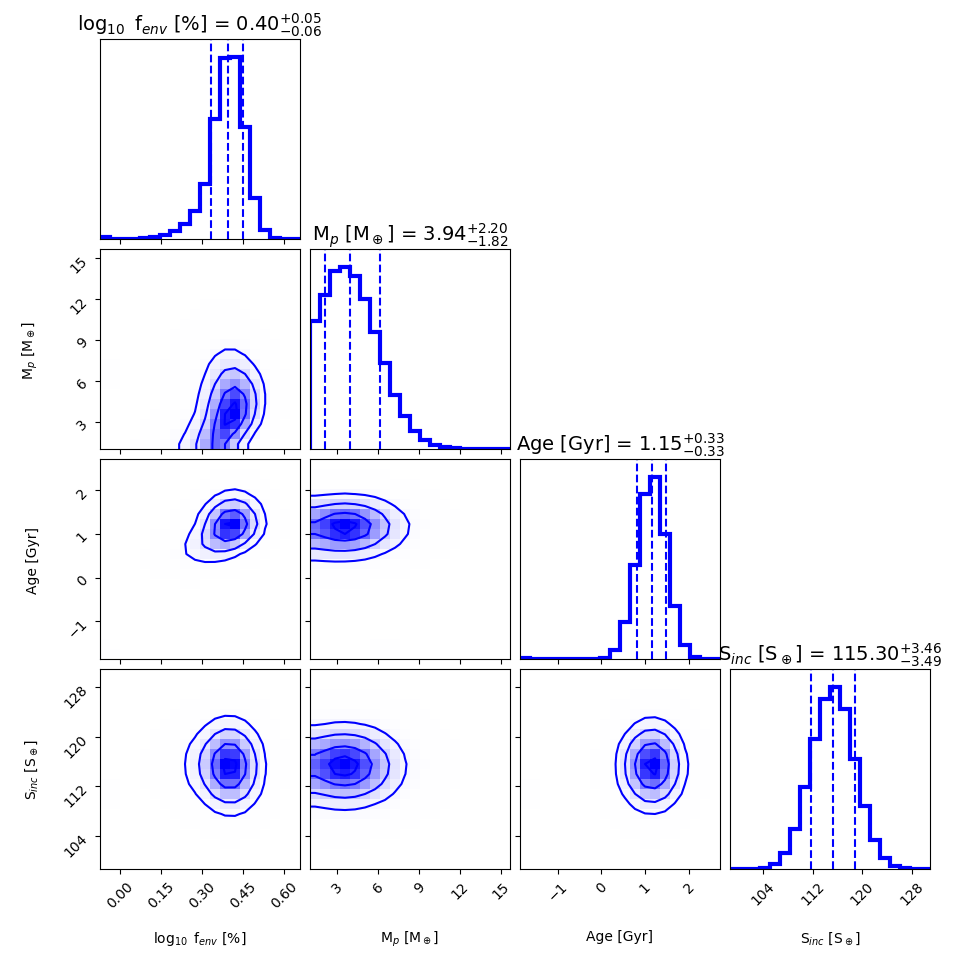}
   \includegraphics[width=\columnwidth]{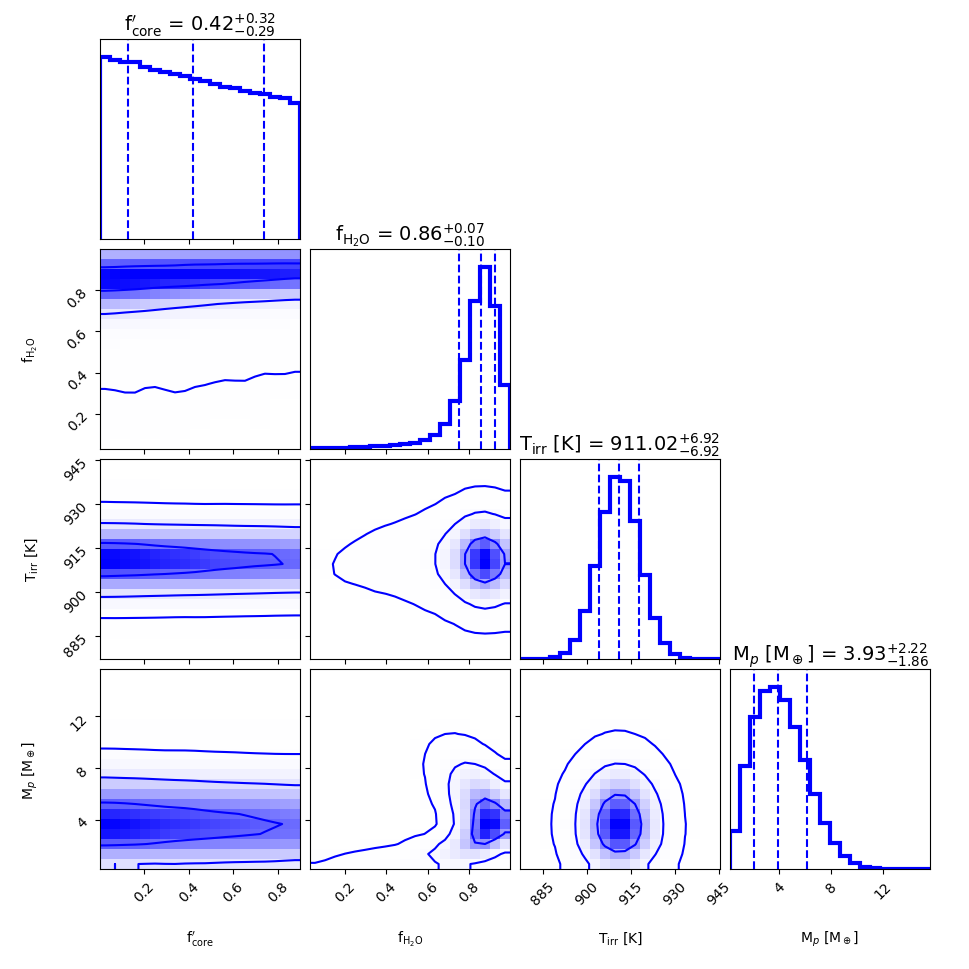}
   \caption{Results of the \texttt{SMINT} tool on parameters of HD 73344b. Top panel corresponds to the H\textsubscript{2}-He envelope case \citepads{Lopez2014}. Bottom panel corresponds to the pure \ce{H2O} envelope case \citepads{2021ApJ...914...84A}.}
   \label{fig:smint-results}
\end{figure}

For the H\textsubscript{2}-He case, the MCMC takes as input Gaussian priors on planet mass, planet radius, incident stellar flux, and age, and produces as output the posterior for the envelope mass fraction $f_{\mathrm{env}}$ that best matches the radius. For the pure H\textsubscript{2}O case, the inputs are planet mass, planet radius, and equilibrium temperature, and the outputs are the water mass fraction $f_{\ce{H2O}}$, and the composition of the core $f^\prime_{\mathrm{core}}$. We found that the properties of HD 73344b are compatible with an interior where $f_{\mathrm{env}} = 2.5\pm 0.3\%$ or $f_{\ce{H2O}} = 86^{+7}_{-10}\%$. The posteriors on all parameters for the two cases are shown in Figure \ref{fig:smint-results}. We tested the case of an envelope with a metallicity 50 times solar \citep[also from][]{Lopez2014}, and found a smaller value $f_{\mathrm{env, 50}} = 2.1\pm 0.4\%$. This is likely due to the fact that higher metallicity planets have greater atmospheric opacity, and therefore cool down (and contract) slower than solar metallicity planets. We also tested the pure water case where we fixed composition of the core to the Earth value $f^\prime_{\mathrm{core}}=0.325$. The results were extremely similar, owing to the fact that the core represents only $\sim 20\%$ of the planet mass (and, consequently, radius), so that its composition has a marginal impact on the total planet radius. We noticed that in all our cases, the posterior on the mass is centered at a mass of $\sim 4~M_\oplus$ instead of the measured $\sim 3~M_\oplus$. This is very likely due to the Gaussian prior on the mass, which would allow masses of $<1~M_\oplus$ at 1-$\sigma$ (limit on the validity range of interior models), and even negative masses at 2-$\sigma$. Such values are discarded from the fit, favoring the higher-end distribution of masses and pushing the mean mass to a slightly greater value. This implies that the volatile content is slightly underestimated. Therefore, from this analysis we conclude that the volatile content of HD 73344b would be $>75\%$ if it was water, and $2-3\%$ if it was gas of solar composition.

 \end{appendix}
 
\end{document}